\documentclass[11pt]{article} 
\usepackage{latexsym}

\def\hybrid{\topmargin -25pt    \oddsidemargin 0in %0pt
	\headheight 0pt \headsep 0pt
	\textwidth 6.5in       % A4 paper
	\textheight 9.5in       % A4 paper
	\marginparwidth .875in
	\parskip 5pt plus 1pt   \jot = 1.5ex}
\hybrid
\def\cQ{{\cal Q}}
\def\cG{{\cal G}}
\def\cL{{\cal L}}

\def\cM{{\cal M}}
\def\cH{{\cal H}}
\def\ket#1{|{#1}\rangle}
\def\noi{\noindent}
\def\half{{1\over2}}
\def\baselinestretch{1.2}

\catcode`\@=11

\def\marginnote#1{}
\def\draftlabel#1{{\@bsphack\if@filesw {\let\thepage\relax
   \xdef\@gtempa{\write\@auxout{\string
      \newlabel{#1}{{\@currentlabel}{\thepage}}}}}\@gtempa
   \if@nobreak \ifvmode\nobreak\fi\fi\fi\@esphack}
	\gdef\@eqnlabel{#1}}
\def\@eqnlabel{}
\def\@vacuum{}
\def\draftmarginnote#1{\marginpar{\raggedright\scriptsize\tt#1}}

\def\draft{\oddsidemargin -.2truein
	\def\@oddfoot{\sl preliminary draft \hfil
	\rm\thepage\hfil\sl\today\quad\militarytime}
	\let\@evenfoot\@oddfoot \overfullrule 3pt
	\let\label=\draftlabel
	\let\marginnote=\draftmarginnote
   \def\@eqnnum{(\theequation)\rlap{\kern\marginparsep\tt\@eqnlabel}%
\global\let\@eqnlabel\@vacuum}  }

\def\preprint{\twocolumn\sloppy\flushbottom\parindent 2em
	\leftmargini 2em\leftmarginv .5em\leftmarginvi .5em
	\oddsidemargin -.5in    \evensidemargin -.5in
	\columnsep .4in \footheight 0pt
	\textwidth 10.in        \topmargin  -.4in
	\headheight 12pt \topskip .4in
	\textheight 6.9in \footskip 0pt
	\def\@oddhead{\thepage\hfil\addtocounter{page}{1}\thepage}
	\let\@evenhead\@oddhead \def\@oddfoot{} \def\@evenfoot{} }

\def\numberbysection{\@addtoreset{equation}{section}
	\def\theequation{\thesection.\arabic{equation}}}

\def\underline#1{\relax\ifmmode\@@underline#1\else
	$\@@underline{\hbox{#1}}$\relax\fi}

\def\titlepage{\@restonecolfalse\if@twocolumn\@restonecoltrue
\onecolumn
     \else \newpage \fi \thispagestyle{empty}\c@page\z@
	\def\thefootnote{\fnsymbol{footnote}} }

\def\endtitlepage{\if@restonecol\twocolumn \else \newpage \fi
	\def\thefootnote{\arabic{footnote}}
	\setcounter{footnote}{0}}  %\c@footnote\z@ }

\catcode`@=12
\relax

\def\figcap{\section*{Figure Captions\markboth
	{FIGURECAPTIONS}{FIGURECAPTIONS}}\list
	{Figure \arabic{enumi}:\hfill}{\settowidth\labelwidth{Figure
999:}
	\leftmargin\labelwidth
	\advance\leftmargin\labelsep\usecounter{enumi}}}
 \relax
\def\tablecap{\section*{Table Captions\markboth
	{TABLECAPTIONS}{TABLECAPTIONS}}\list
	{Table \arabic{enumi}:\hfill}{\settowidth\labelwidth{Table
999:}
	\leftmargin\labelwidth
	\advance\leftmargin\labelsep\usecounter{enumi}}}
 \relax
\def\reflist{\section*{References\markboth
	{REFLIST}{REFLIST}}\list
	{[\arabic{enumi}]\hfill}{\settowidth\labelwidth{[999]}
	\leftmargin\labelwidth
	\advance\leftmargin\labelsep\usecounter{enumi}}}
 \relax
\makeatletter
\newcounter{pubctr}
\def\publist{\@ifnextchar[{\@publist}{\@@publist}}
\def\@publist[#1]{\list
	{[\arabic{pubctr}]\hfill}{\settowidth\labelwidth{[999]}
	\leftmargin\labelwidth
	\advance\leftmargin\labelsep
	\@nmbrlisttrue\def\@listctr{pubctr}
	\setcounter{pubctr}{#1}\addtocounter{pubctr}{-1}}}
\def\@@publist{\list
	{[\arabic{pubctr}]\hfill}{\settowidth\labelwidth{[999]}
	\leftmargin\labelwidth
	\advance\leftmargin\labelsep
	\@nmbrlisttrue\def\@listctr{pubctr}}}
 \relax
\makeatother
\newskip\humongous \humongous=0pt plus 1000pt minus 1000pt

\newif\ifdtup

\font\Scbig=cmss10 scaled\magstep1
\font\Scscr=cmss8 scaled\magstep1
\font\Scscrscr=cmss8
\newfam\Scfam
\textfont\Scfam=\Scbig
\scriptfont\Scfam=\Scscr
\scriptscriptfont\Scfam=\Scscrscr
\def\Sc{\fam\Scfam}

\relax
\def\lvm{\leavevmode\hbox to\parindent{\hfill}}

\def\BE{\begin{equation}}
\def\EE{\end{equation}}
\def\BA{\begin{eqnarray}}
\def\EA{\end{eqnarray}}

\def\D{\Delta}

\def\th{\theta}

\def\tt{\bar\tau}

\def\lvm{\leavevmode\hbox to\parindent{\hfill}}
\def\bar{\overline}
\def\req#1{(\ref{#1})}

\def\L{\left}
\def\R{\right}

\def\BE{\begin{equation}}
\def\EE{\end{equation} \vskip 0.30\baselineskip}
\def\BA{\begin{array}}
\def\EA{\end{array}}

\def\noi{\noindent}

\def\frac#1#2{{\textstyle{{#1}\over{#2}}}}
\def\half{{1\over2}}

\def\Kr#1{\delta_{{#1},0}}
\def\ket#1{|{#1}\rangle}

\def\cA{{\cal A}}
\def\cG{{\cal G}}
\def\cH{{\cal H}}

\def\cL{{\cal L}}

\def\cQ{{\cal Q}}

\def\cU{{\cal U}}

\def\open#1{\mbox{{\bf{#1}}}}

\def\oZ{{\open Z}}

\def\ctop{{\Sc c}}
\def\htop{{\Sc h}}

\def\ie{{\it i.e.}}
\def\svec{singular vector}

\def\Qz{\cQ_0}
\def\Gz{\cG_0}
\def\Qn{$\Qz$}
\def\Gn{$\Gz$}
\def\kc{{\ket{\chi}}}
\def\kcc#1#2#3{{\kc_{#1}^{({#2}){#3}}}}

\newif\ifold \oldtrue \def\new{\oldfalse}
\let\ssection=\section
\def\section{\setcounter{equation}{0}\ssection}

\begin{document}
\renewcommand{\theequation}{\thesection.\arabic{equation}}
\newcommand{\beq}{\begin{equation}}
\newcommand{\eeq}[1]{\label{#1}\end{equation}}
\newcommand{\bea}{\begin{eqnarray}}
\newcommand{\eea}{\end{eqnarray}}
\newcommand{\eer}[1]{\label{#1}\end{eqnarray}}
\begin{titlepage}
\begin{center}

\hfill DAMTP-99-32\\ 
\hfill HUTP-98/A055\\
\hfill IMAFF-FM-99/07\\
\hfill  NIKHEF-99-004\\
\hfill hep-th/9905063
\vskip .4in

{\large \bf Determinant Formula for the Topological N=2 Superconformal 
Algebra}                        
\vskip .4in

{\bf Matthias D\"orrzapf}$^{a}$ {\bf and Beatriz Gato-Rivera}$^{b,c}$ \\
\vskip .3in

${ }^a${\em Department of Applied Mathematics and Theoretical Physics \\ 
University of Cambridge, Silver Street, Cambridge CB3 9EW, UK} 

${ }^b${\em Instituto de Matem\'aticas y F\'\i sica Fundamental, CSIC \\
 Serrano 123, Madrid 28006, Spain} \footnote{e-mail addresses:
M.Doerrzapf@damtp.cam.ac.uk, {\ } bgato@imaff.cfmac.csic.es}\\

${ }^c${\em NIKHEF-H, Kruislaan 409, NL-1098 SJ Amsterdam, The Netherlands}\\

\vskip 1.0in

\end{center}

\begin{center} {\bf ABSTRACT } \end{center}
\begin{quotation}
The Kac determinant for the Topological N=2 superconformal algebra is presented 
as well as a detailed analysis of the singular vectors detected by the roots of 
the determinants. In addition we identify the standard Verma modules containing
`no-label' singular vectors (which are not detected directly by the roots of
the determinants). We show that in standard Verma modules there are (at least) 
four different types of submodules, regarding size and shape. We also review 
the chiral determinant formula, for chiral Verma modules, adding new insights.
Finally we transfer the results obtained to the Verma modules and singular
vectors of the Ramond N=2 algebra, which have been very poorly studied so far. 
This work clarifies several misconceptions and confusing claims appeared in the 
literature about the singular vectors, Verma modules and submodules of the 
Topological N=2 superconformal algebra. 

\end{quotation}
\vskip 1.2cm

May 1999
\end{titlepage}

\def\baselinestretch{1.2}
\baselineskip 16 pt
\section{Introduction and Notation}\lvm

The N=2 superconformal algebras provide the symmetries underlying the
N=2 strings \cite{Ade}\cite{Marcus}. These seem to be related to M-theory 
since many of the basic objects of M-theory are realized in the 
heterotic (2,1) N=2 strings \cite{Marti}. In addition, the topological
version of the algebra is realized in the world-sheet of the bosonic
string \cite{BeSe}, as well as in the world-sheet of the superstrings
\cite{BLNW}.

The Topological N=2 superconformal algebra was written down in 1990 by
Dijkgraaf, Verlinde and Verlinde \cite{DVV}, as the symmetry algebra 
of two-dimensional topological conformal field theory (TCFT). 
As the authors realized, this algebra can also be obtained by twisting
the Neveu-Schwarz N=2 superconformal algebra following the result
\cite{EY}\cite{W-top} that the modification of the stress-energy tensor of a 
N=2 superconformal theory, by adding the derivative of the U(1) current,
leads to a topological theory, procedure known as {\it topological twist}.

A basic tool in the representation theory of infinite dimensional Lie
algebras (and superalgebras) is the determinant formula. This is the
determinant of the matrix of inner products of a basis for the
Verma modules. The zeroes of the determinant formula, which in
the case of the N=2 superconformal algebras are given by vanishing
surfaces involving the conformal weight $\D$, the U(1) charge
$\htop$, and the conformal anomaly $\ctop$, correspond to the 
Verma modules which contain null vectors. The vanishing surfaces
also indicate the borders between regions of positive and negative
signature of the metric and therefore the determinant formula is
crucial to investigate the unitarity and non-unitarity of the
representations \cite{BFK}\cite{Yu}. 

In the middle eighties the determinant formulae for the known N=2 
superconformal algebras; \ie\ the antiperiodic Neveu-Schwarz algebra, 
the periodic Ramond algebra and the twisted algebra, were computed 
by several authors \cite{BFK}\cite{Nam}\cite{KaMa3} (see also 
ref. \cite{Yu}). At that time the 
Topological N=2 algebra had not been discovered yet and the corresponding 
determinant formula has remained so far unpublished\footnote{The 
Neveu-Schwarz, the Ramond and the Topological N=2 superconformal algebras  
are connected by the spectral flows and/or the topological twists. 
However their representation theories are different, what 
has created some confusion in the literature. We thank V. Kac for
discussions on this point}. This has not 
prevented, however, from making substantial progress in the study of the 
topological \svec s. For example, it has been shown \cite{BJI6}\cite{DB2} 
that they can be classified in 29 types in complete Verma modules and 
4 types in chiral Verma modules, taking into account the relative U(1) 
charge and the BRST-invariance properties of the vector itself and of the 
primary on which it is built. In ref. \cite{BJI6} the complete set of 
topological singular vectors was explicitely constructed at level 1,
(28 types in complete Verma modules as one type exists only at level 0),
whereas the rigorous proofs that these types are the only 
possible ones have been given in ref. \cite{DB2} together with the maximal
dimensions of the corresponding singular vector spaces (1, 2, or 3 depending
on the type of singular vector). Furthermore 16 types of topological 
\svec s can be mapped \cite{BJI6}\cite{B2} to the standard \svec s of the 
Neveu-Schwarz N=2 algebra, for which construction formulae are known
\cite{Doerr1}\cite{Doerr2}.

Two years ago the chiral N=2 determinant formulae for chiral 
Verma modules were computed \cite{BJI5} for the Neveu-Schwarz,
the Ramond, and the Topological N=2 algebras. As a bonus subsingular
vectors were discovered for these algebras. The reason is that
chiral Verma modules are incomplete
modules resulting from the quotient of a complete Verma module by 
the submodule generated by a lowest-level singular vector (level zero
for the Topological and Ramond algebras and level 1/2 for the
Neveu-Schwarz algebra). As a consequence, the singular vectors 
in the chiral Verma modules are either the ``surviving" \svec s which 
were not descendant of the lowest-level \svec s that are quotiented out, 
or they are subsingular vectors in the complete Verma module becoming 
singular just in the chiral Verma module. Both origins were traced back 
for the 4 types of \svec s found in chiral Verma modules 
\cite{BJI5}\cite{BJI6}. In ref. \cite{DB2} it was proved that these
four types are the only existing types of \svec s in chiral Verma modules
and their corresponding spaces are always one-dimensional.

In this paper we intend to finish the picture presenting the determinant 
formula for the complete Verma modules of the Topological N=2 algebra. In 
section 2 we review the most basic results regarding the Verma modules
and singular vectors of this algebra. In section 3 we present the 
determinant formulae for the generic (standard) Verma modules and for the
`no-label' Verma modules
and discuss in very much detail the types of \svec s 
detected by the roots of the determinants. In addition we review the chiral 
determinant formula corresponding to chiral Verma modules adding some new
insights. The results obtained are transferred to the Verma modules and 
\svec s of the Ramond N=2 algebra, which have been very poorly analysed
in the literature. Some final remarks are made in section 4. In Appendix A
we identify the generic Verma modules with chiral and no-label 
\svec s, and we give some examples. The latter are not directly 
detected by the roots of the determinants. In Appendix B we write down all 
\svec s at level 1 in Verma modules with zero conformal weight $\D=0$.

\vskip .35in
\noi
{\bf Notation}
\vskip .17in
\noi
{\it Highest weight (h.w.) vectors} denote the states with lowest 
conformal weight. They are necessarily annihilated by all
the positive modes of the generators of the algebra, \ie\
${\ } \cL_{n \geq 1} \kc =  \cH_{n \geq 1} \kc =  {\cG}_{n \geq 1} \kc
=  {\cQ}_{n \geq 1} \kc = 0 {\ }$.

\noi
{\it Primary states} denote h.w. vectors with non-zero norm. 

\noi
{\it Secondary or descendant states} denote states obtained by acting on 
the h.w. vectors with the negative modes of the generators of the algebra
and with the fermionic zero modes \Qn\ and \Gn\ . The fermionic zero 
modes can also interpolate between two h.w. vectors which are on the same
footing (two primary states or two singular vectors).

\noi
{\it Chiral states} $\kc^{G,Q}$ are states
annihilated by both $\cG_0$ and $\cQ_0$. 

\noi
{\it $\cG_0$-closed states} $\kc^G$ are  
states annihilated by $\cG_0$ but not by \Qn\ . 

\noi
{\it $\cQ_0$-closed states} $\kc^Q$ are  
states annihilated by $\cQ_0$ but not by \Gn\ .

\noi
{\it No-label states} $\kc$ denote states  
that cannot be expressed as linear combinations of $\cG_0$-closed,
$\cQ_0$-closed and chiral states.

\noi
{\it The Verma module} associated to a h.w. vector consists of the h.w. 
vector plus the set of secondary states built on it. 

\noi
{\it Null vectors} are states in the kernel of the inner product
which therefore decouple from the whole space of states. 

\noi
{\it Singular vectors} are h.w. null vectors, \ie\ the states with
lowest conformal weight in the null submodules.

\noi
{\it Primitive singular vectors} are the singular vectors that cannot be 
constructed by acting with the generators of the algebra on another 
singular vector. However, the fermionic zero modes \Gn\ and \Qn\
can interpolate between two primitive \svec s at the same level  
(transforming one into each other). 

\noi
{\it Secondary singular vectors} are singular vectors that can be  
constructed by acting with the generators of the algebra on another singular
vector. The level-zero secondary singular vectors cannot `come back' to 
the singular vectors on which they are built by acting with \Gn\ or \Qn\ .

\noi
{\it Subsingular vectors} are non-h.w. null vectors  
 which become singular (\ie\ h.w.) in the quotient 
of the Verma module by a submodule generated by a singular vector.
  
\noi
The singular vectors of the Topological N=2 algebra
will be denoted as {\it topological singular vectors} $\ket{\chi}$.

\noi
The singular vectors of the Ramond N=2 algebra
will be denoted as {\it R singular vectors} $\ket{\chi_R}$.

\section{Verma Modules and Singular Vectors of the Topological N=2 Algebra}

\subsection{The Topological N=2 algebra}\lvm

The two possible topological twists \cite{EY}\cite{W-top} of the  
generators of the Neveu-Schwarz N=2 algebra are:

\BE\new\BA{rclcrcl}
(T_W^{\pm })^{-1}\ \cL^{(\pm)}_m \ T_W^{\pm }&=
&\multicolumn{5}{l}{L_m \pm \half(m+1)H_m\,,}\\
(T_W^{\pm })^{-1}\ \cH^{(\pm)}_m \ T_W^{\pm }&=
&\pm H_m\,,&{}&{}&{}&{}\\
(T_W^{\pm })^{-1}\ \cG^{(\pm)}_m \ T_W^{\pm }&=
&G_{m+\half}^{\pm}\,,&\qquad & (T_W^{\pm })^{-1}\ \cQ_m^{(\pm)}\ T_W^{\pm }&=
&G^{\mp}_{m-\half} \,,\label{twa}\EA\EE

\noi
where $L_m$ and $H_m$ are the bosonic generators corresponding to the
stress-energy tensor (Virasoro generators) and the U(1) current,
respectively, and $G^{\pm}$ are the spin-3/2 fermionic generators.
These twists, which we denote as $T_W^{\pm }$, consist of the modification
of the stress-energy tensor by adding the derivative of the U(1) current.
As a result \cite{DVV} the conformal spins and modes of the 
fermionic fields are also modified in such a manner that the spin-3/2 
generators $G^{\pm}$, with half-integer modes, are traded by spin-1 and 
spin-2 generators $\cQ$ and $\cG$, respectively, with integer modes, the
first ones having the properties of a BRST current\footnote{Let us stress
that the modification of the stress-energy tensor results in the 
modification of the conformal weights and modes of the fermionic fields. 
Therefore there are no spectral flows converting the half-integer modes 
of the Neveu-Schwarz generators into the integer modes of the 
topological generators, as sometimes confused in the literature (the 
spectral flows do not modify the conformal weights).}. Observe that the
twists are mirrored under the interchange $H_m \leftrightarrow -H_m$, 
${\ } G^{+}_r \leftrightarrow G^{-}_r$. 

The Topological N=2 algebra, obtained by twisting in this way  
the Neveu-Schwarz N=2 algebra, reads \cite{DVV}

\BE\new\BA{lclclcl}
\L[\cL_m,\cL_n\R]&=&(m-n)\cL_{m+n}\,,&\qquad&[\cH_m,\cH_n]&=
&{\ctop\over3}m\Kr{m+n}\,,\\
\L[\cL_m,\cG_n\R]&=&(m-n)\cG_{m+n}\,,&\qquad&[\cH_m,\cG_n]&=&\cG_{m+n}\,,
\\
\L[\cL_m,\cQ_n\R]&=&-n\cQ_{m+n}\,,&\qquad&[\cH_m,\cQ_n]&=&-\cQ_{m+n}\,,\\
\L[\cL_m,\cH_n\R]&=&\multicolumn{5}{l}{-n\cH_{m+n}+{\ctop\over6}(m^2+m)
\Kr{m+n}\,,}\\
\L\{\cG_m,\cQ_n\R\}&=&\multicolumn{5}{l}{2\cL_{m+n}-2n\cH_{m+n}+
{\ctop\over3}(m^2+m)\Kr{m+n}\,,}\EA\qquad m,~n\in\oZ\,.\label{topalgebra}
\EE

The eigenvalues of the bosonic zero modes $(\cL_0,\,\cH_0)$ correspond to
the conformal weight and the U(1) charge of the states. These split
conveniently as $(\D+l,\,\htop+q)$ for secondary states, where $l$
and $q$ are the {\it level} and the {\it relative charge} of the state and
$(\D,\,\htop)$ are the conformal weight and the charge of
the primary state on which the secondary is built.
The `topological' central charge $\ctop$ is the central charge 
corresponding to the Neveu-Schwarz N=2 algebra. This algebra is topological 
because the Virasoro generators are BRST-exact, \ie\ can be expressed as
$\cL_m={1\over2}\{\cG_m,\cQ_0\}$, where $\cQ_0$ is the BRST charge. This
implies, as is well known, that the correlators of the fields do not
depend on the two-dimensional metric \cite{NickW}.

An important fact is that the annihilation conditions
$G^{\pm}_{1/2}\, \ket{\chi_{NS}} = 0$ of the Neveu-Schwarz N=2 algebra 
read $\Gz \kc = 0$ after the corresponding twists $T_W^{\pm }$ \req{twa}.
As a result, under $T_W^{+}$ or $T_W^{-}$ any state of the 
Neveu-Schwarz N=2 algebra
annihilated by all the positive modes of the NS generators becomes a state 
of the Topological N=2 algebra annihilated by \Gn\ as well as by all the 
positive modes of the topological generators, as the reader can easily 
verify. Conversely, any topological state annihilated by \Gn\ and by all
the positive modes of the topological generators transforms into a 
NS state annihilated by all the positive modes of the NS generators.
The zero mode \Qn , in turn, corresponds to the negative modes 
$G^{\mp}_{-1/2}$. 
Observe that a topological state not annihilated by \Gn\ is transformed
into a NS state not annihilated by $G^{+}_{1/2}\,$ or by $G^{-}_{1/2}\,$.
Consequently, the Neveu-Schwarz counterpart of the topological primaries 
and singular vectors not annihilated by \Gn\ are not primary states and
singular vectors themselves but rather they are secondary states of no
special type, or they are subsingular vectors. To be precise,
the \Qn-closed topological primaries correspond to NS secondary states
obtained by acting with $G^+_{-1/2}$ or $G^-_{-1/2}$ on the NS primaries,
the \Qn-closed topological \svec s with non-zero conformal weight 
correspond to null descendants of NS singular vectors, and the 
\Qn-closed topological \svec s with zero conformal weight and the no-label 
topological \svec s correspond to NS subsingular vectors \cite{DB1}. As to
the no-label topological primaries, they do not have NS counterpart.

\subsection{Topological Verma modules}

{\bf Highest weight vectors}

\vskip .13in

In a given representation of the Topological N=2 algebra, the primary states,
\ie\ the states with {\it lowest} conformal weight denoted traditionally as
{\it highest} weight vectors, require to be annihilated by all the positive 
modes of the generators (the lowering operators). Therefore the
h.w. conditions can be defined unambiguously as the vanishing conditions:  
${\ } \cL_{n \geq 1} \kc =  \cH_{n \geq 1} \kc =  {\cG}_{n \geq 1} \kc
=  {\cQ}_{n \geq 1} \kc = 0 {\ }$.

The zero modes \Gn\ and \Qn\ 
provide the BRST-invariance properties of the topological states in
the sense that a state annihilated by \Qn\ is BRST-invariant while
a state annihilated by \Gn\ can be regarded as 
anti-BRST-invariant. The states annihilated by both \Gn\
and \Qn\ are called chiral \cite{DVV}, generalizing the fact that 
the chiral and antichiral primaries\footnote{Chiral primary states of the
N=2 superconformal algebras, which were introduced in ref. \cite{LVW},
are of special relevance in physics (see also refs. \cite{DVV} and
\cite{NickW}).} 
of the Neveu-Schwarz N=2 algebra are 
transformed, under the topological twists \req{twa}, into topological 
primaries annihilated by both 
\Gn\ and \Qn : $(G^{\pm}_{1/2},\,G^{\mp}_{-1/2}) \to (\Gz,\,\Qz)$.
In what follows the states annihilated by \Qn\ but not by \Gn\ 
will be called \Qn-closed whereas the states annihilated by 
\Gn\ but not by \Qn\ will be called \Gn-closed.

{\ } From the anticommutator $\{ \cQ_0, \cG_0\} = 2 \cL_0 $ one deduces 
\cite{BJI6} that a topological state (primary or secondary)
 with non-zero conformal weight $\D$ can be either
 \Gn-closed, or \Qn-closed, or a linear combination of both types:
\BE \kc = {1\over 2\D} \Qz \Gz \kc + {1\over 2\D} \Gz \Qz \kc \,. \EE   
One deduces also that \Qn-closed (\Gn-closed) 
topological states with non-zero conformal weight 
 are \Qn-exact (\Gn-exact) as well. The topological
states with zero conformal weight, however, can be \Qn-closed (satisfying 
$\cQ_0\cG_0\kc^Q=0$), or \Gn-closed (satisfying $\cG_0\cQ_0\kc^G=0$), 
or chiral, or no-label (not decomposible into \Gn-closed, \Qn-closed and
chiral states, satisfying $\cQ_0\cG_0\kc=-\cG_0\cQ_0\kc \neq 0$).

Hence one can distinguish three different types of topological
primaries giving rise to complete Verma modules (provided they do not
satisfy additional constraints): \Gn-closed primaries $\ket{\D,\htop}^G$, 
\Qn-closed primaries $\ket{\D,\htop}^Q$, and no-label primaries
$\ket{0,\htop}$. We will not consider primaries $\ket{\D,\htop}$ which are 
linear combinations of two or more of these types. Chiral primaries
$\ket{0,\htop}^{G,Q}$ give rise to incomplete Verma modules because the 
chirality constraint on the primary state is not required 
(just allowed) by the algebra. 

As to the topological secondary states, in particular singular vectors,
they are labelled in addition by the level $l$ and the relative charge $q$.
Hence the topological secondary states are denoted as $\kc_l^{(q)G}$
($\cG_0$-closed), $\kc_l^{(q)Q}$ ($\cQ_0$-closed), 
$\kc_l^{(q)G,Q}$ (chiral), and $\kc_l^{(q)}$ (no-label).
It is convenient also to indicate the conformal weight $\D$,
the charge $\htop$, and the BRST-invariance properties
of the primary state on which the secondary is built. Notice that
the conformal weight and the total U(1) charge of the secondary
states are given by $\D + l$ and $\htop+q$, respectively. 

Now we will give a first description of the different kinds of complete 
Verma modules as well as of the chiral Verma modules. 
   
\vskip .17in
\noi
{\bf Generic Verma modules}

The Verma modules built on $\cG_0$-closed or $\cQ_0$-closed primary states
without additional constraints are called {\it generic} Verma modules
\cite{BJI6}\cite{DB2}. 
They are complete in the sense that the constraints on the primaries 
of being annihilated either by $\cG_0$ or by $\cQ_0$ are required by
the algebra, as we have just discussed. Furthermore for 
non-zero conformal weight $\D\neq0$ the h.w. vector of
any generic Verma module is degenerate, \ie\ there are two primary
states. The reason is that the action of
\Qn\ on $\ket{\D,\htop}^G$ produces another primary state:
$\cQ_0\ket{\D,\htop}^G=\ket{\D,\htop-1}^Q$, and the action of
\Gn\ on $\ket{\D,\htop -1}^Q$ brings the state back to $\ket{\D,\htop}^G$
(up to a constant): $\cG_0\ket{\D,\htop -1}^Q=2 \D \ket{\D,\htop}^G$. 
For $\D=0$, however, $\, \cQ_0\ket{0,\htop}^G$ is not a primary 
state but a level-zero chiral singular vector instead, denoted as
$\kc_{0,\ket{0,\htop}^G}^{(-1)G,Q}$, and similarly 
$\, \cG_0\ket{0,\htop-1}^Q$ is the chiral singular vector
$\kc_{0,\ket{0,\htop-1}^Q}^{(1)G,Q}$. As a consequence, the h.w. vectors
$\ket{0,\htop}^G$ and $\ket{0,\htop-1}^Q$ are located in different
Verma modules $V(\ket{0,\htop}^G)$ and $V(\ket{0,\htop-1}^Q)$, whereas
for $\D \neq 0$
the Verma modules built on $\ket{\D,\htop}^G$ and $\ket{\D,\htop-1}^Q$
coincide: $V(\ket{\D,\htop}^G)=V(\ket{\D,\htop-1}^Q)$ iff $\D \neq 0$.

\vskip .17in
\noi
{\bf No-label Verma modules}

The Verma modules built on no-label primary states are called {\it no-label}
Verma modules $V(\ket{0,\htop})$ \cite{BJI6}\cite{DB2}. 
They are complete, obviously, as the no-label primaries are annihilated 
only by the positive modes of the generators of the algebra. The action
of \Gn\ and \Qn\ on $\ket{0,\htop}$ produce the charged \svec s 
$\cG_0\ket{0,\htop}=\kc_{0,\ket{0,\htop}}^{(1)G}$ and
$\cQ_0\ket{0,\htop}=\kc_{0,\ket{0,\htop}}^{(-1)Q}$, which
cannot `come back' to $\ket{0,\htop}$ as the action of \Qn\ and \Gn ,
respectively, produces an uncharged chiral \svec\ instead:
$\cQ_0\cG_0\ket{0,\htop}= - \cG_0\cQ_0\ket{0,\htop}=
\kc_{0,\ket{0,\htop}}^{(0)G,Q}$. (As a matter of fact, one deduces that
$\cG_0\ket{0,\htop}$ and $\cQ_0\ket{0,\htop}$ are singular vectors
taking into account that $\cG_0\cQ_0\ket{0,\htop}$ is a singular vector).
The level-zero states in a no-label 
Verma module consist therefore of the primary state 
$\ket{0,\htop}$ plus the two charged \svec s $\cG_0\ket{0,\htop}$ and
$\cQ_0\ket{0,\htop}$, and the uncharged chiral \svec\
$\cQ_0\cG_0\ket{0,\htop}$.

\vskip .17in
\noi
{\bf Chiral Verma modules}

The Verma modules built on chiral primary states are called {\it chiral}
Verma modules $V(\ket{0,\htop}^{G,Q})$ \cite{BJI5}\cite{BJI6}\cite{DB2}. 
They are not complete because the primary state being annihilated by
both \Gn\ and \Qn\ amounts to an additional constraint not required
(just allowed) by the algebra. Chiral Verma modules result from the
quotient of generic Verma modules, with $\D=0$, by the submodules generated
by the level-zero \svec s. That is, a chiral Verma module can be expressed
as the quotient $\,V(\ket{0,\htop}^{G,Q})= V(\ket{0,\htop}^G) / 
\Qz \ket{0,\htop}^G \,$ and also can be expressed as the quotient
$\,V(\ket{0,\htop}^{G,Q})= V(\ket{0,\htop}^Q) / \Gz \ket{0,\htop}^Q\,$,
equivalently. Therefore it can be regarded as a complete Verma module 
with a piece `cut off'. The level-zero states in a chiral Verma module 
consist of only the primary state $\ket{0,\htop}^{G,Q}$, obviously.

\subsection{Topological singular vectors}\lvm

The topological \svec s can be classified in 29 different types in
complete Verma modules and 4 different types in chiral Verma modules, 
distinguished by the relative charge $q$ and
the BRST-invariance properties of the \svec s and of the primaries 
on which they are built \cite{BJI6}\cite{DB2}. An important question
is whether the singular vectors with non-zero conformal weight, 
$\D+l \neq 0$, are linear combinations of \Gn-closed and \Qn-closed singular 
vectors. From the anticommutator
$\{\Qz , \Gz \}=2 \cL_0$ one obtains the decomposition

\BE \kc_l= {1\over 2(\D+l)} {\ } \Gz \Qz \, \kc_l + 
   {1\over 2(\D+l)} {\ } \Qz \Gz \, \kc_l = \kc_l^G + \kc_l^Q \,. \EE

\noi 
If $\kc_l$ is a singular vector, \ie\ satisfies the h.w. conditions
${\ } \cL_{n \geq 1} \kc =  \cH_{n \geq 1} \kc =  {\cG}_{n \geq 1} \kc
=  {\cQ}_{n \geq 1} \kc = 0 {\ }$, then $\, \Gz \Qz \kc_l$  
and $\, \Qz \Gz \kc_l$ satisfy the h.w. conditions 
too, as one deduces straightforwardly using the algebra
\req{topalgebra}. Therefore, regarding singular vectors with
non-zero conformal weight, we can restrict ourselves to \Gn-closed and
to \Qn-closed singular vectors. The singular vectors with zero conformal 
weight, $\D+l=0$, can also be chiral or no-label.

In what follows we will see how the different types of \svec s are 
distributed in the generic, no-label and chiral Verma modules.
Then we will discuss the
action of the fermionic zero modes \Gn\ and \Qn\  on the \svec s.

\vskip .17in
\noi
{\bf Generic Verma modules}

The possible existing types of topological singular vectors in generic 
Verma modules are the following \cite{BJI6}\cite{DB2} : 

- Ten types built on $\cG_0$-closed primaries
$\ket{\D,\htop}^G$:

\BE
\begin{tabular}{r|l l l l}
{\ }& $q=-2$ & $q=-1$ & $q=0$ & $q=1$\\
\hline\\
\Gn-closed & $-$ & $\kc_l^{(-1)G}$ & $\kc_l^{(0)G}$ & $\kc_l^{(1)G}$\\
\Qn-closed & $\kc_l^{(-2)Q}$ & $\kc_l^{(-1)Q}$ & $\kc_l^{(0)Q}$ & $-$ \\
chiral & $-$ & $\kc_l^{(-1)G,Q}$ & 
$\kc_l^{(0)G,Q}$ & $-$ \\
no-label & $-$ & $\kc_l^{(-1)}$ &
$\kc_l^{(0)}$ & $-$\\
\end{tabular}
\label{tabl2}
\EE

- Ten types built on $\cQ_0$-closed primaries
$\ket{\D,\htop}^Q$:

\BE
\begin{tabular}{r|l l l l}
{\ }& $q=-1$ & $q=0$ & $q=1$ & $q=2$\\
\hline\\
$\cG_0$-closed & $-$ & $\kc_l^{(0)G}$ & $\kc_l^{(1)G}$ & $\kc_l^{(2)G}$\\
$\cQ_0$-closed & $\kc_l^{(-1)Q}$ & $\kc_l^{(0)Q}$ & $\kc_l^{(1)Q}$ & $-$ \\
chiral & $-$ & $\kc_l^{(0)G,Q}$ & $\kc_l^{(1)G,Q}$ & 
$-$ \\
no-label & $-$ & $\kc_l^{(0)}$ &
$\kc_l^{(1)}$ & $-$\\
\end{tabular}
\label{tabl3}
\EE

\vskip .15in
\noi
with $l=-\D$ in the case of chiral and no-label \svec s.
The maximal dimensions of the corresponding \svec\ spaces \cite{DB2} are
two, for the \svec s of types $\kc_{l,\ket{\D,\htop}^G}^{(0)G}\,$,
$\kc_{l,\ket{\D,\htop}^G}^{(-1)Q}\,$, $\kc_{l,\ket{\D,\htop}^Q}^{(0)Q}\,$
and $\kc_{l,\ket{\D,\htop}^Q}^{(1)G}\,$,
and one for the remaining types\footnote{The maximal dimension $n$ for a 
given \svec\ space puts a theoretical limit on the number of linearly 
independent \svec s of the corresponding type that one can write down 
at the same level in a given Verma module \cite{DB2}. For the \svec s 
given in tables \req{tabl2} and \req{tabl3} the corresponding maximal
dimensions have been verified by the low level computations.}. 
   
For $\D \neq 0$ the singular vectors of table \req{tabl2}
are equivalent to singular vectors of table \req{tabl3} with a shift
on the U(1) charges due to the existence of two primary states in the
Verma module: one \Gn-closed primary $\ket{\D,\htop}^G$ and one
\Qn-closed primary $\ket{\D,\htop-1}^Q$, as was discussed in
subsection 2.2. In particular,
the charged (uncharged) chiral \svec s of table \req{tabl2} are equivalent
to uncharged (charged) chiral \svec s of table \req{tabl3}. 

An useful observation is that chiral singular vectors
$\kc_{l}^{(q)G,Q}\,$ can be regarded as particular cases of \Gn-closed
singular vectors $\kc_{l}^{(q)G}\,$ and/or as particular cases
of \Qn-closed singular vectors $\kc_{l}^{(q)Q}\,$. That is, 
some \Gn-closed and \Qn-closed singular vectors `become' chiral 
when the conformal weight of the singular vector turns out to be zero, 
\ie\ $\D+l=0$. This is always the case for \svec s of types 
$\kc_{l,\ket{\D,\htop}^G}^{(-1)G}\,$, $\kc_{l,\ket{\D,\htop}^G}^{(0)Q}\,$,
$\kc_{l,\ket{\D,\htop}^Q}^{(0)G}\,$ and $\kc_{l,\ket{\D,\htop}^Q}^{(1)Q}\,$, 
as explained in refs. \cite{B2}\cite{DB2} (see below), while this never 
occurs to \svec s of types $\kc_{l,\ket{\D,\htop}^G}^{(1)G}\,$,
$\kc_{l,\ket{\D,\htop}^G}^{(-2)Q}\,$, $\kc_{l,\ket{\D,\htop}^Q}^{(-1)Q}\,$
and $\kc_{l,\ket{\D,\htop}^Q}^{(2)G}\,$,
as there are no chiral \svec s of the corresponding types.

All topological \svec s in tables \req{tabl2} and \req{tabl3} can be 
organized into families \cite{BJI6} involving different Verma modules and
different levels, every member of a family being mapped to any other
member by the {\it topological} spectral flows \cite{BJI3}\cite{B1} and/or
the fermionic zero modes \Gn\ and \Qn . These families follow infinite
many different patterns. Furthermore, 
with the exception of the no-label \svec s, all other 16 types of 
topological \svec s in tables \req{tabl2} and \req{tabl3} can be easily 
mapped to the \svec s of the NS algebra \cite{B2}. 
As a bonus one obtains construction formulae for the 16 types of topological 
singular vectors using the construction formulae for the NS singular vectors
given in refs. \cite{Doerr1}\cite{Doerr2}. 

\vskip .17in
\noi
{\bf No-label Verma modules}

The possible existing types of topological singular vectors in no-label 
Verma modules are the following nine types \cite{BJI6}\cite{DB2} built on
no-label primaries $\ket{0,\htop}$: 

\BE
\begin{tabular}{r|l l l l l}
{\ }& $q=-2$ & $q=-1$ & $q=0$ & $q=1$ & $q=2$\\
\hline\\
$\cG_0$-closed & $-$ & $\kc_l^{(-1)G}$ & $\kc_l^{(0)G}$ & $\kc_l^{(1)G}$ &
$\kc_l^{(2)G}$\\
$\cQ_0$-closed & $\kc_l^{(-2)Q}$ &$\kc_l^{(-1)Q}$ & 
$\kc_l^{(0)Q}$ & $\kc_l^{(1)Q}$ & $-$ \\
chiral & $-$ & $-$ & $\kc_0^{(0)G,Q}$ & $-$ & $-$ \\
\end{tabular}
\label{tabl4}
\EE

\vskip .17in
\noi
The maximal dimensions of the corresponding \svec\ spaces \cite{DB2} are
three\footnote{At levels 1 and 2 there are no 3-dimensional 
singular spaces (there are 2-dimensional ones though). Further 
computations are needed to decide whether or not they actually exist
at higher levels.},
for the \svec s of types $\kc_{l,\ket{0,\htop}}^{(0)G}\,$,
$\kc_{l,\ket{0,\htop}}^{(1)G}\,$, $\kc_{l,\ket{0,\htop}}^{(0)Q}\,$ and
$\kc_{l,\ket{0,\htop}}^{(-1)Q}\,$, and one for the remaining types.
Observe that the chiral type of \svec\ only exists for level zero. It 
is given by $\kc_{0,\ket{0,\htop}}^{(0)G,Q}=\cG_0\cQ_0\ket{0,\htop}$. 
These \svec s can also be organized into families \cite{BJI6} resulting  in
two different kinds of families with a unique pattern each: the box
diagram consisting of four \svec s connected by \Gn , \Qn\ and the 
universal odd spectral flow automorphism\footnote{The topological odd 
spectral flow for $\theta=1$, denoted simply as $\cA$ \cite{BJI3}\cite{B1},
acting as $\cL_m \to \cL_m-m\cH_m$, $\cH_m \to -\cH_m-{\ctop\over3}$,
$\cG_n \to \cQ_n$ and $\cQ_n \to \cG_n$, is 
an involutive automorphism, i.e. $\cA^{-1}=\cA$. It is universal in the 
sense that it transforms every primary state and every singular 
vector back into another primary state and another singular vector. 
In particular it is the only spectral flow operator that maps chiral 
states to chiral states and no-label states to no-label states. Hence
any other spectral flow operators (except the identity) `destroy'
chiral Verma modules as well as no-label Verma modules.} $\cA$.

\vskip .17in
\noi
{\bf Chiral Verma modules}

The possible existing types of topological singular vectors in chiral 
Verma modules are the following four types \cite{BJI6}\cite{DB2} built on
chiral primaries $\ket{0,\htop}^{G,Q}$: 

\BE
\begin{tabular}{r|l l l}
{\ }& $q=-1$ & $q=0$ & $q=1$\\
\hline\\
$\cG_0$-closed & $-$ & $\kc_l^{(0)G}$ & $\kc_l^{(1)G}$ \\
$\cQ_0$-closed & $\kc_l^{(-1)Q}$ & $\kc_l^{(0)Q}$ & $-$ \\
\end{tabular}
\label{tabl1}
\EE

\vskip .17in
\noi
This result was to be expected since a chiral Verma module 
$\,V(\ket{0,\htop}^{G,Q})\,$ can be viewed as a Verma module 
$\,V(\ket{0,\htop}^G)\,$, built on a \Gn-closed primary, with a piece
`cut off', but it can also be viewed as a Verma module 
$\,V(\ket{0,\htop}^Q)\,$, built on a \Qn-closed primary, with a piece
`cut off'. Therefore one might think that the only possible \svec s in
chiral Verma modules should correspond to the types existing in both
table \req{tabl2}, built on \Gn-closed primaries, and  
table \req{tabl3}, built on \Qn-closed primaries, whereas the non-common
types must be projected out belonging to the submodules that are set to
zero. Inspecting tables \req{tabl2} and \req{tabl3} one finds that the
common types of \svec s are precisely the ones given in table \req{tabl1}
plus chiral and no-label \svec s which cannot possibly exist in chiral
Verma modules (one cannot construct chiral or no-label descendant
states on a chiral primary because they would require level zero but
\Gn\ and \Qn\ annihilate the primary). The reasoning we just made is not
completely correct in spite of its apparent success, however, because
in the generic Verma modules $\,V(\ket{0,\htop}^G)\,$ and 
$\,V(\ket{0,\htop}^Q)\,$ there exist {\it subsingular} vectors which
become singular only in the chiral Verma module, \ie\ after setting to zero
the null submodules generated by the level-zero \svec s 
\cite{BJI5}\cite{BJI6}. Nevertheless the resulting types of `new' \svec s 
which are only singular in the chiral Verma modules
are again of the types shown in table \req{tabl1}. 

The maximal dimension of the corresponding \svec\ spaces is one \cite{DB2}.
These \svec s can also be organized into families \cite{BJI5}\cite{BJI6} 
with a unique pattern: the box diagram consisting of four \svec s,
one of each type, connected by \Gn , \Qn\ and the 
spectral flow automorphism $\cA$ (see footnote 6). 

\newpage
% \vskip .17in
\noi
{\bf The fermionic zero modes}

Most topological singular vectors come in pairs at the same level in
the same Verma module, differing by one unit of relative charge and by 
the BRST-invariance properties. The reason is that the fermionic zero 
modes \Gn\ and \Qn\ acting on a singular vector produce another 
singular vector, as can be checked straightforwardly using the algebra 
\req{topalgebra}. Therefore, at least in principle, only chiral singular
vectors can be `alone',
whereas the no-label singular vectors are accompanied by three, rather 
than one, singular vectors at the same level in the same Verma module.
To be precise, inside a given Verma module $V(\Delta,\htop)$ and for 
a given level $l$ the topological singular vectors 
with non-zero conformal weight are connected
by the action of \Qn\ and \Gn\ as:
\begin{eqnarray}  \Qz \, \kcc{l}{q}{G} \to \kcc{l}{q-1}{Q} , 
 \qquad\Gz \,\kcc{l}{q}{Q} \to \kcc{l}{q+1}{G}\, , \label{GQh}
\end{eqnarray}

\noi
where the arrows can be reversed (up to constants), using \Gn\ and
\Qn\ respectively, since the conformal weight of the singular vectors
is different from zero, \ie\ $\D+l\neq0$. Otherwise, on the right-hand side
of the arrows one obtains {\it chiral secondary} singular vectors
which cannot ``come back" to
the singular vectors on the left-hand side:  
\begin{eqnarray}  \Qz \,\kcc{l=-\D}{q}{G} \to \kcc{l=-\D}{q-1}{G,Q} , 
 \qquad\Gz \,\kcc{l=-\D}{q}{Q} \to \kcc{l=-\D}{q+1}{G,Q}\, . \label{GQch}
\end{eqnarray}

Observe that the non-existence of chiral \svec s of types 
$\kc_{l,\ket{\D,\htop}^G}^{(-2)G,Q}\,$, 
$\kc_{l,\ket{\D,\htop}^G}^{(1)G,Q}\,$,
$\kc_{l,\ket{\D,\htop}^Q}^{(2)G,Q}\,$ and
$\kc_{l,\ket{\D,\htop}^Q}^{(-1)G,Q}\,$, as one can see in tables \req{tabl2} 
and \req{tabl3}, implies in turn the absence of \svec s of types
$\kc_{l,\ket{\D,\htop}^G}^{(-1)G}\,$, $\kc_{l,\ket{\D,\htop}^G}^{(0)Q}\,$,
$\kc_{l,\ket{\D,\htop}^Q}^{(0)G}\,$ and $\kc_{l,\ket{\D,\htop}^Q}^{(1)Q}\,$, 
for $\D+l=0$. Therefore for zero conformal weight these types of \svec s
do not exist as such but they always `become' chiral, \ie\ of types
$\kc_{l,\ket{\D,\htop}^G}^{(-1)G,Q}\,$, $\kc_{l,\ket{\D,\htop}^G}^{(0)G,Q}\,$,
$\kc_{l,\ket{\D,\htop}^Q}^{(0)G,Q}\,$ and 
$\kc_{l,\ket{\D,\htop}^Q}^{(1)G,Q}\,$, instead.

Regarding no-label singular vectors 
$\kc_l^{(q)}$, they always satisfy $\D+l=0$. The action of \Gn\ and \Qn\
on a no-label singular vector produce three singular vectors:
 \begin{eqnarray} \Qz \,\kcc{l=-\D}{q}{ } \to \kcc{l=-\D}{q-1}{Q} ,  
 \quad\Gz \,\kcc{l=-\D}{q}{ } \to \kcc{l=-\D}{q+1}{G} , \quad 
 \Gz \, \Qz \,\kcc{l=-\D}{q}{ } \to \kcc{l=-\D}{q}{G,Q} \,. \label {QGnh} 
 \end{eqnarray}

\noi
All three are secondary \svec s which cannot come back to the no-label
\svec\ $\kcc{l=-\D}{q}{ }$ by acting with \Gn\ and \Qn , the chiral
\svec\ being in addition secondary with respect to the \Gn-closed and
the \Qn-closed \svec s.

Summarizing, \Gn\ and \Qn\ interpolate between two singular vectors
with non-zero conformal weight, in both directions, whereas they 
produce {\it secondary} singular vectors when acting on singular vectors
with zero conformal weight.

\section{Determinant Formulae}

In what follows we will write down the determinant formulae for the 
Topological N=2 algebra and we will analyse in much detail the interpretation
of the roots of the determinants in terms of singular vectors, making 
use of the results discussed in the last section. We will also investigate 
the different types of submodules, regarding shape and size, which
appear in the Verma modules. We will start with 
the generic Verma modules, then we will review the results 
for the chiral Verma modules \cite{BJI5}, adding new insights,
and finally we will consider the no-label Verma modules. The results
obtained will be transferred straightforwardly to the Verma modules of
the Ramond N=2 algebra (which have been very poorly studied in the
literature) as these turn out to be isomorphic to the Verma modules of
the Topological N=2 algebra. 

\subsection{Generic Verma modules}\lvm

For all the generic Verma modules -- either with two h.w. vectors  
$\ket{\D,\htop}^G$ and $\ket{\D,\htop-1}^Q$ ($\D \neq 0$) or with 
only one h.w. vector $\ket{0,\htop}^G$ or $\ket{0,\htop-1}^Q$ --
the determinant formula reads
 
\BE
det(\cM_l^T)= 
\prod_{2\leq rs \leq 2l}(f_{r,s})^{2 P(l-{rs\over2})}
\prod_{0 \leq k\leq l}(g_k^+)^{2 P_k(l-k)} 
\prod_{0 \leq k\leq l}(g_k^-)^{2 P_k(l-k)}  {\ } ,
\label{det1}
\EE
\noi
where  
\BE
f_{r,s}(\D,\htop,t) = -2t\D + t\htop - \htop^2 - {1 \over 4} t^2
       + {1 \over 4} (s-tr)^2  \,, {\ \ \ }r\in\oZ^+,{\ }s\in2\oZ^+
       \label{frs} \EE
\noi
and 
\BE
g_k^{\pm}(\D,\htop,t) = 2 \D \mp 2k \htop - tk(k \mp 1) \,, {\ \ \ }
	 0 \leq k \in \oZ \,, \label{gk} \EE
\noi
defining the parameter $t={(3-\ctop) / 3}$.
For $\ctop \neq 3 {\ \ } (t \neq 0) {\ }$ one can factorize $f_{r,s}$ as
\BE
f_{r,s}(\D,\htop,t\neq 0) = -2t (\D-\D_{r,s}) \,, \qquad
\D_{r,s} = - {1\over 2t} (\htop-\htop_{r,s}^{(0)}) (\htop-\hat\htop_{r,s})\,,
\label{Drs} \EE
\noi
with
\BE \htop_{r,s}^{(0)} = {t \over 2}(1+r)-{s \over 2} \ , \ \ \ \ \ 
r \in {\bf Z}^+ , \ \ s \in 2{\bf Z}^+ \,, \label{h0rs} \EE
\BE \hat\htop_{r,s} = {t \over 2}(1-r)+{s \over 2}  \ , \ \ \ \ \ 
r \in {\bf Z}^+ , \ \ s \in 2{\bf Z}^+ \,. \label{hhrs} \EE
\noi
For $\ctop=3 {\ \ } (t=0) {\ }$ one gets however
\BE f_{r,s}(\D,\htop,t=0) = {1 \over 4} s^2 - \htop^2 \,. \label{frs0} \EE
\noi
For all values of $\ctop$ one can factorize $g_k^+$ and $g_k^-$ as
\BE g_k^{\pm}(\D,\htop,t) = 2 (\D-\D_k^{\pm}) \,, \qquad
\D_k^{\pm} = \pm k \, (\htop-\htop_{k}^{\pm}) \,,  \label{Dk} \EE
\noi
with 
\BE \htop_k^{\pm} =  {t \over 2}\,(1 \mp k ) \,, \ \ \ \ \ k\in {\bf Z}^+ 
\label{hpml} \EE

\noi
The partition functions are defined by
\BE
\sum_N P_k(N)x^N={1\over 1+x^k}{\ }\sum_n P(n)x^n=
{1\over 1+x^k}{\ }\prod_{0<r\in {\bf Z},{\ }0<m\in {\bf Z}}{(1+x^r)^2
 \over (1-x^m)^2} .
\label{part}\EE
It is easy to check, by counting of states, that the partitions
$2 P(l-{rs\over2})$, exponents of $f_{r,s}$ in the determinant 
formula, correspond to complete Verma submodules whereas 
$2 P_k(l-k)$, exponents of $g_k^{\pm}$ in the determinant formula,
correspond to incomplete Verma submodules. Furthermore, as we will see, 
taking into account the size and the shape one can distinguish four types
of submodules. The fact that  
$P(0)=P_k(0)=1$ indicates that the \svec s come two by two at the
same level, in the same Verma module. 

The roots of the quadratic vanishing surface $f_{r,s}(\D,\htop,t)=0$
and of the vanishing planes $g_k^{\pm}(\D,\htop,t)=0$
are related to the corresponding roots of the determinant
formula for the Neveu-Schwarz N=2
algebra \cite{BFK}\cite{Nam}\cite{KaMa3} via the topological twists 
\req{twa}. These transform the standard h.w. vectors of the 
Neveu-Schwarz N=2 algebra into \Gn-closed h.w. vectors of the
Topological N=2 algebra, as we discussed in subsection 2.1 (see refs.
\cite{BJI5}\cite{BJI6} for a detailed account of the twisting and 
untwisting of primary states and singular vectors). The exponents 
$P(l-{rs\over2})$ and $P_k(l-k)$ cannot be obtained (easily) from
the exponents corresponding to the determinants of the Neveu-Schwarz N=2
algebra due to the complicated deformations produced by the topological 
twists \req{twa} on the Neveu-Schwarz Verma modules. However, these 
exponents essentially coincide with the exponents corresponding to the 
determinants of the Ramond N=2 algebra, as we will deduce in subsection 3.4.

The interpretation of the roots of $f_{r,s}(\D,\htop,t)=0$ and 
$g_k^{\pm}(\D,\htop,t)=0$ in terms of the singular vectors that one 
encounters in the corresponding Verma modules is as follows.

\vskip 0.17in
\noi 
{\bf Case $f_{r,s}(\D,\htop,t \neq 0)=0$} 

For $\D=\D_{r,s}$, eq. \req{Drs}, the Verma module built on the 
\Gn-closed h.w. vector
$\ket{\D_{r,s},\htop}^G$ has generically one uncharged \svec\ of type 
$\kc_l^{(0)G}$ and one charged \svec\ of type $\kc_l^{(-1)Q}$ at level
$l={rs\over2}$. In the general case $\,\D_{r,s}\neq -l\,$, \Gn\ and \Qn\ 
interpolate between these two \svec s (up to constants), so that they 
are on the same footing, \ie\ the h.w. vector of the null submodule 
is degenerate. In the case $\D_{r,s} = -l$, however, the \svec s have zero 
conformal weight and the action of \Gn\ or \Qn\ on such \svec s produces 
chiral {\it secondary} \svec s which cannot `come back' to the 
{\it primitive} \svec\ by acting with \Gn\ and \Qn , as was explained in
subsection 2.3. As a result, in the case $\D_{r,s} = -l$ the generic
situation is that only one of the 
two \svec s $\kc_l^{(0)G}$ or $\kc_l^{(-1)Q}$ is the h.w. vector of the 
null submodule, the other becoming a chiral secondary \svec\ (\ie\ of
type $\kc_l^{(0)G,Q}$ or $\kc_l^{(-1)G,Q}$ instead).
However, it can even happen
that the two \svec s $\kc_l^{(0)G}$ and $\kc_l^{(-1)Q}$ become chiral 
and thus disconnected from each other, both vectors being the h.w. vectors
of two different (although overlapping) null submodules. In Appendix A
we have deduced the values of $\htop$ and $t$ corresponding to the
Verma modules with $\D_{r,s} = -l$ for the three different possibilities:
$\kc_l^{(0)G}$ becoming chiral, $\kc_l^{(-1)Q}$ becoming chiral, and
both of them becoming chiral. In the first case one finds  
$\htop=\htop^{(-)}_{r,s}$, eqn. \req{curv2}, for $t \neq -{s \over n}, \,
n=1,...,r$. In the second case $\htop=\htop^{(+)}_{r,s}$, eqn. \req{curv1}, 
for $t \neq -{s \over n}, \, n=1,...,r$. Finally, in the third case one
finds $\htop=\htop^{(+)}_{r,s}$ and $\htop=\htop^{(-)}_{r,s}$ for
$t = -{s \over n}, \, n=1,...,r$ (see also table \req{tabfrs}). 
We also give all these \svec s at level 1. 
  
It is important to distinguish between the following two possibilities:
$\D_{r,s} \neq 0$ and $\D_{r,s} = 0$, as shown in table \req{tabfrs},
Fig. I and Fig. II. In the first case there is a 
degeneracy of the ground states, so that the Verma module built on  
$\ket{\D_{r,s}, \htop}^G$ has also a \Qn-closed h.w. vector
$\ket{\D_{r,s}, \htop-1}^Q$, \Gn\ and \Qn\ interpolating between them.
As a result, one can choose to express the descendant states in the
Verma module as built on $\ket{\D_{r,s}, \htop}^G$ or as built on
$\ket{\D_{r,s}, \htop-1}^Q$ (with a corresponding rearrangement of the
relative charges). Consequently, the \svec s of type 
$\kc_{l,\,\ket{\D,\,\htop}^G}^{(-1)Q}$ are equivalent to \svec s of
type $\kc_{l,\,\ket{\D,\,\htop-1}^Q}^{(0)Q}$, and one
can view the pair of \svec s for $\D_{r,s} \neq 0$ at level 
$l={rs\over2}$ as two {\it uncharged} \svec s: 
$\kc_{l,\,\ket{\D,\,\htop}^G}^{(0)G}$ in the `$\cG$-sector' and
$\kc_{l,\,\ket{\D,\,\htop-1}^Q}^{(0)Q}$ in the `$\cQ$-sector' (one 
of them, or both, becoming chiral for the case $\D_{r,s} = -l$), in 
analogy with the $(+)$ and $(-)$ sectors of the Ramond algebra.

\BE
\begin{tabular}{c|c}
{\ }& Singular vectors at level ${rs/2}$ for $f_{r,s}(\D,\htop, t \neq 0)=0$\\
\hline\\
$0 \neq \D_{r,s} \neq -l$  & 
$\kc_{\ket{\D,\htop}^G}^{(0)G}\, = \,\kc_{\ket{\D,\htop-1}^Q}^{(1)G}$,
$\ \,\kc_{\ket{\D,\htop-1}^Q}^{(0)Q}\, =\,\kc_{\ket{\D,\htop}^G}^{(-1)Q}\,$ \\
{\ }& {\ } \\
$0 \neq \D_{r,s}= -l$, $\htop^{(-)}_{r,s} \,, 
t \neq -{s \over n}, n=1,..,r$ &
$\kc_{\ket{\D,\htop}^G}^{(0)G,Q}\, =\,\kc_{\ket{\D,\htop-1}^Q}^{(1)G,Q}$,
$\ \,\kc_{\ket{\D,\htop-1}^Q}^{(0)Q}\, =\,\kc_{\ket{\D,\htop}^G}^{(-1)Q}\,$ \\
{\ }& {\ } \\
$0 \neq \D_{r,s}= -l$, $\htop^{(+)}_{r,s} \,,
t \neq -{s \over n}, n=1,..,r$ &
$\kc_{\ket{\D,\htop}^G}^{(0)G}\, =\,\kc_{\ket{\D,\htop-1}^Q}^{(1)G}$,
$\ \,\kc_{\ket{\D,\htop-1}^Q}^{(0)G,Q}\, =
\,\kc_{\ket{\D,\htop}^G}^{(-1)G,Q}\,$ \\
{\ }& {\ } \\
$0 \neq \D_{r,s}= -l$, $\htop^{(+)}_{r,s}$ and
$\htop^{(-)}_{r,s}\,$, $t = -{s \over n}$ &
$\kc_{\ket{\D,\htop}^G}^{(0)G,Q}\, =\,\kc_{\ket{\D,\htop-1}^Q}^{(1)G,Q}$,
$\ \,\kc_{\ket{\D,\htop-1}^Q}^{(0)G,Q}\, = 
\,\kc_{\ket{\D,\htop}^G}^{(-1)G,Q}\,$ \\
{\ }& {\ } \\
$\D_{r,s}= 0$, $\ \htop^{(0)}_{r,s}$ and 
$\,\hat\htop_{r,s}$  &
$\kc_{\ket{0,\htop}^G}^{(0)G}\ \,, \ \ \,\kc_{\ket{0,\htop}^G}^{(-1)Q}\ \,,$
$\ \ \,\kc_{\ket{0,\htop-1}^Q}^{(0)Q}\ \,,
\ \ \,\kc_{\ket{0,\htop-1}^Q}^{(1)G}$\\
{\ }& {\ } \\
\end{tabular}
\label{tabfrs}
\EE
\vskip .17in

For $\D_{r,s} = 0$ there are two sets of solutions 
for $\htop=\htop_{r,s}$ that annihilate the
determinants: $\htop_{r,s}=\htop_{r,s}^{(0)}$, eq. \req{h0rs}, and
$\htop_{r,s}=\hat\htop_{r,s}$, eq. \req{hhrs}.
The Verma modules built on the h.w. vectors $\ket{0, \htop_{r,s}}^G$
and the Verma modules built on the h.w. vectors $\ket{0, \htop_{r,s}-1}^Q$ 
are different because $\Gz \ket{0, \htop_{r,s}-1}^Q$ and 
$\Qz \ket{0, \htop_{r,s}}^G$ are level-zero chiral charged \svec s 
(corresponding to $g^{\pm}_0(\D,\htop,t)=0$, see below) which 
cannot be interchanged by the 
h.w. vectors $\ket{0, \htop_{r,s}}^G$ and $\ket{0, \htop_{r,s}-1}^Q$,
respectively. Consequently, in these Verma modules the descendant states
cannot be expressed in two ways (with the exception of the states  
inside the null submodules generated by the level-zero \svec s). Hence 
in the Verma modules built on the h.w. vectors $\ket{0, \htop_{r,s}}^G$,
with $\htop_{r,s}$ given either by $\htop_{r,s}^{(0)}$, eq. \req{h0rs},
or by $\hat\htop_{r,s}$, eq. \req{hhrs},
one finds generically one \svec\ of type $\kc_l^{(0)G}$ and one \svec\ 
of type $\kc_l^{(-1)Q}$ at level $l={rs\over2}$, \Gn\ and \Qn\ 
interpolating between them. Moreover, these \svec s are located outside, or
inside, the submodules generated by the level zero \svec s if 
$\htop_{r,s}=\htop_{r,s}^{(0)}$, or $\htop_{r,s}=\hat\htop_{r,s}$, 
respectively (see Fig. I). In Verma modules built on the h.w. vectors 
$\ket{0, \htop_{r,s}-1}^Q$ one finds generically one \svec\ of type
$\kc_l^{(0)Q}$ and one \svec\ of type $\kc_l^{(1)G}$ 
at level $l={rs\over2}$, \Gn\ and \Qn\ interpolating between them. These
\svec s are located outside, or inside, the submodules generated by
the level zero \svec s if $\htop_{r,s}=\hat\htop_{r,s}$ or
$\htop_{r,s}=\htop_{r,s}^{(0)}$, respectively (see Fig. II).
Observe that in the Verma modules with $\D=0$ there are no chiral \svec s,
besides the ones at level zero, because of the condition $\D+l=0$ for
chiral \svec s to exist. 

\vskip .5in
% Spacings:
\def\sk {{\hskip 1 cm}}
\def\bk {{\hskip 0.2 cm}}
\def\bbk{{\hskip 0.2 cm}}
\def\rk {{\hskip 3 cm}}
\def\com{{\hskip 0.2 cm},}
\def\pkt{{\hskip 0.2 cm}.}

\def\cQ{{\cal Q}}
\def\cG{{\cal G}}

\def\abs{\,|\,}                                 % modulus
\def\tilt {\tilde{t}}             % tilde{t}
\def\tils {\tilde{s}}             % tilde{s}

\newcommand{\fig}[1]{{\sc Fig.}\,{\sf #1}}        % Figure reference
\newcommand{\figs}[1]{{\sc Figs.}\,{\sf #1}}      % Figures reference
\newcommand{\ch}[1]{\makebox(0,0)[b]{\scriptsize$#1$}}  % chargeline
\newcommand{\cl}{tt}              % classification label for chap. 5
\newcommand{\case}[1]{\makebox(0,0)[b]{\footnotesize #1}}                
%Title of pictures
%\newcommand{\ket}[1]{\left| {#1} \right\rangle}        % ket

% Picture environment:
\newcounter{pics}
\newcommand{\bpic}[4]{\begin{center}\begin{picture}(#1,#2)(#3,#4)
\refstepcounter{pics}}
\renewcommand{\thepics}{{\sf\Roman{pics}}}
\newcommand{\epic}[1]{\end{picture}\\
\end{center}{ \footnotesize {\sc Fig.} \thepics \bk #1}}
\newcommand{\epicspl}{\end{picture}\\           % Different ending for picture
\addtocounter{pics}{-1}\end{center}}
\baselineskip 12pt              % environment for splitt
% pictures

\vbox{
\bpic{400}{120}{20}{-20}
% chargeline
%\put(0,195){\ch{G sector:}}
%\put(20,195){\ch{-2}}
%\put(50,195){\ch{-1}}
%\put(80,195){\ch{0}}
%\put(110,195){\ch{+1}}
%\put(0,185){\ch{Q sector:}}
%\put(20,185){\ch{-1}}
%\put(50,185){\ch{0}}
%\put(80,185){\ch{+1}}
%\put(110,185){\ch{+2}}
% Diagram 1 
\put(48,20){\framebox(4,3){}}
\put(78,20){\framebox(4,3){}}
\put(54,23){\vector(1,0){22}}
\put(76,20){\vector(-1,0){22}}
\put(47,24){\line(-1,2){50}}
\put(82,24){\line(1,2){50}}
\put(65,14){\makebox(0,0){\scriptsize $\cQ_0$}}
\put(65,28){\makebox(0,0){\scriptsize $\cG_0$}}
\put(47,15){\makebox(0,0)[r]{\scriptsize $\ket{\Delta ,\htop -1}^Q$}}
\put(83,15){\makebox(0,0)[l]{\scriptsize $\ket{\Delta ,\htop}^G$}}

\multiput(50,24) (0,5) {14}{\circle*{.5}}

\multiput(80,24) (0,5) {14}{\circle*{.5}}

\multiput(52,24) (2,5) {14}{\circle*{.5}}

\multiput(78,24) (-2,5) {14}{\circle*{.5}}

%\multiput(50,24) (0,5) {14}{\line(0,1){2}}

%\multiput(80,24) (0,5) {14}{\line(0,1){3}}

%\multiput(52,24) (2,5) {14}{\line(2,5){2}}

%\multiput(78,24) (-2,5) {14}{\line(-2,5){3}}

\put(51,91){\circle*{3}}
\put(79,91){\circle*{3}}
\put(54,89){\vector(1,0){22}}
\put(76,92){\vector(-1,0){22}}
%\put(65,98){\makebox(0,0){\scriptsize $\cQ_0$}}
%\put(65,84){\makebox(0,0){\scriptsize $\cG_0$}}
\put(67,94){\makebox(0,0)[rb]{\scriptsize 
	$\ket{\chi}_{l,\ket{\Delta ,\htop-1}^Q}^{(0)Q}$}}
\put(75,94){\makebox(0,0)[lb]{\scriptsize 
	$\ket{\chi}_{l,\ket{\Delta ,\htop}^G}^{(0)G}$}}
\put(65,-10){\case{$\D = \Delta_{r,s}\not =0$}}

% Diagram 3
\put(350,20){\circle*{3}}
\put(378,20){\framebox(4,3){}}
%\put(354,23){\vector(1,0){22}}
\put(376,20){\vector(-1,0){22}}
\put(347,24){\line(-1,2){50}}
\put(382,24){\line(1,2){50}}
\put(365,14){\makebox(0,0){\scriptsize $\cQ_0$}}
%\put(365,28){\makebox(0,0){\scriptsize $\cG_0$}}
\put(347,15){\makebox(0,0)[r]{\scriptsize $\cQ_0\ket{0, \htop}^G$}}
\put(383,15){\makebox(0,0)[l]{\scriptsize $\ket{0, \htop}^G$}}

\multiput(350,24) (0,5) {14}{\circle*{.5}}

\multiput(380,24) (0,5) {14}{\circle*{.5}}

\multiput(352,24) (2,5) {14}{\circle*{.5}}

\multiput(378,24) (-2,5) {14}{\circle*{.5}}

\put(351,91){\circle*{3}}
\put(380,91){\circle*{3}}
\put(354,89){\vector(1,0){22}}
\put(376,92){\vector(-1,0){22}}
%\put(365,98){\makebox(0,0){\scriptsize $\cQ_0$}}
%\put(365,84){\makebox(0,0){\scriptsize $\cG_0$}}
\put(367,94){\makebox(0,0)[rb]{\scriptsize 
$\ket{\chi}_{l,\ket{0, \htop}^G}^{(-1)Q}$}}
\put(375,94){\makebox(0,0)[lb]{\scriptsize 
$\ket{\chi}_{l,\ket{0, \htop}^G}^{(0)G}$}}
\put(365,-10){\case{$\D = \Delta_{r,s} =0$, $\ \htop=\hat\htop_{r,s}$}}

% Diagram 2 
\put(200,20){\circle*{3}}
\put(228,20){\framebox(4,3){}}
%\put(204,23){\vector(1,0){22}}
\put(226,20){\vector(-1,0){22}}
\put(197,24){\line(-1,2){50}}
\put(232,24){\line(1,2){50}}
\put(215,14){\makebox(0,0){\scriptsize $\cQ_0$}}
%\put(215,28){\makebox(0,0){\scriptsize $\cG_0$}}
\put(197,15){\makebox(0,0)[r]{\scriptsize $\cQ_0\ket{0 ,\htop}^G$}}
\put(233,15){\makebox(0,0)[l]{\scriptsize $\ket{0 ,\htop}^G$}}

\multiput(230,24) (0,5) {14}{\circle*{.5}}

\multiput(228,24) (-2,5) {14}{\circle*{.5}}

\put(201,91){\circle*{3}}
\put(229,91){\circle*{3}}
\put(204,89){\vector(1,0){22}}
\put(226,92){\vector(-1,0){22}}
%\put(215,98){\makebox(0,0){\scriptsize $\cQ_0$}}
%\put(215,84){\makebox(0,0){\scriptsize $\cG_0$}}
\put(217,94){\makebox(0,0)[rb]{\scriptsize $\ket{\chi}_{l,\ket{0,\htop}^G}^{(-1)Q}$}}
\put(225,94){\makebox(0,0)[lb]{\scriptsize $\ket{\chi}_{l,\ket{0,\htop}^G}^{(0)G}$}}
\put(215,-10){\case{$\D = \Delta_{r,s} =0$, $\ \htop=\htop^{(0)}_{r,s}$}}
\epic{
 For $\D=\D_{r,s}\neq 0$ one finds at level $l={rs\over 2}$ two \svec s,
connected by \Qn\ and \Gn , on the vertical lines over the two h.w. vectors 
of the Verma module, the \svec\ on the vertical over the \Gn-closed 
(\Qn-closed) h.w. vector being \Gn-closed (\Qn-closed) itself. 
Therefore the two \svec s can be appropriately described as two uncharged 
\svec s, one in the $\cG$-sector and the other in the $\cQ$-sector. 
(For $\D_{r,s}=-l$ one, at least, of the two \svec s becomes chiral and
consequently the two horizontal arrows, corresponding to \Gn\ and \Qn ,
reduce to one single arrow, or to no arrow at all if both \svec s become
chiral and thus disconnected). For zero conformal
weight $\D=\D_{r,s}= 0$, however, there is only one h.w. vector 
in the Verma module and there is one \svec\ at level zero. 
For the Verma modules with h.w. vectors
$\ket{0, \htop_{r,s}}^G$, for the case $\htop_{r,s}=\htop_{r,s}^{(0)}$
the singular vectors at level $l$ are located outside the submodule
generated by the level zero \svec\ $\cQ_0 \ket{0, \htop_{r,s}^{(0)}}^G$.
Therefore they can only be expressed as type $\kc_l^{(0)G}$ and type
$\kc_l^{(-1)Q}$ descendant states of the h.w. vector 
$\ket{0, \htop_{r,s}^{(0)}}^G$. For the case $\htop_{r,s}=
\hat\htop_{r,s}$, however, the \svec s at level $l$ are located inside
the submodule generated by $\cQ_0 \ket{0, \hat\htop_{r,s}}^G$. Therefore
they can either be expressed as type $\kc_l^{(0)G}$ and type 
$\kc_l^{(-1)Q}$ descendant states of the h.w. vector 
$\ket{0, \hat\htop_{r,s}}^G$,  or they can be expressed as secondary
\svec s built on the primitive \svec\ $\cQ_0 \ket{0, \hat\htop_{r,s}}^G$.}}

\baselineskip 16pt
\vskip 0.17in
\noi 
{\bf Case $f_{r,s}(\D,\htop,t=0)=0$} 

{\ } From expression \req{frs0} one deduces that for $\ctop=3 {\ }$
$(t=0)$ the singular vectors of types $\kc_l^{(0)G}$ and $\kc_l^{(-1)Q}$
at level $l={rs\over2}$, built on the \Gn-closed h.w. vector
$\ket{\D , \htop_s}^G$, exist for all values of $\D$ and $r$ provided
$\htop_s=\pm {s\over2}$. The analysis of the different possibilities for
$\D$ yields similar results as in the previous case for $\ctop \neq 3$,
as one can see in table \req{tabfrs0}.
That is, for $\D \neq 0$ the two singular vectors can be viewed as
two uncharged \svec s, one in the $\cG$-sector and the other in the 
$\cQ$-sector, for $\D = -l$ one of the two \svec s becoming chiral (see table 
\req{tabfrs0} and Appendix A). For $\D=0$, in the Verma modules built on the 
h.w. vectors $\ket{0, \htop_s}^G$ one finds generically one \svec\ 
of type $\kc_l^{(0)G}$ and one \svec\ of type $\kc_l^{(-1)Q}$. These are 
located outside, or inside, the submodules generated by the level zero 
\svec\ $\cQ_0 \ket{0, \htop_s}^G$ if $\htop_s= - {s\over2}$, or 
$\htop_s= {s\over2}$, respectively. In the Verma modules built on the
h.w. vectors $\ket{0, \htop_s-1}^Q$ one finds generically one \svec\ 
of type $\kc_l^{(0)Q}$ and one \svec\ of type $\kc_l^{(1)G}$. These are 
located outside, or inside, the submodules generated by the level zero 
\svec\ $\cG_0 \ket{0, \htop_s-1}^Q$ if $\htop_s= {s\over2}$, or 
$\htop_s= - {s\over2}$, respectively.  

\baselineskip 12pt
\vskip .5in

\vbox{
\bpic{250}{120}{170}{-20}
% chargeline
%\put(0,195){\ch{G sector:}}
%\put(20,195){\ch{-2}}
%\put(50,195){\ch{-1}}
%\put(80,195){\ch{0}}
%\put(110,195){\ch{+1}}
%\put(0,185){\ch{Q sector:}}
%\put(20,185){\ch{-1}}
%\put(50,185){\ch{0}}
%\put(80,185){\ch{+1}}
%\put(110,185){\ch{+2}}

% Diagram 3
\put(348,20){\framebox(4,3){}}
\put(380,20){\circle*{3}}
\put(354,20){\vector(1,0){22}}
%\put(376,20){\vector(-1,0){22}}
\put(347,24){\line(-1,2){50}}
\put(382,24){\line(1,2){50}}
%\put(365,14){\makebox(0,0){\scriptsize $\cQ_0$}}
\put(365,14){\makebox(0,0){\scriptsize $\cG_0$}}
\put(350,15){\makebox(0,0)[r]{\scriptsize $\ket{0,\htop}^Q$}}
\put(383,15){\makebox(0,0)[l]{\scriptsize $\cG_0\ket{0,\htop}^Q$}}

\multiput(350,24) (0,5) {14}{\circle*{.5}}
\multiput(380,24) (0,5) {14}{\circle*{.5}}
\multiput(352,24) (2,5) {14}{\circle*{.5}}
\multiput(378,24) (-2,5) {14}{\circle*{.5}}

\put(351,91){\circle*{3}}
\put(380,91){\circle*{3}}
\put(354,89){\vector(1,0){22}}
\put(376,92){\vector(-1,0){22}}
%\put(365,98){\makebox(0,0){\scriptsize $\cQ_0$}}
%\put(365,84){\makebox(0,0){\scriptsize $\cG_0$}}
\put(367,94){\makebox(0,0)[rb]{\scriptsize $\ket{\chi}_{l,\ket{0,\htop}^Q}^{(0)Q}$}}
\put(370,94){\makebox(0,0)[lb]{\scriptsize $\ket{\chi}_{l,\ket{0,\htop}^Q}^{(1)G}$}}
\put(365,-10){\case{$\D = \Delta_{r,s} =0$, $\ \htop=\htop^{(0)}_{r,s} - 1$}}

% Diagram 2 
\put(198,20){\framebox(4,3){}}
\put(230,20){\circle*{3}}
\put(204,20){\vector(1,0){22}}
%\put(226,20){\vector(-1,0){22}}
\put(197,24){\line(-1,2){50}}
\put(232,24){\line(1,2){50}}
%\put(215,14){\makebox(0,0){\scriptsize $\cQ_0$}}
\put(215,14){\makebox(0,0){\scriptsize $\cG_0$}}
\put(197,15){\makebox(0,0)[r]{\scriptsize $\ket{0, \htop}^Q$}}
\put(228,15){\makebox(0,0)[l]{\scriptsize $\cG_0\ket{0, \htop}^Q$}}

\multiput(200,24) (0,5) {14}{\circle*{.5}}

%\multiput(230,24) (0,5) {14}{\circle*{.5}}

\multiput(202,24) (2,5) {14}{\circle*{.5}}

%\multiput(228,24) (-2,5) {14}{\circle*{.5}}

\put(201,91){\circle*{3}}
\put(229,91){\circle*{3}}
\put(204,89){\vector(1,0){22}}
\put(226,92){\vector(-1,0){22}}
%\put(215,98){\makebox(0,0){\scriptsize $\cQ_0$}}
%\put(215,84){\makebox(0,0){\scriptsize $\cG_0$}}
\put(217,94){\makebox(0,0)[rb]{\scriptsize
 $\ket{\chi}_{l,\ket{0, \htop}^Q}^{(0)Q}$}}
\put(220,94){\makebox(0,0)[lb]{\scriptsize
 $\ket{\chi}_{l,\ket{0, \htop}^Q}^{(1)G}$}}
\put(215,-10){\case{$\D=\Delta_{r,s} =0$, $\ \htop=\hat\htop_{r,s} - 1$}}
\epic{ 
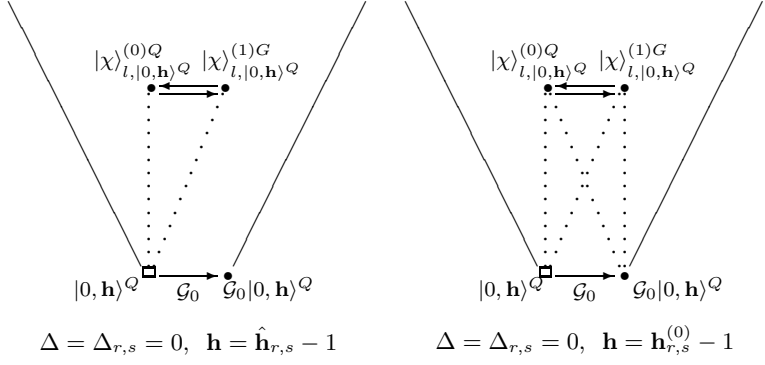
For the Verma modules with h.w. vectors $\ket{0, \htop_{r,s}-1}^Q$, 
for the case $\htop_{r,s}=\hat\htop_{r,s}$
the singular vectors at level $l$ are located outside the submodule
generated by the level zero \svec\ $\cG_0 \ket{0, \hat\htop_{r,s}-1}^Q$.
Therefore they can only be expressed as type $\kc_l^{(1)G}$ and type
$\kc_l^{(0)Q}$ descendant states of the h.w. vector 
$\ket{0, \hat\htop_{r,s}-1}^Q$. For the case $\htop_{r,s}=
\htop^{(0)}_{r,s}$, however, the \svec s at level $l$ are located inside
the submodule generated by $\cG_0 \ket{0, \htop^{(0)}_{r,s}-1}^Q$. Therefore
they can either be expressed as type $\kc_l^{(1)G}$ and type $\kc_l^{(0)Q}$ 
descendant states of the h.w. vector $\ket{0, \htop^{(0)}_{r,s}-1}^Q$, 
or they can be expressed as secondary \svec s
built on the primitive \svec\ $\cG_0 \ket{0, \htop^{(0)}_{r,s}-1}^Q$.}}

\baselineskip 16pt
\vskip .17in

\BE
\begin{tabular}{c|c}
{\ }& Singular vectors at level ${rs/2}$ for $f_{r,s}(\D, \htop, t=0)=0$\\
\hline\\
$0 \neq \D \neq -l$, $\,\htop_{s} = \pm {s \over 2}$  & 
$\kc_{\ket{\D,\htop}^G}^{(0)G}\, = \,\kc_{\ket{\D,\htop-1}^Q}^{(1)G}$,
$\ \,\kc_{\ket{\D,\htop-1}^Q}^{(0)Q}\, =\,\kc_{\ket{\D,\htop}^G}^{(-1)Q}\,$ \\
{\ }& {\ } \\
$0 \neq \D = -l$, $\,\htop_{s} = {-{s \over 2}}$ &
$\kc_{\ket{\D,\htop}^G}^{(0)G,Q}\, =\,\kc_{\ket{\D,\htop-1}^Q}^{(1)G,Q}$,
$\ \,\kc_{\ket{\D,\htop-1}^Q}^{(0)Q}\, =\,\kc_{\ket{\D,\htop}^G}^{(-1)Q}\,$ \\
{\ }& {\ } \\
$0 \neq \D = -l$, $\,\htop_{s} = {s \over 2}$ &
$\kc_{\ket{\D,\htop}^G}^{(0)G}\, =\,\kc_{\ket{\D,\htop-1}^Q}^{(1)G}$,
$\ \,\kc_{\ket{\D,\htop-1}^Q}^{(0)G,Q}\, =
\,\kc_{\ket{\D,\htop}^G}^{(-1)G,Q}\,$ \\
{\ }& {\ } \\
$\D = 0$, $\,\htop_{s} = \pm {s \over 2}$ &
$\kc_{\ket{0,\htop}^G}^{(0)G}\ \,, \ \ \,\kc_{\ket{0,\htop}^G}^{(-1)Q}\ \,,$
$\ \ \,\kc_{\ket{0,\htop-1}^Q}^{(0)Q}\ \,,
\ \ \,\kc_{\ket{0,\htop-1}^Q}^{(1)G}$\\
{\ }& {\ } \\
\end{tabular}
\label{tabfrs0}
\EE

\vskip 0.17in
\noi 
{\bf Case $g_k^{+}(\D,\htop,t)=0$}

For $\D=\D_k^+$, eq. \req{Dk}, the Verma module built on the 
\Gn-closed h.w. vector
$\ket{\D_k^+, \htop}^G$ has generically one charged \svec\ of type
$\kc_l^{(1)G}$ and one uncharged \svec\ of type $\kc_l^{(0)Q}$ at  
level $l=k$. In the case $\D_k^+=-l$ the \svec\ of type $\kc_l^{(0)Q}$
always becomes chiral because there are no chiral \svec s of type
$\kc_l^{(1)G,Q}$ built on \Gn-closed primaries, as shown in table 
\req{tabl1}, which otherwise would be produced as $\,\cG_0\kc_l^{(0)Q}$. 
For $\D_k^+ \neq 0$ these \svec s can be viewed as two $q=1$ 
{\it charged} \svec s: $\kc_{l,\,\ket{\D,\,\htop}^G}^{(1)G}$ 
in the $\cG$-sector and $\kc_{l,\,\ket{\D,\,\htop-1}^Q}^{(1)Q}$ 
in the $\cQ$-sector, as shown in Fig. III, 
due to the existence of the two h.w. vectors $\ket{\D_k^+, \htop}^G$
and $\ket{\D_k^+, \htop-1}^Q$ in the Verma module, the second type
becoming chiral, \ie\ of type $\kc_{l,\,\ket{-l,\,\htop-1}^Q}^{(1)G,Q}$ 
in the case $\D_k^+=-l$. 

For $\D_k^+ = 0$, however, there is only one h.w. vector in the Verma module 
and one chiral charged singular vector at level zero, corresponding to
the solution $k=0$, for any value of $\htop$:  $\cQ_0 \ket{0, \htop}^G$ or
$\cG_0 \ket{0, \htop-1}^Q$ depending on which is the h.w. vector.
In addition, in Verma modules built on primaries $\ket{0,\htop_k^+}^G$, 
with $\htop_k^+$ satisfying eq. \req{hpml}, one finds generically 
one charged \svec\ of type $\kc_l^{(1)G}$ and one uncharged \svec\ 
of type $\kc_l^{(0)Q}$ at level $l=k$, \Gn\ and \Qn\ interpolating
between them. These \svec s
are located outside the submodule generated by the level zero \svec\ .
In Verma modules built on primaries $\ket{0,\htop_k^+-1}^Q$
one finds generically two positively charged \svec s at level $l=k$, one
of type $\kc_l^{(1)Q}$ and one of type $\kc_l^{(2)G}$, 
\Gn\ and \Qn\ interpolating between them. Moreover, these two \svec s
are located inside the submodule generated by the level zero \svec\  (see
Fig. III). 

\vskip .5in
\baselineskip 12pt                              
\vbox{
\bpic{400}{120}{20}{-20}
% chargeline
%\put(0,195){\ch{G sector:}}
%\put(20,195){\ch{-2}}
%\put(50,195){\ch{-1}}
%\put(80,195){\ch{0}}
%\put(110,195){\ch{+1}}
%\put(0,185){\ch{Q sector:}}
%\put(20,185){\ch{-1}}
%\put(50,185){\ch{0}}
%\put(80,185){\ch{+1}}
%\put(110,185){\ch{+2}}
% Diagram 1 
\put(48,20){\framebox(4,3){}}
\put(78,20){\framebox(4,3){}}
\put(54,23){\vector(1,0){22}}
\put(76,20){\vector(-1,0){22}}
\put(47,24){\line(-1,2){50}}
\put(82,24){\line(1,2){50}}
\put(65,14){\makebox(0,0){\scriptsize $\cQ_0$}}
\put(70,28){\makebox(0,0){\scriptsize $\cG_0$}}
\put(47,15){\makebox(0,0)[r]{\scriptsize $\ket{\Delta, \htop-1}^Q$}}
\put(83,15){\makebox(0,0)[l]{\scriptsize $\ket{\Delta, \htop}^G$}}
%\put(50,24){\line(0,1){65}}

\multiput(80,24) (0,5) {14}{\circle*{.5}}
\multiput(52,24) (2,5) {14}{\circle*{.5}}
\multiput(81,24) (2,5) {14}{\circle*{.5}}
\multiput(54,24) (4,5) {14}{\circle*{.5}}

\put(79,91){\circle*{3}}
\put(108,91){\circle*{3}}
\put(82,89){\vector(1,0){22}}
\put(104,92){\vector(-1,0){22}}
%\put(65,98){\makebox(0,0){\scriptsize $\cQ_0$}}
%\put(65,84){\makebox(0,0){\scriptsize $\cG_0$}}
\put(87,94){\makebox(0,0)[rb]{\scriptsize
 $\ket{\chi}_{l,\ket{\Delta, \htop-1}^Q}^{(1)Q}$}}
\put(116,80){\makebox(0,0)[lb]{\scriptsize
 $\ket{\chi}_{l,\ket{\Delta, \htop}^G}^{(1)G}$}}
\put(65,-10){\case{$\D = \Delta_l^+\not =0$}}
% Diagram 2 
\put(200,20){\circle*{3}}
\put(228,20){\framebox(4,3){}}
%\put(204,23){\vector(1,0){22}}
\put(226,20){\vector(-1,0){22}}
\put(197,24){\line(-1,2){50}}
\put(232,24){\line(1,2){50}}
\put(215,14){\makebox(0,0){\scriptsize $\cQ_0$}}
%\put(220,28){\makebox(0,0){\scriptsize $\cG_0$}}
\put(197,15){\makebox(0,0)[r]{\scriptsize $\cQ_0\ket{0, \htop}^G$}}
\put(233,15){\makebox(0,0)[l]{\scriptsize $\ket{0, \htop}^G$}}
%\put(200,24){\line(0,1){65}}

\multiput(230,24) (0,5) {14}{\circle*{.5}}
\multiput(231,24) (2,5) {14}{\circle*{.5}}

\put(229,91){\circle*{3}}
\put(258,91){\circle*{3}}
\put(232,89){\vector(1,0){22}}
\put(254,92){\vector(-1,0){22}}
%\put(215,98){\makebox(0,0){\scriptsize $\cQ_0$}}
%\put(215,84){\makebox(0,0){\scriptsize $\cG_0$}}
\put(237,94){\makebox(0,0)[rb]{\scriptsize
 $\ket{\chi}_{l,\ket{0, \htop}^G}^{(0)Q}$}}
\put(266,80){\makebox(0,0)[lb]{\scriptsize
 $\ket{\chi}_{l,\ket{0, \htop}^G}^{(1)G}$}}
\put(215,-10){\case{$\D = \Delta_l^+ = 0$, $\ \,\htop = \htop_l^+$}}
% Diagram 3
\put(349,20){\framebox(4,3){}}
\put(380,21){\circle*{3}}
\put(355,20){\vector(1,0){22}}
%\put(376,20){\vector(-1,0){22}}
\put(347,24){\line(-1,2){50}}
\put(382,24){\line(1,2){50}}
%\put(365,14){\makebox(0,0){\scriptsize $\cQ_0$}}
\put(365,14){\makebox(0,0){\scriptsize $\cG_0$}}
\put(347,15){\makebox(0,0)[r]{\scriptsize $\ket{0, \htop}^Q$}}
\put(383,15){\makebox(0,0)[l]{\scriptsize $\cG_0\ket{0, \htop}^Q$}}
%\put(350,24){\line(0,1){65}}

\multiput(380,24) (0,5) {14}{\circle*{.5}}
\multiput(352,24) (2,5) {14}{\circle*{.5}}
\multiput(381,24) (2,5) {14}{\circle*{.5}}
\multiput(354,24) (4,5) {14}{\circle*{.5}}

\put(379,91){\circle*{3}}
\put(408,91){\circle*{3}}
\put(382,89){\vector(1,0){22}}
\put(404,92){\vector(-1,0){22}}
%\put(365,98){\makebox(0,0){\scriptsize $\cQ_0$}}
%\put(365,84){\makebox(0,0){\scriptsize $\cG_0$}}
\put(387,94){\makebox(0,0)[rb]{\scriptsize
 $\ket{\chi}_{l,\ket{0, \htop}^Q}^{(1)Q}$}}
\put(416,80){\makebox(0,0)[lb]{\scriptsize
 $\ket{\chi}_{l,\ket{0, \htop}^Q}^{(2)G}$}}
\put(365,-10){\case{$\D = \Delta_l^+ = 0$, $\ \,\htop = \htop_l^+ - 1$}}
\epic{. 
For $\D=\D_l^+ \neq 0$ one finds at level $l$ two \svec s,
connected by \Qn\ and \Gn , located one 
unit to the right of the vertical lines over the two h.w. vectors of the
Verma module, the \svec\ to the right of
the vertical over the \Gn-closed (\Qn-closed)
h.w. vector being \Gn-closed (\Qn-closed) itself. Therefore the two \svec s
can be appropriately described as two $q=1$ charged \svec s, one in the
$\cG$-sector and the other in the $\cQ$-sector. For $\D_l^+=-l$ 
the \Qn-closed \svec\ always becomes chiral whereas the \Gn-closed \svec\
never becomes chiral. Consequently the two arrows reduce to the arrow
corresponding to \Qn .  For zero conformal
weight $\D=\D_l^+ = 0$, however, there is only one h.w. vector 
in the Verma module and there is one \svec\ at level zero.
For the Verma modules with \Gn-closed h.w. vectors $\ket{0, \htop_l^+}^G$
the singular vectors at level $l$ are located outside the submodule
generated by the level zero \svec\ $\cQ_0 \ket{0, \htop_l^+}^G$.
Therefore they can only be expressed as type $\kc_l^{(1)G}$ and type
$\kc_l^{(0)Q}$ descendant states of the h.w. vector $\ket{0,\htop_l^+}^G$.
For the Verma modules with \Qn-closed h.w. vectors $\ket{0,\htop_l^+-1}^Q$,
however, the \svec s at level $l$ are located inside the submodule 
generated by the \svec\ $\cG_0 \ket{0, \htop_l^+-1}^Q$. Therefore
they can either be expressed as type $\kc_l^{(2)G}$ and type 
$\kc_l^{(1)Q}$ descendant states of the h.w. vector 
$\ket{0, \htop_l^+-1}^Q$,  or they can be expressed as secondary
\svec s built on the primitive \svec\ $\cG_0 \ket{0, \htop_l^+-1}^Q$.}}

\baselineskip 16pt
\vskip .17in
\BE
\begin{tabular}{c|c}
{\ }& Singular vectors at level $l>0$ for $g_l^{+}(\D, \htop, t)=0$\\
\hline\\
$0 \neq \D_l^+ \neq -l$  & 
$\kc_{\ket{\D,\htop}^G}^{(1)G}\, = \,\kc_{\ket{\D,\htop-1}^Q}^{(2)G}$,
$\ \,\kc_{\ket{\D,\htop-1}^Q}^{(1)Q}\, =\,\kc_{\ket{\D,\htop}^G}^{(0)Q}\,$ \\
{\ }& {\ } \\
$0 \neq \D_l^+ = -l$ &
$\kc_{\ket{\D,\htop}^G}^{(1)G}\, = \,\kc_{\ket{\D,\htop-1}^Q}^{(2)G}$,
$\ \,\kc_{\ket{\D,\htop-1}^Q}^{(1)G,Q}\, =
\,\kc_{\ket{\D,\htop}^G}^{(0)G,Q}\,$ \\
{\ }& {\ } \\
$\D_l^+ = 0$, $\,\htop = \htop_l^+$ &
$\kc_{\ket{0,\htop}^G}^{(1)G}\ \,, \ \ \,\kc_{\ket{0,\htop}^G}^{(0)Q}\ \,,$
$\ \ \,\kc_{\ket{0,\htop-1}^Q}^{(1)Q}\ \,,
\ \ \,\kc_{\ket{0,\htop-1}^Q}^{(2)G}$\\
{\ }& {\ } \\
\end{tabular}
\label{tabgkp}
\EE

\vskip 0.17in
\noi 
{\bf Case $g_k^{-}(\D,\htop,t)=0$}

For $\D=\D_k^-$, eq. \req{Dk},
the Verma module built on the \Gn-closed h.w. vector
$\ket{\D_k^-, \htop}^G$ has generically two 
negatively charged \svec s at level $l=k$, one of type
$\kc_l^{(-1)G}$ and one of type $\kc_l^{(-2)Q}$.  
In the case $\D_k^-=-l$ the \svec\ of type $\kc_l^{(-1)G}$
always becomes chiral because there are no chiral \svec s of type
$\kc_l^{(-2)G,Q}$, as shown in table \req{tabl1}, which otherwise would 
be produced as $\,\cQ_0\kc_l^{(-1)G}$. For $\D_k^- \neq 0$
these \svec s can be viewed as two $q=-1$ {\it charged} \svec s:
$\kc_{l,\,\ket{\D,\,\htop}^G}^{(-1)G}$ in the $\cG$-sector and 
$\kc_{l,\,\ket{\D,\,\htop-1}^Q}^{(-1)Q}$ in the $\cQ$-sector, as shown
in Fig. IV, due to the existence of the two h.w. vectors 
$\ket{\D_k^-, \htop}^G$ and $\ket{\D_k^-, \htop-1}^Q$ in the Verma module, 
the first type becoming chiral, \ie\ of type 
$\kc_{l,\,\ket{-l,\,\htop}^G}^{(-1)G,Q}$ in the case $\D_k^-=-l$.

For $\D_k^- = 0$, as in the previous case, there is only one h.w. vector 
in the Verma module and one chiral charged \svec\ at level zero for any
$\htop$. In addition, in Verma modules built on primaries 
$\ket{0,\htop_k^-}^G$, with $\htop_k^-$ satisfying eq. \req{hpml}, 
one finds generically one charged \svec\ of type $\kc_l^{(-1)G}$ and 
one charged \svec\ of type $\kc_l^{(-2)Q}$ at level $l=k$,
\Gn\ and \Qn\ interpolating between them. 
These \svec s are located inside the submodule 
generated by the level zero \svec\ $\cQ_0 \ket{0,\htop_k^-}^G$.
In Verma modules built on primaries $\ket{0,\htop_k^--1}^Q$
one finds generically one charged \svec\ of type $\kc_l^{(-1)Q}$ 
and one uncharged \svec\ of type $\kc_l^{(0)G}$, 
\Gn\ and \Qn\ interpolating between them .
These \svec s are located outside the submodule generated 
by the level zero \svec\ $\cG_0 \ket{0,\htop_k^--1}^Q$ (see Fig. IV).

\vskip .5in
\baselineskip 12pt
			 
\vbox{
\bpic{400}{120}{20}{-20}
% chargeline
%\put(0,195){\ch{G sector:}}
%\put(20,195){\ch{-2}}
%\put(50,195){\ch{-1}}
%\put(80,195){\ch{0}}
%\put(110,195){\ch{+1}}
%\put(0,185){\ch{Q sector:}}
%\put(20,185){\ch{-1}}
%\put(50,185){\ch{0}}
%\put(80,185){\ch{+1}}
%\put(110,185){\ch{+2}}
% Diagram 1 
\put(48,20){\framebox(4,3){}}
\put(78,20){\framebox(4,3){}}
\put(54,23){\vector(1,0){22}}
\put(76,20){\vector(-1,0){22}}
\put(47,24){\line(-1,2){50}}
\put(82,24){\line(1,2){50}}
\put(65,14){\makebox(0,0){\scriptsize $\cQ_0$}}
\put(60,28){\makebox(0,0){\scriptsize $\cG_0$}}
\put(47,15){\makebox(0,0)[r]{\scriptsize $\ket{\Delta, \htop-1}^Q$}}
\put(83,15){\makebox(0,0)[l]{\scriptsize $\ket{\Delta, \htop}^G$}}

\multiput(50,24) (0,5) {14}{\circle*{.5}}
\multiput(49,24) (-2,5) {14}{\circle*{.5}}
\multiput(78,24) (-2,5) {14}{\circle*{.5}}
\multiput(76,24) (-4,5) {14}{\circle*{.5}}

\put(22,91){\circle*{3}}
\put(51,91){\circle*{3}}
\put(26,89){\vector(1,0){22}}
\put(47,92){\vector(-1,0){22}}
%\put(65,98){\makebox(0,0){\scriptsize $\cQ_0$}}
%\put(65,84){\makebox(0,0){\scriptsize $\cG_0$}}
\put(16,80){\makebox(0,0)[rb]{\scriptsize 
$\ket{\chi}_{l,\ket{\Delta, \htop-1}^Q}^{(-1)Q}$}}
\put(50,94){\makebox(0,0)[lb]{\scriptsize
 $\ket{\chi}_{l,\ket{\Delta, \htop}^G}^{(-1)G}$}}
\put(65,-10){\case{$\D = \Delta_l^-\not = 0$}}
% Diagram 2 
\put(200,20){\circle*{3}}
\put(228,20){\framebox(4,3){}}
%\put(204,23){\vector(1,0){22}}
\put(226,20){\vector(-1,0){22}}
\put(197,24){\line(-1,2){50}}
\put(232,24){\line(1,2){50}}
\put(215,14){\makebox(0,0){\scriptsize $\cQ_0$}}
%\put(210,28){\makebox(0,0){\scriptsize $\cG_0$}}
\put(197,15){\makebox(0,0)[r]{\scriptsize $\cQ_0\ket{0, \htop}^G$}}
\put(233,15){\makebox(0,0)[l]{\scriptsize $\ket{0, \htop}^G$}}

\multiput(200,24) (0,5) {14}{\circle*{.5}}
\multiput(199,24) (-2,5) {14}{\circle*{.5}}
\multiput(228,24) (-2,5) {14}{\circle*{.5}}
\multiput(226,24) (-4,5) {14}{\circle*{.5}}

\put(172,91){\circle*{3}}
\put(201,91){\circle*{3}}
\put(176,89){\vector(1,0){22}}
\put(197,92){\vector(-1,0){22}}
%\put(215,98){\makebox(0,0){\scriptsize $\cQ_0$}}
%\put(215,84){\makebox(0,0){\scriptsize $\cG_0$}}
\put(166,80){\makebox(0,0)[rb]{\scriptsize
 $\ket{\chi}_{l,\ket{0, \htop}^G}^{(-2)Q}$}}
\put(200,94){\makebox(0,0)[lb]{\scriptsize
 $\ket{\chi}_{l,\ket{0, \htop}^G}^{(-1)G}$}}
\put(215,-10){\case{$\D = \Delta_l^- =0$, $\ \, \htop = \htop_l^-$}}
% Diagram 3
\put(348,20){\framebox(4,3){}}
\put(380,20){\circle*{3}}
\put(354,20){\vector(1,0){22}}
%\put(376,20){\vector(-1,0){22}}
\put(347,24){\line(-1,2){50}}
\put(382,24){\line(1,2){50}}
%\put(365,14){\makebox(0,0){\scriptsize $\cQ_0$}}
\put(365,14){\makebox(0,0){\scriptsize $\cG_0$}}
\put(347,15){\makebox(0,0)[r]{\scriptsize $\ket{0, \htop}^Q$}}
\put(383,15){\makebox(0,0)[l]{\scriptsize $\cG_0\ket{0, \htop}^Q$}}

\multiput(350,24) (0,5) {14}{\circle*{.5}}
\multiput(349,24) (-2,5) {14}{\circle*{.5}}

\put(322,91){\circle*{3}}
\put(350,91){\circle*{3}}
\put(325,89){\vector(1,0){22}}
\put(346,92){\vector(-1,0){22}}
%\put(365,98){\makebox(0,0){\scriptsize $\cQ_0$}}
%\put(365,84){\makebox(0,0){\scriptsize $\cG_0$}}
\put(316,80){\makebox(0,0)[rb]{\scriptsize 
$\ket{\chi}_{l,\ket{0, \htop}^Q}^{(-1)Q}$}}
\put(350,94){\makebox(0,0)[lb]{\scriptsize
 $\ket{\chi}_{l,\ket{0, \htop}^Q}^{(0)G}$}}
\put(365,-10){\case{$\D = \Delta_l^-=0$, $\ \, \htop = \htop_l^- - 1$}}
\epic{. 
For $\D=\D_l^- \neq 0$ one finds at level $l$ two \svec s,
connected by \Qn\ and \Gn , located one 
unit to the left of the vertical lines over the two h.w. vectors of the
Verma module, the \svec\ to the left of
the vertical over the \Gn-closed (\Qn-closed)
h.w. vector being \Gn-closed (\Qn-closed) itself.
Therefore the two \svec s can be appropriately described as two $q=-1$ 
charged \svec s, one in the $\cG$-sector and the other in the $\cQ$-sector. 
For $\D_l^-=-l$ the \Gn-closed \svec\ always becomes chiral whereas the
\Qn-closed \svec\ never becomes chiral. Consequently the two arrows 
reduce to the arrow corresponding to \Gn . For zero conformal
weight $\D=\D_l^- = 0$, however, there is only one h.w. vector 
in the Verma module and there is one \svec\ at level zero.
For the Verma modules with \Gn-closed h.w. vectors $\ket{0,\htop_l^+}^G$
the \svec s at level $l$ are located inside the submodule 
generated by the \svec\ $\cQ_0 \ket{0, \htop_l^+}^G$. Therefore
they can either be expressed as type $\kc_l^{(-1)G}$ and type 
$\kc_l^{(-2)Q}$ descendant states of the h.w. vector 
$\ket{0, \htop_l^+}^G$, or they can be expressed as secondary
\svec s built on the primitive \svec\ $\cQ_0 \ket{0, \htop_l^+}^G$. 
For the Verma modules with \Qn-closed h.w. vectors $\ket{0, \htop_l^--1}^Q$,
however, 
the singular vectors at level $l$ are located outside the submodule
generated by the level zero \svec\ $\cG_0 \ket{0, \htop_l^--1}^Q$.
Therefore they can only be expressed as type $\kc_l^{(0)G}$ and type
$\kc_l^{(-1)Q}$ descendant states of the h.w. vector $\ket{0,\htop_l^--1}^Q$.}}

\vskip .17in
\baselineskip 16pt
\BE
\begin{tabular}{c|c}
{\ }& Singular vectors at level $l>0$ for $g_l^{-}(\D, \htop, t)=0$\\
\hline\\
$0 \neq \D_l^- \neq -l$  & 
$\kc_{\ket{\D,\htop}^G}^{(-1)G}\, = \,\kc_{\ket{\D,\htop-1}^Q}^{(0)G}$,
$\ \,\kc_{\ket{\D,\htop-1}^Q}^{(-1)Q}\, =\,\kc_{\ket{\D,\htop}^G}^{(-2)Q}\,$ 
\\ {\ }& {\ } \\
$0 \neq \D_l^- = -l$ &
$\kc_{\ket{\D,\htop}^G}^{(-1)G,Q}\, = \,\kc_{\ket{\D,\htop-1}^Q}^{(0)G,Q}$,
$\ \,\kc_{\ket{\D,\htop-1}^Q}^{(-1)Q}\, =\,\kc_{\ket{\D,\htop}^G}^{(-2)Q}\,$ 
\\{\ }& {\ } \\
$\D_l^- = 0$, $\,\htop = \htop_l^-$ &
$\kc_{\ket{0,\htop}^G}^{(-1)G}\ \,, \ \ \,\kc_{\ket{0,\htop}^G}^{(-2)Q}\ \,,$
$\ \ \,\kc_{\ket{0,\htop-1}^Q}^{(-1)Q}\ \,,
\ \ \,\kc_{\ket{0,\htop-1}^Q}^{(0)G}$\\
{\ }& {\ } \\
\end{tabular}
\label{tabgkm}
\EE

\vskip 0.17in
\noi 
{\bf No-label \svec s}

Apart from the \svec s that we have analysed,
`detected' by the determinant formula, there may still exist other
primitive \svec s, besides subsingular vectors, hidden by the first ones. 
Such \svec s, denoted as {\it isolated}, have not been
investigated properly for any N=2 superconformal algebra.
Furthermore, it can even happen that some \svec s
predicted by the determinant formula are secondary with respect to an 
isolated overlooked \svec\ at the same level, the latter being the 
generator of the detected \svec s, by acting on it with some zero 
modes of the algebra.

In the case of the N=2 superconformal algebras subsingular vectors have been
discovered and partially analysed in refs.
\cite{BJI5}\cite{BJI6}\cite{DB1}\cite{DB3}. 
As to the `invisible' isolated \svec s, in the Topological algebra
(as well as in the Ramond algebra) there are certainly such \svec s, 
at least the ones which are in addition generators of `visible'
secondary \svec s detected by the determinant formula. 
Namely, all the no-label \svec s in generic 
Verma modules, given in tables \req{tabl2} and \req{tabl3}, belong 
to such category. No-label \svec s, written down at levels 1 and 2
in refs. \cite{BJI6} and \cite{DB1}, are very scarce as they only exist
for specific values of $\ctop$. As deduced in Appendix A, these values are
$\ctop = {3r-6 \over r}\,$, corresponding to $t={2 \over r}$. The reason
why no-label \svec s are undetected by the determinant formula is that
they only appear in certain Verma modules, for discrete values of
$\D, \htop, t$, where there are intersections,
at the same level $l$, of \svec s corresponding to the series 
$f_{r,s}(\D, \htop, t)=0$ with \svec s corresponding to one of the series
$g_k^{\pm}(\D, \htop, t)=0$, with ${rs \over 2} = k = l$ and $\D = -l$
(see Appendix A for the details). To be precise, as was explained in
section 2, no-label \svec s generate three secondary \svec s at the same
level, just by the action of \Gn\ and \Qn , which cannot `come back' to
the no-label \svec\ (see Fig. V). It happens that one of these \svec s 
corresponds to the series $f_{r,s}(\D, \htop, t)=0$, another one 
corresponds to the series $g_k^{\pm}(\D, \htop, t)=0$, and the remaining 
one corresponds to both series.
For example, in the case of an uncharged no-label \svec\ 
$\ket{\chi}_{l}^{(0)}$ the three secondary \svec s are of the types:
$\ket{\chi}_{l}^{(1)G} = \cG_0 \ket{\chi}_{l}^{(0)}$,
$\ket{\chi}_{l}^{(-1)Q} = \cQ_0 \ket{\chi}_{l}^{(0)}$ and
$\ket{\chi}_{l}^{(0)G,Q} = \cQ_0 \cG_0 \ket{\chi}_{l}^{(0)}
  = - \cG_0 \cQ_0 \ket{\chi}_{l}^{(0)}$ (see Fig. V), whereas in the case
of a charged no-label \svec\ $\ket{\chi}_{l}^{(-1)}$ the three secondary
\svec s are of the types:
$\ket{\chi}_{l}^{(0)G} = \cG_0 \ket{\chi}_{l}^{(-1)}$,
$\ket{\chi}_{l}^{(-2)Q} = \cQ_0 \ket{\chi}_{l}^{(-1)}$ and
$\ket{\chi}_{l}^{(-1)G,Q} = \cQ_0 \cG_0 \ket{\chi}_{l}^{(-1)}$. In 
Appendix A we show the no-label \svec\ $\ket{\chi}_{1}^{(0)}$, at level 1, 
built on the \Gn-closed h.w. vector $\ket{-1,\htop}^G$ together 
with the three secondary \svec s at level 1: $\ket{\chi}_{1}^{(1)G}$,
$\ket{\chi}_{1}^{(-1)Q}$ and $\ket{\chi}_{1}^{(0)G,Q}$.

\vskip .5in
\baselineskip 12pt
\vbox{
\bpic{100}{120}{20}{-20}
% chargeline
%\put(0,195){\ch{G sector:}}
%\put(20,195){\ch{-2}}
%\put(50,195){\ch{-1}}
%\put(80,195){\ch{0}}
%\put(110,195){\ch{+1}}
%\put(0,185){\ch{Q sector:}}
%\put(20,185){\ch{-1}}
%\put(50,185){\ch{0}}
%\put(80,185){\ch{+1}}
%\put(110,185){\ch{+2}}
% Diagram 1 
\put(48,20){\framebox(4,3){}}
\put(78,20){\framebox(4,3){}}
\put(54,24){\vector(1,0){22}}
\put(76,19){\vector(-1,0){22}}
\put(47,24){\line(-1,2){50}}
\put(82,24){\line(1,2){50}}
\put(65,14){\makebox(0,0){\scriptsize $\cQ_0$}}
\put(65,28){\makebox(0,0){\scriptsize $\cG_0$}}
\put(47,15){\makebox(0,0)[r]{\scriptsize $\ket{-l,\htop-1}^Q$}}
\put(83,15){\makebox(0,0)[l]{\scriptsize $\ket{-l,\htop}^G$}}

\multiput(80,24) (0,5) {14}{\circle*{.5}}
\multiput(78,24) (-2,5) {14}{\circle*{.5}}
\multiput(81,24) (2,5) {14}{\circle*{.5}}

\put(51,91){\circle*{3}}
\put(80,89){\circle*{3}}
\put(80,93){\circle*{3}}
\put(108,91){\circle*{3}}
\put(76,89){\vector(-1,0){22}}
\put(54,93){\vector(1,0){22}}
\put(84,89){\vector(1,0){22}}
\put(106,93){\vector(-1,0){22}}
%\put(65,98){\makebox(0,0){\scriptsize $\cQ_0$}}
%\put(65,84){\makebox(0,0){\scriptsize $\cG_0$}}
\put(101,76){\makebox(0,0)[rb]{\scriptsize $\ket{\chi}_{l}^{(0)}$}}
\put(50,89){\makebox(0,0)[rb]{\scriptsize 
	$\ket{\chi}_{l}^{(-1)Q}$}}
\put(116,82){\makebox(0,0)[lb]{\scriptsize 
	$\ket{\chi}_{l}^{(1)G}$}}
\put(60,94){\makebox(0,0)[lb]{\scriptsize 
	$\ket{\chi}_{l}^{(0)G,Q}$}}
%\put(65,-10){\case{no-label singular vector $\ket{\chi}_l^{(0)}$}}
\epic{. 
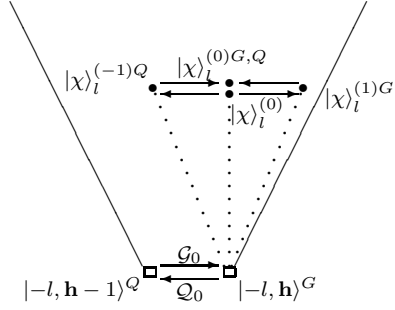
The uncharged no-label \svec\ $\ket{\chi}_{l}^{(0)}$ at level $l$, built
on the h.w. vector $\ket{-l,\htop}^G$, is the primitive \svec\ generating
the three secondary \svec s at level $l$:
$\ket{\chi}_{l}^{(1)G} = \cG_0 \ket{\chi}_{l}^{(0)}$,
$\ket{\chi}_{l}^{(-1)Q} = \cQ_0 \ket{\chi}_{l}^{(0)}$ and  
$\ket{\chi}_{l}^{(0)G,Q} = \cQ_0 \cG_0 \ket{\chi}_{l}^{(0)}
= - \cG_0 \cQ_0 \ket{\chi}_{l}^{(0)}$. These cannot generate the no-label
\svec\ by acting with the algebra. However, they are the \svec s detected
by the determinant formula, corresponding to the series 
$f_{r,s}(\D, \htop, t)=0$ ($\ket{\chi}_{l}^{(-1)Q}$ and
$\ket{\chi}_{l}^{(0)G,Q}$) and the series $g_{k}^+(\D, \htop, t)=0$
($\ket{\chi}_{l}^{(1)G}$ and $\ket{\chi}_{l}^{(0)G,Q}$). }}

\baselineskip 16pt

\vskip 0.17in
\noi 
{\bf Types of submodules}

Now let us identify the different types of submodules
that one finds in the generic Verma modules built on the
h.w. vectors $\ket{\D,\htop}^G$ and/or $\ket{\D,\htop-1}^Q$. We will take
into account the size of the submodules as well as their shape at the
bottom. The \svec s
corresponding to the series $f_{r,s}(\D, \htop, t)=0$ belong to two 
different types of submodules, as shown in Fig. VI, with the same
partition functions $P(l-{rs\over2})$
corresponding to complete Verma submodules, but with
drastically different shapes at the bottom of the submodules. In
the most general case the bottom of the submodule consists of two \svec s
connected by one or two horizontal arrows corresponding to
\Qn\ and/or \Gn\ (only one arrow if one of the \svec s is chiral). However,
it can also happen that the two \svec s at the bottom of the submodule
are chiral both, and therefore disconnected from each other. An important
observation is that a given Verma submodule may not be completely generated
by the \svec s at the bottom. These could generate the submodule only
partially, in which case one or more subsingular vectors generate the 
missing parts. 

\vskip .3in
\baselineskip 12pt

\vbox{
\bpic{400}{120}{20}{-20}
% chargeline
%\put(0,195){\ch{G sector:}}
%\put(20,195){\ch{-2}}
%\put(50,195){\ch{-1}}
%\put(80,195){\ch{0}}
%\put(110,195){\ch{+1}}
%\put(0,185){\ch{Q sector:}}
%\put(20,185){\ch{-1}}
%\put(50,185){\ch{0}}
%\put(80,185){\ch{+1}}
%\put(110,185){\ch{+2}}
% Diagram 1

\put(48,20){\framebox(4,3){}}
\put(78,20){\framebox(4,3){}}
\put(54,23){\vector(1,0){22}}
\put(76,20){\vector(-1,0){22}}
\put(47,24){\line(-1,2){50}}
\put(82,24){\line(1,2){50}}

%\put(65,14){\makebox(0,0){\scriptsize $\cQ_0$}}
%\put(65,14){\makebox(0,0){\scriptsize $\cG_0$}}
%\put(47,15){\makebox(0,0)[r]{\scriptsize $\ket{\Delta_{r,s},h-1}^Q$}}
%\put(78,15){\makebox(0,0)[l]{\scriptsize $\cG_0\ket{\Delta_{r,s},h-1}^Q$}}
%\put(50,24){\line(0,1){65}}
%\put(80,24){\line(0,1){65}}
%\put(52,24){\line(2,5){26}}
%\put(78,24){\line(-2,5){26}}

\put(51,51){\circle*{3}}
\put(79,51){\circle*{3}}
\put(54,49){\vector(1,0){22}}
\put(76,52){\vector(-1,0){22}}
\put(49,53){\line(-1,2){30}}
\put(81,53){\line(1,2){30}}

%\put(65,98){\makebox(0,0){\scriptsize $\cQ_0$}}
%\put(65,84){\makebox(0,0){\scriptsize $\cG_0$}}
%\put(67,94){\makebox(0,0)[rb]{\scriptsize $\ket{\chi}_{l,\ket{0,h-1}^Q}^{(0)Q}$}}
%\put(70,94){\makebox(0,0)[lb]{\scriptsize $\ket{\chi}_{l,\ket{0,h-1}^Q}^{(1)G}$}}
\put(65,-10){\case{$f_{r,s}(\D,\htop,t)=0$, $\Delta \neq - l$}}

% Diagram 2 

\put(198,20){\framebox(4,3){}}
\put(228,20){\framebox(4,3){}}
\put(204,23){\vector(1,0){22}}
\put(226,20){\vector(-1,0){22}}
\put(197,24){\line(-1,2){50}}
\put(232,24){\line(1,2){50}}
%\put(215,14){\makebox(0,0){\scriptsize $\cQ_0$}}
%\put(215,14){\makebox(0,0){\scriptsize $\cG_0$}}
%\put(197,15){\makebox(0,0)[r]{\scriptsize $\ket{\Delta_{r,s},h-1}^Q$}}
%\put(228,15){\makebox(0,0)[l]{\scriptsize $\cG_0\ket{\Delta_{r,s},h-1}^Q$}}
%\put(200,24){\line(0,1){65}}
%\put(230,24){\line(0,1){65}}
%\put(202,24){\line(2,5){26}}
%\put(228,24){\line(-2,5){26}}
\put(201,51){\circle*{3}}
\put(229,51){\circle*{3}}
%\put(204,49){\vector(1,0){22}}
\put(226,51){\vector(-1,0){22}}
\put(199,53){\line(-1,2){30}}
\put(231,53){\line(1,2){30}}
%\put(215,98){\makebox(0,0){\scriptsize $\cQ_0$}}
%\put(215,84){\makebox(0,0){\scriptsize $\cG_0$}}
%\put(217,94){\makebox(0,0)[rb]{\scriptsize $\ket{\chi}_{l,\ket{0,h-1}^Q}^{(0)Q}$}}
%\put(220,94){\makebox(0,0)[lb]{\scriptsize $\ket{\chi}_{l,\ket{0,h-1}^Q}^{(1)G}$}}
\put(215,-10){\case{$f_{r,s}(\D, \htop, t)=0$, $\Delta = - l$, 
$t \neq - {s \over n}$}}

% Diagram 3
\put(348,20){\framebox(4,3){}}
\put(378,20){\framebox(4,3){}}
\put(354,23){\vector(1,0){22}}
\put(376,20){\vector(-1,0){22}}
\put(347,24){\line(-1,2){50}}
\put(382,24){\line(1,2){50}}
%\put(365,14){\makebox(0,0){\scriptsize $\cQ_0$}}
%\put(365,14){\makebox(0,0){\scriptsize $\cG_0$}}
%\put(347,15){\makebox(0,0)[r]{\scriptsize $\ket{\Delta_{r,s},h-1}^Q$}}
%\put(378,15){\makebox(0,0)[l]{\scriptsize $\cG_0\ket{\Delta_{r,s},h-1}^Q$}}
%\put(350,24){\line(0,1){65}}
%\put(380,24){\line(0,1){65}}
%\put(352,24){\line(2,5){26}}
%\put(378,24){\line(-2,5){26}}
\put(351,51){\circle*{3}}
\put(379,51){\circle*{3}}
%\put(354,49){\vector(1,0){22}}
%\put(376,52){\vector(-1,0){22}}
\put(349,53){\line(-1,2){30}}
\put(381,53){\line(1,2){30}}
\put(353,53){\line(1,2){30}}
\put(377,53){\line(-1,2){30}}
%\put(365,98){\makebox(0,0){\scriptsize $\cQ_0$}}
%\put(365,84){\makebox(0,0){\scriptsize $\cG_0$}}
%\put(367,94){\makebox(0,0)[rb]{\scriptsize $\ket{\chi}_{l,\ket{0,h-1}^Q}^{(0)Q}$}}
%\put(370,94){\makebox(0,0)[lb]{\scriptsize $\ket{\chi}_{l,\ket{0,h-1}^Q}^{(1)G}$}}
\put(365,-10){\case{$f_{r,s}(\D, \htop, t)=0$, $\Delta = - l$, 
$t = - {s \over n}$ }}
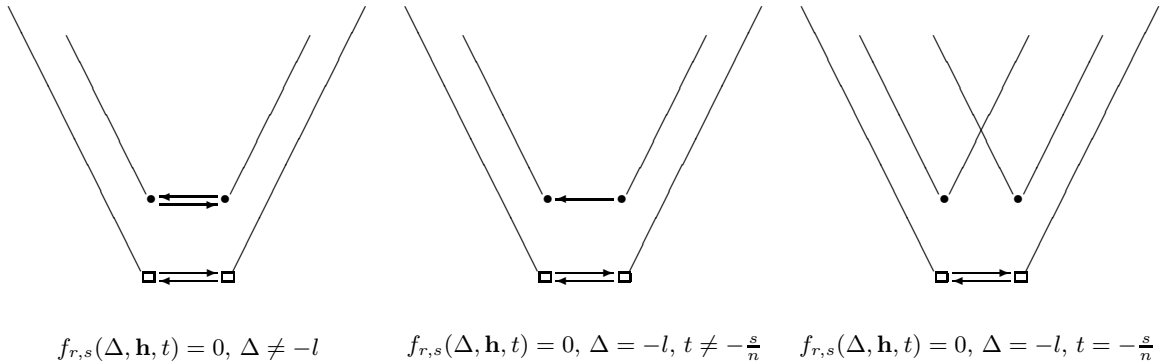
\epic{. The \svec s corresponding to the series $f_{r,s}(\D, \htop, t)=0$
belong to two different types of submodules of the same size (complete
Verma submodules). In the first type the two \svec s at the bottom of
the submodules are connected by \Gn\ and/or \Qn , depending on whether
$\D \neq -l$ or $\D = -l, \, t \neq - {s \over n}, \, n=1,..,r $ 
(for which one of the
\svec s is chiral). In the second type, corresponding to 
$\D = -l, \, t = - {s \over n}, \, n=1,..,r $, the two \svec s 
are chiral and therefore disconnected from each other. }}

\baselineskip 16pt

The \svec s
corresponding to the series $g_{k}^{\pm}(\D, \htop, t)=0$ belong to only
one type of submodule, as shown in Fig. VII, with partition functions
$P_k(l-k)$
corresponding to incomplete Verma submodules. The two \svec s at the 
bottom of the submodule are always connected by \Gn\ and/or \Qn\ since
at most one of these \svec s can be chiral. Finally, the no-label \svec s
generate their own no-label submodules. As shown in Fig. VII, at the
bottom of the no-label submodule one finds four \svec s: the primitive
no-label \svec\ and the three secondary \svec s generated by acting 
on this one with \Qn\ and \Gn . No-label submodules are therefore
much wider than the other types of submodules. 

\vskip .5in

\baselineskip 12pt
\vbox{
\bpic{400}{120}{20}{-20}
% chargeline
%\put(0,195){\ch{G sector:}}
%\put(20,195){\ch{-2}}
%\put(50,195){\ch{-1}}
%\put(80,195){\ch{0}}
%\put(110,195){\ch{+1}}
%\put(0,185){\ch{Q sector:}}
%\put(20,185){\ch{-1}}
%\put(50,185){\ch{0}}
%\put(80,185){\ch{+1}}
%\put(110,185){\ch{+2}}
% Diagram 1

\put(48,20){\framebox(4,3){}}
\put(78,20){\framebox(4,3){}}
\put(54,23){\vector(1,0){22}}
\put(76,20){\vector(-1,0){22}}
\put(47,24){\line(-1,2){50}}
\put(82,24){\line(1,2){50}}

%\put(65,14){\makebox(0,0){\scriptsize $\cQ_0$}}
%\put(65,14){\makebox(0,0){\scriptsize $\cG_0$}}
%\put(47,15){\makebox(0,0)[r]{\scriptsize $\ket{\Delta_{r,s},h-1}^Q$}}
%\put(78,15){\makebox(0,0)[l]{\scriptsize $\cG_0\ket{\Delta_{r,s},h-1}^Q$}}
%\put(50,24){\line(0,1){65}}
%\put(80,24){\line(0,1){65}}
%\put(52,24){\line(2,5){26}}
%\put(78,24){\line(-2,5){26}}

\put(81,81){\circle*{3}}
\put(99,81){\circle*{3}}
\put(84,79){\vector(1,0){12}}
\put(96,82){\vector(-1,0){12}}
\put(79,83){\line(-1,2){20}}
\put(101,83){\line(1,2){20}}
%\put(111,103){\line(1,3){7}}

%\put(65,98){\makebox(0,0){\scriptsize $\cQ_0$}}
%\put(65,84){\makebox(0,0){\scriptsize $\cG_0$}}
%\put(67,94){\makebox(0,0)[rb]{\scriptsize 
%$\ket{\chi}_{l,\ket{0,h-1}^Q}^{(0)Q}$}}
%\put(70,94){\makebox(0,0)[lb]{\scriptsize 
%$\ket{\chi}_{l,\ket{0,h-1}^Q}^{(1)G}$}}
\put(65,-10){\case{$g^{\pm}_k(\Delta,\htop,t)=0$, $\D \neq -l$}}
% Diagram 2 
\put(198,20){\framebox(4,3){}}
\put(228,20){\framebox(4,3){}}
\put(204,23){\vector(1,0){22}}
\put(226,20){\vector(-1,0){22}}
\put(197,24){\line(-1,2){50}}
\put(232,24){\line(1,2){50}}

%\put(215,14){\makebox(0,0){\scriptsize $\cQ_0$}}
%\put(215,14){\makebox(0,0){\scriptsize $\cG_0$}}
%\put(197,15){\makebox(0,0)[r]{\scriptsize $\ket{\Delta_{r,s},h-1}^Q$}}
%\put(228,15){\makebox(0,0)[l]{\scriptsize $\cG_0\ket{\Delta_{r,s},h-1}^Q$}}
%\put(200,24){\line(0,1){65}}
%\put(230,24){\line(0,1){65}}
%\put(202,24){\line(2,5){26}}
%\put(228,24){\line(-2,5){26}}

\put(231,81){\circle*{3}}
\put(249,81){\circle*{3}}
\put(234,79){\vector(1,0){12}}
%\put(246,82){\vector(-1,0){12}}
\put(229,83){\line(-1,2){20}}
\put(251,83){\line(1,2){20}}

%\put(215,98){\makebox(0,0){\scriptsize $\cQ_0$}}
%\put(215,84){\makebox(0,0){\scriptsize $\cG_0$}}
%\put(217,94){\makebox(0,0)[rb]{\scriptsize 
%$\ket{\chi}_{l,\ket{0,h-1}^Q}^{(0)Q}$}}
%\put(220,94){\makebox(0,0)[lb]{\scriptsize 
%$\ket{\chi}_{l,\ket{0,h-1}^Q}^{(1)G}$}}
\put(215,-10){\case{$g^{\pm}_k(\Delta,\htop,t)=0$, $\D = -l$}}

% Diagram 3 
\put(348,20){\framebox(4,3){}}
\put(378,20){\framebox(4,3){}}
\put(354,23){\vector(1,0){22}}
\put(376,20){\vector(-1,0){22}}
\put(347,24){\line(-1,2){50}}
\put(382,24){\line(1,2){50}}

\put(361,80){\circle*{3}}
\put(380,78){\circle*{3}}
\put(380,82){\circle*{3}}
\put(399,80){\circle*{3}}
\put(384,78){\vector(1,0){12}}
\put(396,82){\vector(-1,0){12}}
\put(364,82){\vector(1,0){12}}
\put(376,78){\vector(-1,0){12}}
\put(359,83){\line(-1,2){20}}
\put(401,83){\line(1,2){20}}

\put(365,-10){\case{no-label submodules}}
\epic{. 
The \svec s
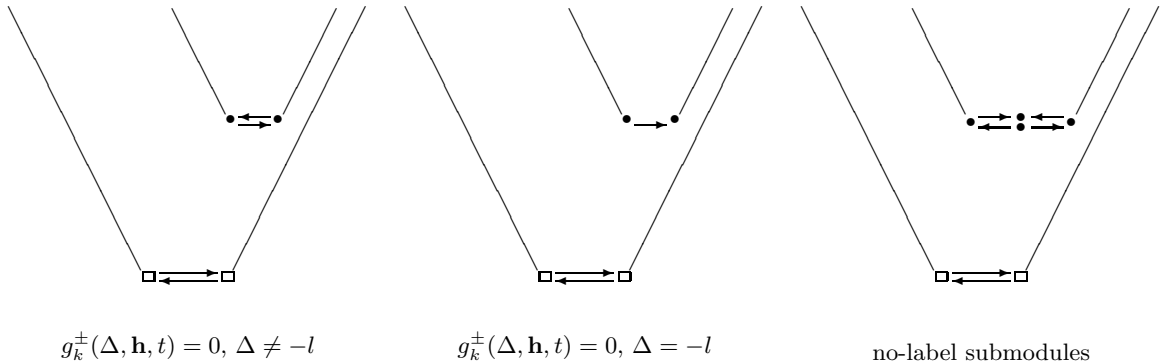
corresponding to the series $g_{k}^{\pm}(\D, \htop, t)=0$ belong to only
one type of submodules (incomplete Verma submodules). 
The two \svec s at the bottom of the submodules are connected 
by \Gn\ and/or \Qn , depending on whether $\D \neq -l$ or $\D=-l$.
The no-label \svec s generate the widest submodules with four
\svec s at the bottom. }}

\vskip .2in

\baselineskip 16pt

We see therefore that there are four different types of submodules
that may appear in generic Verma modules, distinguished by their size
and/or the shape at the bottom of the submodule\footnote{In the literature
there are claims (without proofs) that there are only two different types 
of submodules 
in generic (standard) Verma modules of the Topological N=2 algebra. 
In particular, the chiral-chiral submodules on the right of Fig. VI 
and the no-label submodules on the right of Fig. VII do not fit
into that classification.}. A more accurate classification of the
submodules should take into account also the shape of the whole  
submodule, including the possible existence of subsingular 
vectors \cite{DB5}.

\subsection{Chiral Verma modules}\lvm

Chiral Verma modules $V(\ket{0,\htop}^{G,Q})$ result from the quotient of
generic Verma modules of types $V(\ket{0,\htop}^Q)$
and $V(\ket{0,\htop}^G)$, by the submodules generated by the level-zero
singular vectors $\,\Gz\ket{0,\htop}^Q$ and 
$\,\Qz\ket{0,\htop}^G$, respectively. Therefore there are two possible
origins for the \svec s in chiral Verma modules, as shown in Fig. VIII:
they are either the `surviving' \svec s in $V(\ket{0,\htop}^G)$ or
$V(\ket{0,\htop}^Q)$ which do not belong to
the submodules set to zero, or they appear when subsingular vectors 
in $V(\ket{0,\htop}^G)$ or $V(\ket{0,\htop}^Q)$ are converted into 
\svec s in $V(\ket{0,\htop}^{G,Q})$ as a consequence of the quotient 
procedure\footnote{We stress the fact that these two are the only possible
origins for the \svec s in chiral Verma modules (by definition of
subsingular vectors). The failure to see this 
has lead to very confusing statements in the literature where \svec s
in chiral Verma modules are viewed as objects of different nature than
the singular and subsingular vectors in the generic Verma modules.}. In
the first case one deduces straightforwardly, from the results of tables
\req{tabl2} and \req{tabl3}, that the only possible types of such surviving 
\svec s are the four types shown in table \req{tabl1}: 
$\,\ket{\chi}_l^{(0)G}$, $\ket{\chi}_l^{(1)G}$, 
$\ket{\chi}_l^{(0)Q}$ and $\ket{\chi}_l^{(-1)Q}$. 
In the second case, when the \svec s in chiral Verma modules originate from
subsingular vectors, one cannot deduce their possible types from tables
\req{tabl2} and \req{tabl3}. However, we know that the only possible types 
of such \svec s are again the four types given in table \req{tabl1} 
since we have
proved, in ref. \cite{DB2}, that the space associated to any other type of
`would-be' \svec\ in chiral Verma modules has maximal dimension zero (\ie\
no \svec s of such types exist). Observe that in chiral Verma modules
there are no chiral neither no-label \svec s. As a consequence the action 
of either \Gn\ or \Qn\ on a \svec\ produces always another \svec . As a 
result the \svec s in chiral Verma modules 
come always two by two at the same level in the 
same Verma module, \Gn\ and \Qn\ interpolating between them.

\vskip .5in
\baselineskip 12pt
\vbox{
\bpic{250}{120}{20}{-20}
% chargeline
%\put(0,195){\ch{G sector:}}
%\put(20,195){\ch{-2}}
%\put(50,195){\ch{-1}}
%\put(80,195){\ch{0}}
%\put(110,195){\ch{+1}}
%\put(0,185){\ch{Q sector:}}
%\put(20,185){\ch{-1}}
%\put(50,185){\ch{0}}
%\put(80,185){\ch{+1}}
%\put(110,185){\ch{+2}}
% Diagram 1 
\put(50,20){\circle*{.3}}
\put(78,20){\framebox(4,3){}}
%\put(54,23){\vector(1,0){22}}
\put(76,20){\vector(-1,0){22}}
\put(47,24){\line(-1,2){50}}
\put(82,24){\line(1,2){50}}
\put(65,14){\makebox(0,0){\scriptsize $\cQ_0$}}
%\put(60,28){\makebox(0,0){\scriptsize $\cG_0$}}
\put(47,15){\makebox(0,0)[r]{\scriptsize $\cQ_0\ket{0,\htop}^{G}\equiv 0$}}
\put(83,15){\makebox(0,0)[l]{\scriptsize $\ket{0,\htop}^{G}$}}

\multiput(80,24) (0,5) {14}{\circle*{.5}}

\put(51,24){\line(1,6){16}}
\put(49,24){\line(-1,3){32}}
%\multiput(51,24)(.5,3){32}{\circle*{.3}}
\put(45,38){\line(2,-1){7}}
\put(40,53){\line(2,-1){13}}
\put(35,68){\line(2,-1){20}}
\put(30,83){\line(2,-1){27}}
\put(25,98){\line(2,-1){34}}
\put(40,103){\line(2,-1){21}}
%\put(60,120){\line(2,-1){8}}
%\put(78,24){\line(-2,5){26}}
%\put(76,24){\line(-4,5){52}}
\put(67,116){\makebox(0,0)[rb]{\scriptsize $V(\cQ_0\ket{0,\htop}^G)$}}
\put(50,106){\makebox(0,0)[rb]{\scriptsize $\equiv \{ 0\}$}}
\put(80,91){\circle*{3}}
\put(110,94){\makebox(0,0)[rb]{\scriptsize $\ket{\chi}_{l,\ket{0,\htop}^G}$}}
\put(65,-10){\case{singular vector}}
% Diagram 2 
\put(200,20){\circle*{.3}}
\put(228,20){\framebox(4,3){}}
%\put(224,23){\vector(1,0){22}}
\put(226,20){\vector(-1,0){22}}
\put(197,24){\line(-1,2){50}}
\put(232,24){\line(1,2){50}}
\put(215,14){\makebox(0,0){\scriptsize $\cQ_0$}}
%\put(210,28){\makebox(0,0){\scriptsize $\cG_0$}}
\put(197,15){\makebox(0,0)[r]{\scriptsize $\cQ_0\ket{0,\htop}^{G}\equiv 0$}}
\put(233,15){\makebox(0,0)[l]{\scriptsize $\ket{0,\htop}^{G}$}}
%\put(230,24){\line(0,1){65}}
\put(201,24){\line(1,6){16}}
\put(199,24){\line(-1,3){32}}
%\multiput(201,24)(.5,3){32}{\circle*{.3}}
\put(195,38){\line(2,-1){7}}
\put(190,53){\line(2,-1){13}}
\put(185,68){\line(2,-1){20}}
\put(180,83){\line(2,-1){27}}
\put(175,98){\line(2,-1){34}}
\put(190,103){\line(2,-1){21}}
%\put(210,120){\line(2,-1){8}}
%\put(228,24){\line(-2,5){26}}
%\put(226,24){\line(-4,5){52}}
\put(217,116){\makebox(0,0)[rb]{\scriptsize $V(\cQ_0\ket{0,\htop}^G)$}}
\put(200,106){\makebox(0,0)[rb]{\scriptsize $\equiv \{ 0\}$}}
\put(230,91){\circle*{3}}
\put(229,89){\vector(-1,-1){33}}
%\put(215,70){\makebox(0,0)[lb]{\scriptsize $A^+$}}
\put(260,94){\makebox(0,0)[rb]{\scriptsize $\ket{\Upsilon}_{l,
\ket{0,\htop}^G}$}}
\put(215,-10){\case{subsingular vector}}
\epic{. When the \svec\ $\cQ_0\ket{0,\htop}^G$ is set to zero, the generic
Verma module $V(\ket{0,\htop}^G)$
built on the h.w. vector $\ket{0,\htop}^G$ is
divided by the submodule $V(\cQ_0\ket{0,\htop}^G)$. As a consequence, the
\Gn-closed h.w. vector $\ket{0,\htop}^G$ becomes the chiral h.w. vector 
$\ket{0,\htop}^{G,Q}$ and the resulting Verma module is an incomplete, 
chiral Verma module $V(\ket{0,\htop}^{G,Q})$. 
The \svec s in the generic Verma module that are not
located inside the submodule $V(\cQ_0\ket{0,\htop}^G)$ remain after the
quotient procedure, they are the `surviving' \svec s. But it can also happen
that there are subsingular vectors outside the submodule
$V(\cQ_0\ket{0,\htop}^G)$, but descending to it by acting with the 
generators of the algebra. These subsingular vectors become singular
in the resulting chiral Verma module $V(\ket{0,\htop}^{G,Q})$. 
}}

\vskip .18in
\baselineskip 16pt

The determinant formulae for chiral Verma modules of the Topological,
Neveu-Schwarz and Ramond N=2 superconformal algebras were written down 
in ref. \cite{BJI5}. They were deduced from the determinant formulae for 
the complete generic Verma modules by imposing the ansatz that the roots
of the determinants are the same in both cases (for the values of the
conformal weights $\Delta$ corresponding to the chiral Verma modules).
Although no rigorous proofs were presented for this ansatz, the 
corresponding expressions were checked until level 4. For the case of 
the Topological algebra the chiral determinant formula is given as
function of the U(1) charges $\htop$ as the topological chiral Verma 
modules have zero conformal weight $\D=0$. It reads 
\BE
det(\cM^{T-ch}_l) = {\rm cst.} \prod_{2\leq rs \leq 2l}
(\htop-\htop_{r,s}^{(0)})^{2P_r(l-{rs\over2})}{\ \ }
(\htop-\htop_{r,s}^{(1)})^{2P_r(l-{rs\over2})}{\ \ \ }
r\in\oZ^+,{\ }s\in2\oZ^+ \,, \label{Tdet}
\EE
\noi
with the roots $\htop_{r,s}^{(0)}$ and $\htop_{r,s}^{(1)}$, given by
\BE \htop_{r,s}^{(0)} = {t \over 2}(1+r)-{s \over 2} \ , \ \ \ \ \ 
r \in {\bf Z}^+ , \ \ s \in 2{\bf Z}^+ \,, \label{hrs0} \EE
\BE \htop_{r,s}^{(1)} = {t \over 2}(1-r)+{s \over 2} -1 \ , \ \ \ \ \ 
r \in {\bf Z}^+ , \ \ s \in 2{\bf Z}^+  \,. \label{h1rs} \EE

\noi
Observe that $\htop_{r,s}^{(0)}$ is exactly the same as in eq. \req{h0rs}
(we write it here again for convenience). These roots satisfy 
$\htop_{r,s}^{(1)}=-\htop_{r,s}^{(0)}-\ctop/3$, what indicates that the
corresponding Verma modules are related by the odd spectral flow 
automorphism $\cA$ \cite{BJI5}\cite{B1} (see footnote 6).

\noi
The partition functions $P_r$ are defined as in eqn. \req{part}:
\BE
\sum_N P_r(N)x^N= {1\over 1+x^r}{\ } 
\prod_{0<l\in {\bf Z},{\ }0<m\in {\bf Z}} {(1+x^l)^2 \over (1-x^m)^2}\,.
\label{PT}
\EE
\noi
Therefore they correspond to incomplete Verma submodules, as we pointed 
out before. The factors 2 in the exponents of the determinants 
\req{Tdet} show explicitly again that the singular vectors 
come two by two at the same level in the same Verma module, as $P_r(0)=1$.  
One can also see in the determinant formula \req{Tdet}
that the \svec s corresponding to $\htop=\htop^{(0)}_{r,s}$ 
and the \svec s corresponding to $\htop=\htop^{(1)}_{r,s}$ both
belong to the same types of submodules (incomplete Verma submodules as the
chiral Verma modules where they are embedded).
 
The interpretation of the roots of the determinants 
in terms of singular vectors is much simpler than for generic Verma
modules. First, in chiral Verma modules there is only one h.w. vector
and therefore only one way to express the \svec s. Second, in chiral 
Verma modules there are no chiral \svec s neither no-label \svec s. 
The interpretation of the roots is as follows, as shown in Fig. IX.

\vskip .5in
\baselineskip 12pt
\vbox{
\bpic{300}{120}{10}{-20}
% chargeline
%\put(0,195){\ch{G sector:}}
%\put(20,195){\ch{-2}}
%\put(50,195){\ch{-1}}
%\put(80,195){\ch{0}}
%\put(110,195){\ch{+1}}
%\put(0,185){\ch{Q sector:}}
%\put(20,185){\ch{-1}}
%\put(50,185){\ch{0}}
%\put(80,185){\ch{+1}}
%\put(110,185){\ch{+2}}
% Diagram 1
%\put(50,20){\circle*{3}}
\put(78,20){\framebox(4,3){}}
%\put(54,23){\vector(1,0){22}}
%\put(76,20){\vector(-1,0){22}}
\put(77,24){\line(-1,2){50}}
\put(82,24){\line(1,2){50}}
%\put(65,14){\makebox(0,0){\scriptsize $\cQ_0$}}
%\put(65,28){\makebox(0,0){\scriptsize $\cG_0$}}
%\put(47,15){\makebox(0,0)[r]{\scriptsize $\cQ_0\ket{0,\htop}^G$}}
\put(70,15){\makebox(0,0)[l]{\scriptsize $\ket{0,\htop}^{G,Q}$}}
%\put(50,24){\line(0,1){65}}

\multiput(80,24) (0,5) {14}{\circle*{.5}}
\multiput(78,24) (-2,5) {14}{\circle*{.5}}
%\multiput(81,24) (2,5) {14}{\circle*{.5}}

%\put(80,24){\line(0,1){65}}
%\put(52,24){\line(2,5){26}}
%\put(78,24){\line(-2,5){26}}
\put(51,91){\circle*{3}}
\put(80,91){\circle*{3}}
\put(54,89){\vector(1,0){22}}
\put(76,92){\vector(-1,0){22}}

%\put(65,98){\makebox(0,0){\scriptsize $\cQ_0$}}
%\put(65,84){\makebox(0,0){\scriptsize $\cG_0$}}
\put(47,73){\makebox(0,0)[rb]{\scriptsize $\ket{\chi}_{l}^{(-1)Q}$}}
\put(70,94){\makebox(0,0)[lb]{\scriptsize $\ket{\chi}_{l}^{(0)G}$}}
\put(80,-10){\case{$\htop=\htop^{(0)}_{r,s}$}}
% Diagram 2
%\put(200,20){\circle*{3}}
\put(228,20){\framebox(4,3){}}
%\put(204,23){\vector(1,0){22}}
%\put(226,20){\vector(-1,0){22}}
\put(227,24){\line(-1,2){50}}
\put(232,24){\line(1,2){50}}
%\put(215,14){\makebox(0,0){\scriptsize $\cQ_0$}}
%\put(215,28){\makebox(0,0){\scriptsize $\cG_0$}}
%\put(197,15){\makebox(0,0)[r]{\scriptsize $\cQ_0\ket{0,h}^G$}}
\put(220,15){\makebox(0,0)[l]{\scriptsize $\ket{0,\htop}^{G,Q}$}}
%\put(200,24){\line(0,1){65}}

\multiput(230,24) (0,5) {14}{\circle*{.5}}
\multiput(232,24) (2,5) {14}{\circle*{.5}}

%\put(230,24){\line(0,1){65}}
%\put(202,24){\line(2,5){26}}
%put(232,24){\line(2,5){26}}
\put(259,91){\circle*{3}}
\put(230,91){\circle*{3}}
\put(234,89){\vector(1,0){22}}
\put(256,92){\vector(-1,0){22}}

%\put(215,98){\makebox(0,0){\scriptsize $\cQ_0$}}
%\put(215,84){\makebox(0,0){\scriptsize $\cG_0$}}
\put(254,94){\makebox(0,0)[rb]{\scriptsize $\ket{\chi}_{l}^{(0)Q}$}}
\put(263,73){\makebox(0,0)[lb]{\scriptsize $\ket{\chi}_{l}^{(1)G}$}}
\put(230,-10){\case{$\htop=\htop^{(1)}_{r,s}$}}
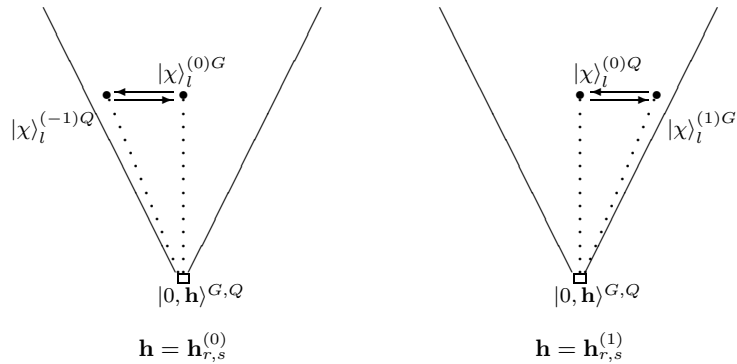
\epic{. In chiral Verma modules built on chiral h.w. vectors 
$\ket{0,\htop}^{G,Q}$, for $\htop=\htop^{(0)}_{r,s}$ one finds at level
$l={rs \over 2}$ one uncharged \svec\ of type $\ket{\chi}_l^{(0)G}$ 
and one charged \svec\ of type $\ket{\chi}_l^{(-1)Q}$, connected by 
\Gn\ and \Qn . For $\htop=\htop^{(1)}_{r,s}$ one finds at level
$l={rs \over 2}$ one uncharged \svec\ of type $\ket{\chi}_l^{(0)Q}$ 
and one charged \svec\ of type $\ket{\chi}_l^{(1)G}$, connected by 
\Gn\ and \Qn .}}

\baselineskip 16pt

\vskip 0.4in
\noi 
{\bf Case $\htop=\htop^{(0)}_{r,s}$}

For $\htop=\htop^{(0)}_{r,s}$ the Verma module built on the chiral h.w.
vector $\ket{0,\htop^{(0)}_{r,s}}^{G,Q}$ has (at least) one uncharged
\svec\ of type $\ket{\chi}_l^{(0)G}$ and one charged \svec\ of type
$\ket{\chi}_l^{(-1)Q}$ at level $l={rs\over2}$, \Gn\ and \Qn\ interpolating
between them. 

\vskip 0.17in
\noi 
{\bf Case $\htop=\htop^{(1)}_{r,s}$}

For $\htop=\htop^{(1)}_{r,s}$ the Verma module built on the chiral h.w.
vector $\ket{0,\htop^{(1)}_{r,s}}^{G,Q}$ has (at least) one uncharged
\svec\ of type $\ket{\chi}_l^{(0)Q}$ and one charged \svec\ of type
$\ket{\chi}_l^{(1)G}$ at level $l={rs\over2}$, \Gn\ and \Qn\ interpolating
between them.

\vskip 0.2in
Now let us analyse the origin of the different \svec s in the chiral Verma
modules $V(\ket{0,\htop^{(0)}_{r,s}}^{G,Q})$ and
$V(\ket{0,\htop^{(1)}_{r,s}}^{G,Q})$
with respect to the singular and subsingular vectors in the generic  
Verma modules. In other words, let us investigate which of 
the \svec s in the generic Verma modules are lost in the quotient,
which are not quotiented out, resulting in 
\svec s in the chiral Verma modules,  
and which are `new' \svec s in the chiral Verma modules
that were not singular in the generic Verma modules but subsingular, 
as a consequence. For this purpose we will consider the two possible
quotients which give rise to the same chiral Verma module, \ie\
$V(\ket{0,\htop}^{G,Q})= V(\ket{0,\htop}^G) / V(\Qz \ket{0,\htop}^G)$ and 
$V(\ket{0,\htop}^{G,Q})= V(\ket{0,\htop}^Q) / V(\Gz \ket{0,\htop}^Q)$.

Let us start with the realization of the chiral Verma modules as the quotient 
$V(\ket{0,\htop}^{G,Q})= V(\ket{0,\htop}^G) / V(\Qz \ket{0,\htop}^G)$. 
As was discussed in the last subsection, setting $\D=0$ in the determinant
formula \req{det1} one finds the generic Verma modules of type
$V(\ket{0,\htop}^G$ which contain \svec s as follows. From 
$f_{r,s}(0,\htop,t\neq0)=0$ one gets the solutions 
$\htop=\htop^{(0)}_{r,s}$ and $\htop=\hat\htop_{r,s}$, 
eqns. \req{h0rs} and \req{hhrs}, whereas from $f_{r,s}(0,\htop,t=0)=0$  
one obtains $\htop=\pm {s\over 2}$. For all these solutions 
the corresponding \svec s are pairs of types $\ket{\chi}_l^{(0)G}$ 
and $\ket{\chi}_l^{(-1)Q}$ at level $l={rs\over2}$. From 
$g_k^+(0,\htop,t)=0$ one gets the solution $\htop=\htop^+_k$, eqn.
\req{hpml}, the corresponding \svec s being pairs of types 
$\ket{\chi}_l^{(1)G}$ and $\ket{\chi}_l^{(0)Q}$ at level $l=k$. 
Finally from $g_k^-(0,\htop,t)=0$ 
one obtains $\htop=\htop^-_k$, eqn. \req{hpml},
the corresponding \svec s being pairs of types 
$\ket{\chi}_l^{(-1)G}$ and $\ket{\chi}_l^{(-2)Q}$ at level $l=k$.
After the quotient $V(\ket{0,\htop}^G) / V(\Qz \ket{0,\htop}^G)$ is 
performed, by setting $\Qz \ket{0,\htop}^G \equiv 0$, one obtains the chiral
Verma modules $V(\ket{0,\htop}^{G,Q})$ which contain \svec s only  for
$\htop=\htop^{(0)}_{r,s}$ and $\htop=\htop^{(1)}_{r,s}$, eqns. \req{h0rs}
and \req{h1rs}, the corresponding \svec s being pairs of types  
$\ket{\chi}_l^{(0)G}$, $\ket{\chi}_l^{(-1)Q}$ and $\ket{\chi}_l^{(1)G}$, 
$\ket{\chi}_l^{(0)Q}$, respectively, at level $l={rs\over2}$. A simple
analysis of these facts, taking into account that
$\htop^+_l=\htop^{(1)}_{r,2}$ and $\hat\htop_{r,s}=\htop^{(1)}_{r,s+2}$,
allows us to deduce the following results\footnote{Some of these results
were already given in refs. \cite{BJI6} and \cite{BJI5}.}:
\vskip .2in

i) In the chiral Verma modules $V(\ket{0,\htop^{(0)}_{r,s}}^{G,Q})$ all the
original pairs of \svec s in $V(\ket{0,\htop^{(0)}_{r,s}}^{G})$, of types
$\ket{\chi}_l^{(0)G}$ and $\ket{\chi}_l^{(-1)Q}$, with $l={rs\over2}$,
remain after the quotient.
Therefore they were located outside the submodule generated by the 
level zero \svec\ $\Qz \ket{0,\htop^{(0)}_{r,s}}^G$, as shown in Fig. I.

ii) In the chiral Verma modules $V(\ket{0,\hat\htop_{r,s}}^{G,Q})=
V(\ket{0,\htop^{(1)}_{r,s+2}}^{G,Q})$
all the original pairs of \svec s in 
$V(\ket{0,\hat\htop_{r,s}}^{G})$, of types $\ket{\chi}_l^{(0)G}$ and 
$\ket{\chi}_l^{(-1)Q}$, with $l={rs\over2}$, have disappeared after the 
quotient. Therefore they were located inside the submodule generated by the 
level zero \svec\ $\Qz \ket{0,\hat\htop_{r,s}}^G$, as shown in Fig. I. 
Moreover, in their place
new \svec s appear, of types $\ket{\chi}_l^{(1)G}$ and $\ket{\chi}_l^{(0)Q}$,
with $l={r(s+2)\over2}$, which were not present before the quotient. These 
\svec s are therefore subsingular in $V(\ket{0,\hat\htop_{r,s}}^{G})$.

iii) For $t=0$ in the chiral Verma modules 
$V(\ket{0,\htop=-{s\over2}}^{G,Q}) = 
V(\ket{0,\htop^{(0)}_{r,s,t=0}}^{G,Q})$ all the original pairs of \svec s 
in $V(\ket{0,\htop=-{s\over2}}^{G})$, of types
$\ket{\chi}_l^{(0)G}$ and $\ket{\chi}_l^{(-1)Q}$, with $l={rs\over2}$,
remain after the quotient.
Therefore they were located outside the submodule generated by the 
level zero \svec\ $\Qz \ket{0,\htop=-{s\over2}}^G$.

iv) For $t=0$ in the chiral Verma modules 
$V(\ket{0,\htop={s\over2}}^{G,Q})=  
V(\ket{0,\htop^{(1)}_{r,s+2,t=0}}^{G,Q})$ 
all the original pairs of \svec s in 
$V(\ket{0,\htop={s\over2}}^{G})$, of types $\ket{\chi}_l^{(0)G}$ and 
$\ket{\chi}_l^{(-1)Q}$, with $l={rs\over2}$, have disappeared after the 
quotient. Therefore they were located inside the submodule generated by the 
level zero \svec\ $\Qz \ket{0,\htop={s\over2}}^G$. Moreover, in their place
new \svec s appear, of types $\ket{\chi}_l^{(1)G}$ and $\ket{\chi}_l^{(0)Q}$,
with $l={r(s+2)\over2}$, which were not present before the quotient. These 
\svec s are therefore subsingular in $V(\ket{0,\htop={s\over2}}^{G})$.

v) In the chiral Verma modules $V(\ket{0,\htop^+_l}^{G,Q}) = 
V(\ket{0,\htop^{(1)}_{r,2}}^{G,Q})$ all the
original pairs of \svec s in $V(\ket{0,\htop^+_l}^{G})$, of types
$\ket{\chi}_l^{(1)G}$ and $\ket{\chi}_l^{(0)Q}$,
remain after the quotient.
Therefore they were located outside the submodule generated by the 
level zero \svec\ $\Qz \ket{0,\htop^+_l}^G$, as shown in Fig. III.

vi) In the chiral Verma modules $V(\ket{0,\htop^-_l}^{G,Q})$ all the
original pairs of \svec s in $V(\ket{0,\htop^-_l}^{G})$, of types
$\ket{\chi}_l^{(-1)G}$ and $\ket{\chi}_l^{(-2)Q}$, have disappeared after 
the quotient. Therefore they were located inside the submodule generated by 
the level zero \svec\ $\Qz \ket{0,\htop^-_l}^G$, as shown in Fig. IV. 
No new \svec s appear corresponding 
to possible subsingular vectors in $V(\ket{0,\htop^-_l}^{G})$.

\vskip .2in

Let us now investigate the realization of the chiral Verma modules as the 
quotient 
$V(\ket{0,\htop}^{G,Q})= V(\ket{0,\htop}^Q) / V(\Gz \ket{0,\htop}^Q)$. 
The generic Verma modules of type $V(\ket{0,\htop}^Q)$ which contain \svec s 
are the following. From $f_{r,s}(0,\htop,t\neq0)=0$ one gets the 
solutions $\htop=\htop^{(0)}_{r,s}-1$ and $\htop=\hat\htop_{r,s}-1$, where 
$\htop^{(0)}_{r,s}$ and  $\hat\htop_{r,s}$ are given by
eqns. \req{h0rs} and \req{hhrs}, whereas from 
$f_{r,s}(0,\htop,t=0)=0$  one obtains $\htop=\pm {s\over 2}-1$. For all
these solutions the corresponding \svec s are pairs of types 
$\ket{\chi}_l^{(0)Q}$ and $\ket{\chi}_l^{(1)G}$ at 
level $l={rs\over2}$. From $g_k^+(0,\htop,t)=0$ one gets the 
solution $\htop=\htop^+_k-1$, with 
$\htop^+_k$ given by eqn. \req{hpml}. The corresponding 
\svec s are pairs of types $\ket{\chi}_l^{(1)Q}$ and 
$\ket{\chi}_l^{(2)G}$ at level $l=k$. Finally from 
$g_k^-(0,\htop,t)=0$ one obtains $\htop=\htop^-_k-1$, 
with $\htop^-_k$ given by eqn. \req{hpml},
the corresponding \svec s being pairs of types 
$\ket{\chi}_l^{(-1)Q}$ and $\ket{\chi}_l^{(0)G}$ at level $l=k$.
After the quotient $V(\ket{0,\htop}^Q) / V(\Gz \ket{0,\htop}^Q)$ is 
performed, by setting $\Gz \ket{0,\htop}^Q \equiv 0$, one obtains as 
in the previous case the chiral
Verma modules $V(\ket{0,\htop}^{G,Q})$ which contain \svec s only for
$\htop=\htop^{(0)}_{r,s}$ and $\htop=\htop^{(1)}_{r,s}$, eqns. \req{h0rs}
and \req{h1rs}. Repeating the same analysis as before, taking into
account that $\htop^{(0)}_{r,s}-1=\htop^{(0)}_{r,s+2}$, 
$\,\hat\htop_{r,s}-1=\htop^{(1)}_{r,s}\,$ and 
$\,\htop^-_l-1=\htop^{(0)}_{r,2}$, one finds the following results:

\vskip .2in

i) In the chiral Verma modules $V(\ket{0,\hat\htop_{r,s}-1}^{G,Q})=
V(\ket{0,\htop^{(1)}_{r,s}}^{G,Q})$ all the
original pairs of \svec s in $V(\ket{0,\hat\htop_{r,s}-1}^{Q})$, of types
$\ket{\chi}_l^{(0)Q}$ and $\ket{\chi}_l^{(1)G}$, with $l={rs\over2}$,
remain after the quotient.
Therefore they were located outside the submodule generated by the 
level zero \svec\ $\Gz \ket{0,\hat\htop_{r,s}-1}^Q$, as shown in Fig. II.

ii) In the chiral Verma modules $V(\ket{0,\htop^{(0)}_{r,s}-1}^{G,Q}) =
V(\ket{0,\htop^{(0)}_{r,s+2}}^{G,Q})$
all the original pairs of \svec s in 
$V(\ket{0,\htop^{(0)}_{r,s}-1}^{Q})$, of types $\ket{\chi}_l^{(0)Q}$ and 
$\ket{\chi}_l^{(1)G}$, with $l={rs\over2}$, have disappeared after the 
quotient. Therefore they were located inside the submodule generated by the 
level zero \svec\ $\Gz \ket{0,\htop^{(0)}_{r,s}-1}^Q$, as shown in Fig. II. 
Moreover, in their place new \svec s appear, 
of types $\ket{\chi}_l^{(0)G}$ and $\ket{\chi}_l^{(-1)Q}$,
with $l={r(s+2)\over2}$, which were not present before the quotient. These 
\svec s are therefore subsingular in $V(\ket{0,\htop^{(0)}_{r,s}-1}^{Q})$.

iii) For $t=0$ in the chiral Verma modules  
$V(\ket{0,\htop={s\over2}-1}^{G,Q})= 
V(\ket{0,\htop^{(1)}_{r,s,t=0}}^{G,Q})$ 
all the original pairs of \svec s in $V(\ket{0,\htop={s\over2}-1}^{Q})$, of 
types $\ket{\chi}_l^{(0)Q}$ and $\ket{\chi}_l^{(1)G}$, with $l={rs\over2}$,
remain after the quotient.
Therefore they were located outside the submodule generated by the 
level zero \svec\ $\Gz \ket{0,\htop={s\over2}-1}^Q$.

iv) For $t=0$ in the chiral Verma modules
$V(\ket{0,\htop=-{s\over2}-1}^{G,Q}) =  
V(\ket{0,\htop^{(0)}_{r,s+2,t=0}}^{G,Q})$ 
all the original pairs of \svec s in 
$V(\ket{0,\htop=-{s\over2}-1}^{Q})$, of types $\ket{\chi}_l^{(0)Q}$ and 
$\ket{\chi}_l^{(1)G}$, with $l={rs\over2}$, have disappeared after the 
quotient. Therefore they were located inside the submodule generated by the 
level zero \svec\ $\Gz \ket{0,\htop=-{s\over2}-1}^Q$. 
Moreover, in their place new \svec s appear, 
of types $\ket{\chi}_l^{(0)G}$ and $\ket{\chi}_l^{(-1)Q}$,
with $l={r(s+2)\over2}$, which were not present before the quotient. These 
\svec s are therefore subsingular in $V(\ket{0,\htop=-{s\over2}-1}^{Q})$.

v) In the chiral Verma modules $V(\ket{0,\htop^-_l-1}^{G,Q}) = 
V(\ket{0,\htop^{(0)}_{r,2}}^{G,Q})$ all the
original pairs of \svec s in $V(\ket{0,\htop^-_l-1}^{Q})$, of types
$\ket{\chi}_l^{(-1)Q}$ and $\ket{\chi}_l^{(0)G}$,
remain after the quotient.
Therefore they were located outside the submodule generated by the 
level zero \svec\ $\Gz \ket{0,\htop^-_l-1}^Q$, as shown in Fig. IV.

vi) In the chiral Verma modules $V(\ket{0,\htop^+_l-1}^{G,Q})$ all the
original pairs of \svec s in $V(\ket{0,\htop^+_l-1}^{Q})$, of types
$\ket{\chi}_l^{(1)Q}$ and $\ket{\chi}_l^{(2)G}$, have disappeared after 
the quotient. Therefore they were located inside the submodule generated by 
the level zero \svec\ $\Gz \ket{0,\htop^+_l-1}^Q$, as shown in Fig. III. 
No new \svec s appear corresponding to possible subsingular vectors in 
$V(\ket{0,\htop^+_l-1}^{Q})$.

\vskip .2in
An interesting observation now \cite{BJI6} is that two series of \svec s in 
chiral Verma modules correspond to both \svec s and subsingular vectors
depending on the specific generic Verma modules. Namely, 
the \svec s of types $\ket{\chi}_l^{(0)G}$ and $\ket{\chi}_l^{(-1)Q}$
in the chiral Verma modules $V(\ket{0,\htop^{(0)}_{r,s}}^{G,Q})$
correspond to \svec s in $V(\ket{0,\htop^{(0)}_{r,s}}^{G})$ but to 
subsingular vectors in $V(\ket{0,\htop^{(0)}_{r,s}-1}^{Q})$ for $s>2$. 
Similarly,
the \svec s of types $\ket{\chi}_l^{(0)Q}$ and $\ket{\chi}_l^{(1)G}$
in the chiral Verma modules $V(\ket{0,\htop^{(1)}_{r,s}}^{G,Q})$
correspond to \svec s in $V(\ket{0,\hat\htop_{r,s}-1}^{Q})$ but to 
subsingular vectors in $V(\ket{0,\hat\htop_{r,s}}^{G})$ for $s>2$. 
These subsingular vectors do not have, a priori, well defined 
BRST-invariance properties, but become \Gn-closed or \Qn-closed when they 
become singular. Let us note that this symmetry between singular/subsingular
vectors in the \Gn-closed/\Qn-closed Verma modules $V(\ket{0,\htop}^{G})$ 
and $V(\ket{0,\htop-1}^{Q})$ is broken for the Neveu-Schwarz N=2 algebra
since the \Qn-closed topological h.w. vectors do not 
correspond to h.w. vectors
(but secondary states) of that algebra, as explained in subsection 2.1,
and consequently there is no NS counterpart for the \Qn-closed Verma 
modules $V(\ket{0,\htop-1}^{Q})$. (For $\D\neq0$ there is NS counterpart
for the Verma modules $V(\ket{\D,\htop-1}^{Q})$, however, because they
coincide with the Verma modules $V(\ket{\D,\htop}^{G})$, as explained 
in subsection 2.2).

\subsection{No-label Verma modules}\lvm

No-label Verma modules $V(\ket{0,\htop})$ built on no-label h.w. vectors
appear as submodules in some generic Verma modules for $t={2\over r}$
($\ctop={3r-6 \over r} $). They have a counterpart in the Ramond N=2
algebra -- no-helicity Verma modules -- as we will see in next subsection.
However, they do not have a counterpart in the 
Neveu-Schwarz N=2 algebra as the NS counterpart of no-label h.w. vectors
are non-h.w. NS vectors. In fact, the no-label \svec s in generic Verma 
modules correspond to NS subsingular vectors, as was proved in ref. 
\cite{DB1}. 

In principle, we could define no-label Verma modules also for $\D \neq 0$.
In this case, however, the no-label h.w. vector $\ket{\D,\htop}$ can be
decomposed into a \Gn-closed h.w. vector and a \Qn-closed h.w. vector so
that $V(\ket{\D,\htop})$ is simply a direct sum of the
generic Verma module built on $\cG_0 \ket{\D,\htop} = \ket{\D,\htop+1}^G$ 
and the generic Verma module built on 
$\cQ_0 \ket{\D,\htop}= \ket{\D,\htop-1}^Q$. Therefore, the corresponding
determinant formula for $\D \neq 0$ is simply the product of the 
two generic determinant formulae and its study does not lead to anything
more than what we have found for the generic Verma modules. 

For the only interesting no-label case, i.e. for $\D = 0$, the no-label 
Verma module is no longer a direct sum of two generic Verma modules, 
however. In fact, the level-zero vectors $\cG_0 \ket{0,\htop}$, 
$\cQ_0 \ket{0,\htop}$, and $\cG_0 \cQ_0 \ket{0,\htop}$
become singular for $\D=0$ and therefore the no-label determinant
formula for $\D=0$ vanishes for all levels. Nevertheless, we can still use 
the roots of the determinants of the no-label Verma modules in the
neighbourhood of $\Delta = 0$ in order to find the singular vectors for 
no-label Verma modules with $\D = 0$. For this purpose
we take the product of the determinant formulae for the generic Verma
modules $V(\ket{\D,\htop+1}^G)$ and $V(\ket{\D,\htop-1}^Q)$ and factorise
their roots in the limit $\D \to 0$. The terms that vanish in this limit
we combine in a factor $\alpha \, (\D=0)=0$ which is responsible for the fact 
that the no-label determinant vanishes at each level. Taking into account
the terms that do not vanish in this limit one finally finds
the following expression
\bea
det(\cM_l^{T-nl}) &=& \alpha \, (\D=0)\, \prod_{2\leq rs \leq 2l} 
(\htop-\htop^{(0)}_{r,s})^{2 P(l-{rs \over 2})} {\ \ }
(\htop-\htop^{(1)}_{r,s})^{2 P(l-{rs \over 2})} {\ \ }   
(\htop-\htop^{(0)}_{r,s}+1)^{2 P(l-{rs \over 2})} \nonumber \\
&& (\htop-\hat\htop_{r,s})^{2 P(l-{rs \over 2})} {\ \ }  
\prod_{0 \leq k\leq l}  \ (\htop-\htop^+_k)^{2 P_k(l-k)} {\ \ }
(\htop-\htop^-_k)^{2 P_k(l-k)} {\ \ } \nonumber \\
&& (\htop-\htop^+_k+1)^{2 P_k(l-k)}
{\ \ } (\htop-\htop^-_k+1)^{2 P_k(l-k)} \,, 
 \label{detnl} \eea
\noi
with $\htop^{(0)}_{r,s}$, $\htop^{(1)}_{r,s}$, $\hat\htop_{r,s}$,
$\htop^+_k$ and $\htop^-_k$ given by eqns. \req{h0rs}, \req{h1rs},
\req{hhrs} and \req{hpml}. The partition functions $P$ and $P_k$ 
are defined in eqn. \req{part}. As was explained in subsection 3.1,
the partitions $2 P(l-{rs \over 2})$ correspond to complete Verma
submodules of generic type, whereas the partitions $2 P_k(l-k)$ correspond 
to incomplete Verma submodules.  
Once again we see that the \svec s come two by two at the same 
level, in the same Verma module, as $ P(0) =  P_k(0)= 1$. 

The factor $\alpha \, (\D=0)=0$ indicates that 
the determinant
always vanishes for any value of $\htop$, as happens for the generic
Verma modules with $\D=0$ (with one \svec\ at level zero). Similarly 
as in that case, the other roots of the determinants identify the
no-label Verma modules with \svec s other than the ones at level zero
(which is an important information in order to study embedding patterns).
Like for chiral Verma modules, in no-label Verma modules there are no 
chiral \svec s nor no-label \svec s.
The interpretation of the roots in terms of the \svec s  
that one finds in the no-label Verma modules is as follows .

\vskip 0.17in
\noi 
{\bf Case $\htop=\htop^{(0)}_{r,s}$}

For $\htop=\htop^{(0)}_{r,s}$ the Verma module built on the no-label h.w.
vector $\ket{0,\htop^{(0)}_{r,s}}$ has (at least) one uncharged
\svec\ of type $\ket{\chi}_l^{(0)G}$ and one charged \svec\ of type
$\ket{\chi}_l^{(-1)Q}$ at level $l={rs\over2}$, \Gn\ and \Qn\ interpolating
between them. These \svec s are inside the chiral submodule generated by the
level zero chiral \svec\ $\Gz \Qz \ket{0,\htop^{(0)}_{r,s}}$.

\vskip 0.17in
\noi 
{\bf Case $\htop=\htop^{(1)}_{r,s}$}

For $\htop=\htop^{(1)}_{r,s}$ the Verma module built on the no-label h.w.
vector $\ket{0,\htop^{(1)}_{r,s}}$ has (at least) one uncharged
\svec\ of type $\ket{\chi}_l^{(0)Q}$ and one charged \svec\ of type
$\ket{\chi}_l^{(1)G}$ at level $l={rs\over2}$, \Gn\ and \Qn\ interpolating
between them. These \svec s are inside the chiral submodule generated by the
level zero chiral \svec\ $\Gz \Qz \ket{0,\htop^{(1)}_{r,s}}$.
 
\vskip 0.17in
\noi 
{\bf Case $\htop=\htop^{(0)}_{r,s}-1$}

For $\htop=\htop^{(0)}_{r,s}-1$ the Verma module built on the no-label h.w.
vector $\ket{0,\htop^{(0)}_{r,s}-1}$ has (at least) one uncharged
\svec\ of type $\ket{\chi}_l^{(0)Q}$ and one charged \svec\ of type
$\ket{\chi}_l^{(1)G}$ at level $l={rs\over2}$, \Gn\ and \Qn\ interpolating
between them. These \svec s are inside the submodule generated by the
level zero \svec\ $\Gz \ket{0,\htop^{(0)}_{r,s}-1}$ (outside the chiral
submodule).
 
\vskip 0.17in
\noi 
{\bf Case $\htop=\hat\htop_{r,s}$}

For $\htop=\hat\htop_{r,s}$ the Verma module built on the no-label h.w.
vector $\ket{0,\hat\htop_{r,s}}$ has (at least) one uncharged
\svec\ of type $\ket{\chi}_l^{(0)G}$ and one charged \svec\ of type
$\ket{\chi}_l^{(-1)Q}$ at level $l={rs\over2}$, \Gn\ and \Qn\ interpolating
between them. These \svec s are inside the submodule generated by the
level zero \svec\ $\Qz \ket{0,\hat\htop_{r,s}}$ (outside the chiral
submodule).
 
\vskip 0.17in
\noi 
{\bf Case $\htop=\htop^-_l$}

For $\htop=\htop^-_l$ the Verma module built on the no-label h.w.
vector $\ket{0,\htop^-_l}$ has (at least) one charged
\svec\ of type $\ket{\chi}_l^{(-1)G}$ and one charged \svec\ of type
$\ket{\chi}_l^{(-2)Q}$ at level $l={rs\over2}$, \Gn\ and \Qn\ interpolating
between them. These \svec s are inside the submodule generated by the
level zero \svec\ $\Qz \ket{0,\htop^-_l}$ (outside the chiral
submodule).
 
\vskip 0.17in
\noi 
{\bf Case $\htop=\htop^+_l-1$}

For $\htop=\htop^+_l-1$ the Verma module built on the no-label h.w.
vector $\ket{0,\htop^+_l-1}$ has (at least) one charged
\svec\ of type $\ket{\chi}_l^{(1)Q}$ and one charged \svec\ of type
$\ket{\chi}_l^{(2)G}$ at level $l={rs\over2}$, \Gn\ and \Qn\ interpolating
between them. These \svec s are inside the submodule generated by the
level zero \svec\ $\Gz \ket{0,\htop^+_l-1}$ (outside the chiral
submodule).
 
\vskip 0.4in
\noi 
{\bf Cases $\htop=\htop^+_l$ and $\htop=\htop^-_l-1$}

These roots do not give new \svec s as $\htop^+_l=\htop^{(1)}_{l,2}$ and
$\htop^-_l-1=\htop^{(0)}_{l,2}$.

\vskip .5in
\baselineskip 12pt

\vbox{
\bpic{180}{120}{20}{-20}
\put(41,22){\circle*{3}}
\put(78,18){\framebox(4,3){}}
\put(80,25){\circle*{3}}
\put(119,22){\circle*{3}}
\put(44,25){\vector(1,0){32}}
\put(76,20){\vector(-1,0){32}}
\put(116,25){\vector(-1,0){32}}
\put(84,20){\vector(1,0){32}}
\put(39,24){\line(-2,3){75}}
\put(121,24){\line(2,3){75}}
\put(78,27){\line(-1,2){50}}
\put(82,27){\line(1,2){50}}
\put(80,8){\makebox(0,0)[cb]{\scriptsize $\ket{0,h}$}}
\put(18,15){\makebox(0,0)[cb]{\scriptsize $\ket{0,h-1}^Q$}}
\put(142,15){\makebox(0,0)[cb]{\scriptsize $\ket{0,h+1}^G$}}
\put(103,27){\makebox(0,0)[cb]{\scriptsize $\ket{0,h}^{G,Q}$}}
\put(80,-10){\case{no-label Verma modules}}
\epic{
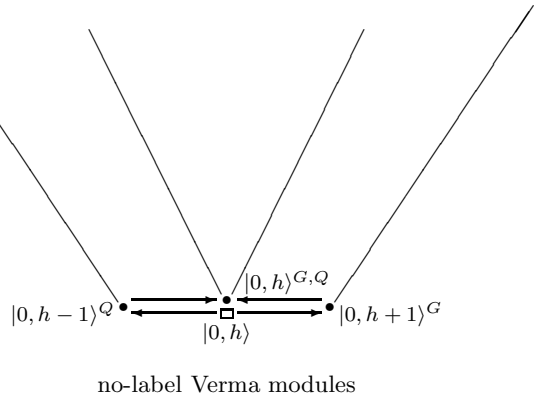
The no-label Verma module built on the no-label h.w. vector $\ket{0,\htop}$
has three singular vectors at level zero: $\ket{0,h+1}^G=\Gz\ket{0,\htop}$, 
$\ket{0,h-1}^Q=\Qz\ket{0,\htop}$ and $\ket{0,h}^{G,Q}=\Gz\Qz\ket{0,\htop}$.
The submodules generated by $\ket{0,h+1}^G$ and $\ket{0,h-1}^Q$ are
generic Verma modules intersecting in the chiral Verma module built on
$\ket{0,h}^{G,Q}$. }}

\baselineskip 16pt

One important remark is that all the \svec s `detected' by the determinant
formula \req{detnl}, that belong to the series that we have just 
described, are located inside the submodules generated by the level
zero \svec s $\Gz \ket{0,\htop}$, $\Qz \ket{0,\htop}$ and
$\Gz \Qz \ket{0,\htop}$. There must exist, however, singular or
subsingular vectors outside these submodules. The reason is that 
after dividing the no-label Verma module by these submodules a chiral
Verma module is left with the corresponding \svec s for 
$\htop=\htop^{(0)}_{r,s}$ and $\htop=\htop^{(1)}_{r,s}$, as discussed
in subsection 3.2. Our conjecture is that outside the submodules 
built on the level zero \svec s there are generically only subsingular
vectors, which are singular just for some discrete values of $t$ 
(i.e. of $\ctop$). At level 1 this is the case for $t=-2$ ($\ctop=9$),
as one can see in Appendix B, and at level 2 this is the case for  
$t=-1$ ($\ctop=6 $), $t=-2$ ($\ctop=9 $) and $t=2$ ($\ctop=-3 $).

\subsection{Verma modules and singular vectors of the 
Ramond N=2 algebra}\lvm

The determinant formula for the standard Verma modules of the Ramond N=2
superconformal algebra was computed in the middle eighties 
\cite{BFK}\cite{KaMa3}. In spite of this, the Verma modules and \svec s
of the Ramond N=2 algebra have not been given much attention in the
literature so far (see however refs. \cite{Kir1}, \cite{DB1} and \cite{DB2}). 
More recently the determinant formula for the `chiral' Verma modules (built 
on the Ramond ground states) has been computed in ref. \cite{BJI5}. As a
bonus, subsingular vectors were discovered for this algebra. Our
purpose now is to show that the Verma modules of the Ramond N=2 algebra
are isomorphic to the Verma modules of the Topological N=2 algebra.
In particular one can construct a one-to-one mapping between every
Ramond \svec\ and every topological \svec\ at the same levels and with
the same relative charges. As a consequence all the results we have
obtained in this section for the Topological N=2 algebra, summarized in 
the four tables \req{tabfrs} -- \req{tabgkm},
and the ten figures Fig. I -- Fig. X, can be transferred straightforwardly
to the Ramond N=2 algebra.

To start let us say a few words about the Ramond N=2 superconformal algebra
given by \cite{Ade}\cite{BFK}\cite{PDiV}\cite{Kir1}  
\BE\new\BA{lclclcl}
\L[L_m,L_n\R]&=&(m-n)L_{m+n}+{\ctop\over12}(m^3-m)\Kr{m+n}
\,,&\qquad&[H_m,H_n]&=
&{\ctop\over3}m\Kr{m+n}\,,\\
\L[L_m,G_r^\pm
\R]&=&\L({m\over2}-r\R)G_{m+r}^\pm
\,,&\qquad&[H_m,G_r^\pm]&=&\pm G_{m+r}^\pm\,,
\\
\L[L_m,H_n\R]&=&{}-nH_{m+n}\\
\L\{G_r^-,G_s^+\R\}&=&\multicolumn{5}{l}{2L_{r+s}-(r-s)H_{r+s}+
{\ctop\over3}(r^2-\frac{1}{4})
\Kr{r+s}\,,}\EA\label{N2algebra}
\EE

\noi
where all the generators, bosonic $L_n$, $H_n$, and fermionic 
$G^{\pm}_r$, are integer-moded. The fermionic zero modes 
characterize the states as being $G_0^+$-closed or $G_0^-$-closed, 
as the anticommutator $\{G_0^+,G_0^-\} = 2L_0 - {\ctop\over12}{\ }$ shows.
We will denote these states as `helicity'-(+) and `helicity'-$(-)$ states.
However, for the conformal weight $\Delta = {\ctop\over24}$ the ground 
states are annihilated by both $G_0^+$ and $G_0^-$ (we will call them 
chiral), and there also exist indecomposable `no-helicity' states which 
cannot be expressed as linear combinations of  helicity-(+), helicity-$(-)$
and chiral states. These no-helicity states, \svec s in particular, 
have been reported very recently for the first time in ref. \cite{DB1}, 
and they were completely overlooked in the early literature.

We already see that the case $\D = {c\over 24}$ for the Ramond N=2 algebra 
is the equivalent to the case $\D = 0$ for the Topological N=2 algebra.
Furthermore, the helicity-(+) and helicity-$(-)$ R states are
analogous to the \Gn-closed $(G)$ and \Qn-closed $(Q)$ topological states,
and the chiral $(+-)$ and no-helicity R states
are analogous to the chiral $(G,Q)$ and no-label topological states,
respectively. 

In order to simplify the analysis that follows we will
define the U(1) charge for the states of the
Ramond N=2 algebra in the same way as for the states of the Topological
and Neveu-Schwarz N=2 algebras. Namely, the U(1) charge of the states
will be denoted by $\htop$ (instead of $\htop \pm \half$, as was the 
standard notation in the past \cite{BFK}), and the relative charge $q$ 
of a secondary state will be defined as the difference between the 
$H_0$-eigenvalue of the state and the $H_0$-eigenvalue of the primary on 
which it is built. Therefore, the relative charges of the R states
are defined to be integer, like for the other N=2 algebras. We will 
denote the R singular vectors as $\ket{\chi_R}_l^{(q)+}$,
$\ket{\chi_R}_l^{(q)-}$, $\ket{\chi_R}_l^{(q)+-}$ or $\ket{\chi_R}_l^{(q)}$, 
where, in addition to the level $l$ and the relative charge $q$, 
the helicities indicate that the vector is annihilated by
$G_0^+$ or $G_0^-$, or both, or none, respectively.  

Now we will construct a one-to-one mapping between the R \svec s
and the topological \svec s that preserves the level $l$ and the relative
charge $q$. For this let us compose the topological twists \req{twa},
which transform the Topological N=2 algebra into the Neveu-Schwarz N=2
algebra, with the spectral flows, which transform the latter into the
Ramond N=2 algebra. Let us consider the odd spectral flow 
\cite{BJI4}\cite{B1}, which is the only fundamental 
spectral flow, as explained in ref. \cite{B1}. It is given by
the one-parameter family of transformations
\BE\new\BA{rclcrcl}
\cA_\th \, L_m \, \cA_\th&=& L_m
 +\th H_m + {\ctop\over 6} \th^2 \delta_{m,0}\,,\\
\cA_\th \, H_m \, \cA_\th&=&- H_m - {\ctop\over3} \th \delta_{m,0}\,,\\
\cA_\th \, G^+_r \, \cA_\th&=&G_{r-\th}^-\,,\\
\cA_\th \, G^-_r \, \cA_\th&=&G_{r+\th}^+\,,\
\label{ospfl} \EA\EE
\noi
satisfying $\cA_{\th}^{-1} = \cA_{\th}$ and giving rise to isomorphic N=2 
algebras. (It is therefore an involution). If we denote by 
$(\Delta, \htop)$ the $(L_0, H_0)$ eigenvalues of any given state, then 
the eigenvalues of the transformed state $\cA_{\th} \kc$ are
$(\Delta+\th \htop +{c\over6} \th^2, - \htop - {c\over3} \th)$.
If the state $\kc$ is a level-$l$ secondary state with relative
charge $q$, then one deduces straightforwardly
that the level of the transformed state $\cA_{\th} \kc$ changes
to $l + \th q$ while the relative charge $q$ reverses its sign.
For half-integer values of $\theta$ the spectral flows interpolate between 
the Neveu-Schwarz N=2 algebra and the Ramond N=2 algebra. 

By analysing the composition of the topological twists \req{twa} and
the spectral flows \req{ospfl}, combining all possibilities, one obtains
that the only mapping that transforms the topological states 
into R states, preserving the level and the relative charge, 
is the composition $\cA_{-1/2}\, (T_W^-)^{-1}$:
\BE  \ket{\chi_R}^{(q)}_l = \cA_{-1/2}\, (T_W^-)^{-1} \kc^{(q)}_l \,. 
\label{compo} \EE
Any other combination of the topological twists \req{twa} and the spectral 
flows \req{ospfl} changes either the level or the relative charge of
the states, or both, as the reader can easily verify\footnote{See also a 
related discussion in ref. \cite{DB1} (subsection 3.3).}. Furthermore
this mapping is one-to-one because it transforms every topological state
into a R state, and the other way around:
$\kc^{(q)}_l =  (T_W^-) \, \cA_{-1/2}\, \ket{\chi_R}^{(q)}_l \,$.

Now let us show that if $\kc^{(q)}_l$ is singular, i.e. satisfies the 
topological h.w. conditions 
${\ } \cL_{n \geq 1} \kc^{(q)}_l =  \cH_{n \geq 1} \kc^{(q)}_l =  
{\cG}_{n \geq 1} \kc^{(q)}_l =  {\cQ}_{n \geq 1} \kc^{(q)}_l = 0 {\ }$,
then $\ket{\chi_R}^{(q)}_l$ is also singular, satisfying in turn
the R h.w. conditions 
${\ } L_{n \geq 1} \ket{\chi_R}^{(q)}_l = 
H_{n \geq 1} \ket{\chi_R}^{(q)}_l =  
{G^+}_{n \geq 1} \ket{\chi_R}^{(q)}_l =  
{G^-}_{n \geq 1} \ket{\chi_R}^{(q)}_l = 0 {\ }$. To see this we have to
study first the transformation, under $\cA_{-1/2}\, (T_W^-)^{-1}$, of the
h.w. vectors of the topological Verma modules. Thus we have four cases 
to analyse, corresponding to the topological h.w. vectors being \Gn-closed
$\ket{\D, \htop}^G$, \Qn-closed $\ket{\D, \htop}^Q$, chiral
$\ket{0, \htop}^{G,Q}$, and no-label $\ket{0, \htop}$. By carefully keeping
track of the transformation of the positive and zero modes of the 
topological generators one obtains the following results:

\vskip .2in

i) The \Gn-closed topological h.w. vectors $\ket{\D, \htop}^G$ are
mapped by $\cA_{-1/2}\, (T_W^-)^{-1}$ to helicity-$(+)$ R h.w. vectors 
$\ket{\D_R, \htop_R}^+_R=\ket{\D+{\ctop\over24},\htop+{\ctop\over6}}^+_R$.

ii) The \Qn-closed topological h.w. vectors $\ket{\D, \htop}^Q$ are
mapped by $\cA_{-1/2}\, (T_W^-)^{-1}$ to helicity-$(-)$ R h.w. vectors 
$\ket{\D_R, \htop_R}^-_R=\ket{\D+{\ctop\over24},\htop+{\ctop\over6}}^-_R$.

iii) The chiral topological h.w. vectors $\ket{0, \htop}^{G,Q}$ are
mapped by $\cA_{-1/2}\, (T_W^-)^{-1}$ to chiral R h.w. vectors 
$\ket{\D_R, \htop_R}^{+-}_R=
\ket{{\ctop\over24},\htop+{\ctop\over6}}^{+-}_R$, usually denoted as
Ramond ground states.

iv) The no-label topological h.w. vectors $\ket{0, \htop}$ are
mapped by $\cA_{-1/2}\, (T_W^-)^{-1}$ to no-helicity R h.w. vectors 
$\ket{\D_R, \htop_R}_R=\ket{{\ctop\over24},\htop+{\ctop\over6}}_R$.

\vskip .2in
Now by taking into account that \svec s are just particular cases of
h.w. vectors (secondary states that satisfy the h.w. conditions, to be
precise) one deduces that the topological \svec s are transformed under
$\cA_{-1/2}\, (T_W^-)^{-1}$ into R \svec s, and   
with the helicities determined by the
exchange $\, G \to + \,, Q \to - \,$. That is, the \Gn-closed topological
\svec s are mapped to helicity-$(+)$ R \svec s, the \Qn-closed
topological \svec s are mapped to helicity-$(-)$ R \svec s, the 
chiral topological \svec s are mapped to chiral R \svec s and the 
no-label topological \svec s are mapped to no-helicity R \svec s.
In addition, as we pointed out, these R \svec s have the same 
level $l$ and relative charge $q$ as the topological \svec s. 
Therefore all the results we have obtained for the 
topological singular vectors and Verma modules, summarized in tables
\req{tabfrs} -- \req{tabgkm} and figures Fig. I -- Fig. X are also valid 
for the \svec s and Verma modules of the Ramond N=2 algebra, simply by 
taking into account that $\D_R=\D+{\ctop\over24}$, 
$\htop_R=\htop+{\ctop\over6}$, and exchanging the labels $G \to +\,$ and  
$\,Q \to -\,$.

These results imply that the standard classification of the Ramond 
\svec s in two sectors, the $+$ sector and the $-$ sector, where the
\svec s and the h.w. vectors on which they are built have the same 
helicities, is not complete. The reason is that there also exist: i) 
\svec s with both helicities, like the level-zero \svec s 
$\ket{\chi_R}^{(1)+-}_{0,\ket{{\ctop\over24},\htop}^-}$ and 
$\ket{\chi_R}^{(-1)+-}_{0,\ket{{\ctop\over24},\htop}^+}$ for 
$\D={\ctop\over24}$, ii) indecomposable \svec s with no 
helicity\footnote{Some examples of no-helicity R \svec s were given 
in \cite{DB1}.}, and iii) \svec s with the opposite 
helicity than the h.w. vector of the Verma module.
The first and the second possibilities can only 
occur if $\D_R+l={\ctop\over24}$. In addition, no-helicity \svec s
only exist for $t={2\over r}\,$ $(\ctop={3r-6\over r})$, like no-label
topological \svec s, as the spectral flows and topogical twists 
do not modify the value of $\ctop$.
The third possibility occurs for $\D={\ctop\over24}$ due to the fact 
that the h.w. vectors $\ket{{\ctop\over24}, \htop}^+_R$ and
$\ket{{\ctop\over24}, \htop-1}^-_R$ are in different Verma modules, 
i.e. there is only one h.w. vector in the Verma modules (plus one
level-zero chiral \svec ), so that
one cannot chose the helicity of the h.w. vector in order to build the
\svec\ of a given helicity. In addition in these Verma modules the 
number of \svec s with helicity (+) is the same as the number of \svec s
with helicity $(-)$. 

As to the no-helicity h.w. vectors and Verma modules, they always have
level-zero \svec s. After dividing the Verma modules by these one is
left with a chiral Verma module built on a Ramond ground state 
annihilated by both $G_0^+$ and $G_0^-$.
   
To finish,
an interesting observation is that one gets exactly the same results by 
constructing the mapping using the even spectral flow $\cU_{\th}$
\cite{SS}\cite{LVW} instead of the odd spectral flow $\cA_{\th}$. In 
that case one finds that the only mapping that preserves the level and
the relative charge of the states is $\cU_{-1/2}\,(T_W^+)^{-1}$. But the
transformation of the topological h.w. vectors is again
$\,\ket{\D,\htop}^G \to \ket{\D+{\ctop\over24},\htop+{\ctop\over6}}^+_R$,
$\,\ket{\D,\htop}^Q \to \ket{\D+{\ctop\over24},\htop+{\ctop\over6}}^-_R$.
Hence this mapping is exactly equivalent to the mapping produced by
$\cA_{-1/2}\, (T_W^-)^{-1}$. Notice that the composition of the topological
twists and the spectral flows does not give any transformation exchanging
the labels $G \to -\,$, $Q \to +\,$ while preserving the level and the
relative charge of the states.

\section{Conclusions and Final Remarks}\lvm

In this paper we have presented the determinant formulae for the Verma 
modules of the Topological N=2 superconformal algebra: the generic
(standard) Verma modules, the chiral Verma modules and the no-label
Verma modules. These were the
last N=2 determinant formulae which remained unpublished. (The absence
of these formulae has created some confusion in the recent literature,
in fact). Then we have analysed in very much detail the interpretation 
of the roots of the determinants in terms of \svec s. As a bonus we 
have obtained a first classification of the different types of 
submodules, regarding size and shape, finding four different types
(at least), in contradiction with some claims in the literature that there 
are only two. A complete classification of the different types of 
submodules requires a deeper understanding of the shape of these,
which in turn requires a classification of the subsingular vectors that
is still lacking \cite{DB5}.

We have also identified the Verma modules which contain chiral and
no-label \svec s. The later are indecomposable primitive \svec s which are
not directly detected by the determinant formulae (but their level-zero
secondaries are) and only exist in generic Verma modules.
As to the no-label Verma modules, built on no-label h.w. vectors, 
they have two-dimensional singular spaces already at level 1, and 
they reduce to chiral Verma modules once the quotient
by the level-zero \svec s is performed. We have found the interesting
result that the \svec s of these chiral Verma modules are mostly 
subsingular vectors in the original no-label Verma module.

Finally we have transferred our analysis and results to the Verma modules 
and \svec s of the Ramond N=2 algebra, which have been very insufficiently 
studied so far. In order to do this we have found a one-to-one mapping 
between the topological \svec s and the R \svec s which preserves the 
grading (level and relative charge). Under this mapping, which is a
composition of the topological twists and the spectral flows, the
topological \svec s are transformed into R \svec s in the following way:
$\kc^{(q)G}_l \to \ket{\chi_R}_l^{(q)+}$,
$\kc^{(q)Q}_l \to \ket{\chi_R}_l^{(q)-}$,
$\kc^{(q)G,Q}_l \to \ket{\chi_R}_l^{(q)+-}$ and
$\kc^{(q)}_l \to \ket{\chi_R}_l^{(q)}$.
Our results imply that the standard classification of the Ramond 
\svec s in two sectors, the $+$ sector and the $-$ sector, where the
\svec s and the h.w. vectors on which they are built have the same 
helicities, is not complete. The reason is that there also exist: \svec s 
with both helicities (chiral), indecomposable \svec s with no helicity, and
\svec s with the opposite helicity than the h.w. vector of the Verma module.
In particular, we have identified the standard Verma modules with chiral 
and indecomposable no-helicity \svec s, which have been overlooked until 
very recently in the literature.
The no-helicity Verma modules and submodules, built on no-helicity h.w. 
vectors, have also been overlooked consequently.

\vskip 1cm

\centerline{\bf Acknowledgements}

We are very grateful to A. Kent and A.N. Schellekens for many explanations
and suggestions about determinant formulae and related issues.
We also thank V. Kac for illuminating discussions on the representation 
theory of N=2 superconformal algebras. M. D. thanks Harvard University
for hospitality, and B. G.-R. thanks CERN for hospitality, where part
of this work was accomplished. M. D. is indebted to the Deutsche
Forschungsgemeinschaft for financial support.

\vskip .17in
\setcounter{equation}{0}
\def\theequation{A.\arabic{equation}}

\subsection*{Appendix A}\lvm

In this appendix we will identify the generic Verma modules 
$V(\ket{\D,\htop}^G)$ which contain chiral \svec s of types 
$\kc_{l}^{(0)G,Q}\,$ or $\kc_{l}^{(-1)G,Q}\,$ (or both), and those 
which contain no-label \svec s $\kc_{l}^{(0)}\,$ or $\kc_{l}^{(-1)}\,$.
Zero conformal weight, $\D+l=0$, is necessary for chiral and no-label 
\svec s to exist. Therefore a first requirement for the generic 
Verma modules to contain chiral or no-label \svec s at level
$l$ is $\D =-l$. 
 
\vskip .17in
\noi
{\bf Chiral singular vectors}

\vskip .15in

As explained in section 3, the determinant formula \req{det1} implies the 
existence of singular vectors
of the types $\kc_{l,\ket{\D,\htop}^G}^{(0)G}\,$
and $\kc_{l,\ket{\D,\htop}^G}^{(-1)Q}\,$ at level $l=\frac{rs}{2}$ built 
on \Gn-closed h.w. vectors $\ket{\D,\htop}^G$ with conformal weight 
$\D=\Delta_{r,s}$, eq. \req{Drs}, for $\ctop \neq 3$,
or with any conformal weight $\D$ for $\htop_s=\pm{s\over2}$, for
$\ctop=3$.
For simplicity in what follows we will denote these \svec s as
$\kc_{r,s}^{(0)G}\,$ and $\kc_{r,s}^{(-1)Q}\,$.
Our analysis showed that $\cG_0$ and $\cQ_0$ interpolate between 
$\kc_{r,s}^{(0)G}\,$ and 
$\kc_{r,s}^{(-1)Q}\,$, except in the cases when these vectors
have zero conformal weight, that is $\Delta_{r,s}+\frac{rs}{2}=0$, 
for which this interpolation breaks down. Namely, at least one of 
the \svec s $\kc_{r,s}^{(0)G}\,$ or $\kc_{r,s}^{(-1)Q}\,$ 
necessarily becomes chiral and therefore is annihilated by both 
\Gn\ and \Qn . {\it A priori} it is not possible to tell which one
of the two singular vectors becomes chiral, or if both become
chiral, but by analysing some of their coefficients
we will in the following give an answer to this question. 

The condition of zero conformal weight $\Delta+\frac{rs}{2}=0$ yields two 
curves of solutions for the charge $\htop$ as functions of $r,s$ and 
$t = {3-\ctop \over 3}$:
\bea
{\cal C}^+: && \htop^{(+)}_{r,s}(t)= \frac{s+(r+1)t}{2} \,, \label{curv1} \\
{\cal C}^-: && \htop^{(-)}_{r,s}(t)=-\frac{s+(r-1)t}{2} \,. \label{curv2}
\eea

In ref. \cite{DB2} we showed that singular vectors 
can be identified by their coefficients with respect to the terms of the
ordering kernel. If two singular vectors agree in the coeffcients of the
ordering kernel then they are identical. In particular, 
if the coefficients with respect to the ordering kernel
all vanish then the vector has to be trivial. The advantage of this
procedure is that the ordering kernel has at most two elements 
in our case, which means that we can identify singular vectors 
and decide if they are trivial just by comparing only two coefficients. 
In addition we can also deduce if the vectors are \Gn-closed, 
\Qn-closed or chiral, as we will see. The ordering kernel for the uncharged
\Gn-closed singular vectors $\kc_{r,s}^{(0)G}\,$ at 
level $l$ is given, for the general case $\ctop \neq 3$ $\, (t \neq 0)$,  
by \cite{DB2} $\{\cL_{-1}^l,\cL_{-1}^{l-1}\cG_{-1}\cQ_0\}$. 
The coefficients of these vectors 
can easily be obtained from the coefficients of the uncharged \svec s
of the Neveu-Schwarz N=2 algebra\footnote{The uncharged \Gn-closed 
\svec s built on 
\Gn-closed h.w. vectors correspond to the uncharged \svec s of the  
Neveu-Schwarz N=2 algebra by performing the topological twists.} given 
in eq. (3.11) of ref. \cite{Doerr2}. For
\bea
\kc_{r,s}^{(0)G}\, &=& \Bigl\{ \alpha \cL_{-1}^{\frac{rs}{2}} + 
\beta \cL_{-1}^{\frac{rs}{2}-1} \cG_{-1}\cQ_0
+ \gamma \cH_{-1}\cL_{-1}^{\frac{rs}{2}-1} +
\delta \cL_{-1}^{\frac{rs}{2}-2} \cG_{-1}\cQ_{-1} + \ldots
\Bigr\} \ket{\Delta_{r,s}, \htop}^G ,
\label{X0rs}
\eea 
we use the notation $\kc_{r,s}^{(0)G}=(\alpha,\beta,\gamma,\delta)^{0,G}$ 
(for convenience we show two other coefficients besides the coefficients 
with respect to the ordering kernel) finding
\bea
\kc_{r,s}^{(0)G}=(1,\frac{1}{2}(\Pi_{r,s}-1),\frac{rs}{2} 
\frac{\htop+1}{t},\frac{rs}{2} \frac{\htop+1-t}{2t}-\frac{1}{2}(\Pi_{r,s}
-1)\Delta_{r,s})^{0,G} \,, \label{eq:psirs0G}
\eea 
where
\bea
\Pi_{r,s} = \prod_{n=1}^{r} \frac{\htop+H(n)}{\htop-H(n)} \,, \qquad
H(n) = \frac{s}{2}-\frac{t}{2}(r+1)+nt \,.
\eea 

Now we act with $\cQ_0$ on $\kc_{r,s}^{(0)G}$ obtaining a \Qn-closed \svec\ 
$\kc_{r,s}^{(-1)Q}$ and consider the fact that according to the ordering 
kernel for this type of \svec s, given in ref. \cite{DB2},  only the 
coefficients of $\cL_{-1}^{\frac{rs}{2}}\cQ_0$ and 
$\cL_{-1}^{\frac{rs}{2}-1}\cQ_{-1}$ decide whether or not 
$\ \kc_{r,s}^{(-1)Q}=\cQ_0\kc_{r,s}^{(0)G}\, $ is trivial. 
One finds that $\cQ_0\kc_{r,s}^{(0)G}\,$ is trivial, and $\kc_{r,s}^{(0)G}$ 
is chiral as a consequence, if and only if 
the following two conditions for 
$\kc_{r,s}^{(0)G}=(\alpha,\beta,\gamma,\delta)^{0,G}$ hold:
\bea
\alpha + 2\beta = 0 \,, \qquad \gamma + 2\delta = 0 \,. 
\label{eq:chiral0}
\eea
It is easy to see that following the curve ${\cal C}^-$ the 
singular vectors $\kc_{r,s}^{(0)G}$ satisfy these two conditions and 
therefore become chiral, i.e. of type $\kc_{r,s}^{(0)G,Q}$ instead.  
For the curve ${\cal C}^+$, however, $\kc_{r,s}^{(0)G}$
does generically not become chiral. 
The only exceptions to this rule are the discrete points
$t_{s,n}=-\frac{s}{n}$ for $n=1,\ldots ,r$ on the curve 
${\cal C}^+$ where $\kc_{r,s}^{(0)G}$  also becomes chiral.
For $\ctop=3$ $\, (t = 0)$ the ordering kernels of the \svec s 
have not been computed yet and therefore these arguments do not apply.
However, in the limit $t \rightarrow 0$ along curve ${\cal C}^-$ the \svec s 
$\kc_{r,s}^{(0)G}$ are well defined, and chiral also as a consequence.  

In the same way we analyse the \svec s $\kc_{r,s}^{(-1)Q}$, which are 
obtained generically from the \svec s $\kc_{r,s}^{(0)G}$ 
under the action of $\cQ_0$. Dividing out overall factors
one then obtains the leading coefficients for $\kc_{r,s}^{(-1)Q}$. 
Using the notation
$\kc_{r,s}^{(-1)Q}=(\alpha,\beta,\gamma,\delta)^{-1,G}$ for
\bea
\kc_{r,s}^{(-1)Q} &=& \Bigl\{ \alpha \cL_{-1}^{\frac{rs}{2}} \cQ_0 
+ \beta \cL_{-1}^{\frac{rs}{2}-1} \cQ_{-1}
+ \gamma \cH_{-1}\cL_{-1}^{\frac{rs}{2}-1}\cQ_0 
+\delta \cL_{-1}^{\frac{rs}{2}-2} \cG_{-1}\cQ_{-1}\cQ_0 + \ldots
\Bigr\} \ket{\Delta_{r,s},\htop}^G \nonumber \,,
\eea 
one finds
\bea
\kc_{r,s}^{(-1)Q} &=& (\Pi_{r-1,s},\frac{1}{2t}\{(\htop+H(0))(\htop-H(1))
\Pi_{r-1,s}-(\htop-H(r+1))(\htop-H(r))\},  \nonumber \\
&& \frac{rs(\htop-1)}{2t}\Pi_{r-1,s},-\frac{1}{4t}(\htop-H(r+1))(\htop-H(r))-
\frac{1}{2}\Pi_{r-1,s}\{\Delta_{r,s}+
\frac{rs(\htop-1)}{2t}\})^{-1,Q} \,. \nonumber
\eea
Again, there are two conditions which have to be satisfied, for the general
case $\ctop \neq 3$ $\, (t \neq 0)$, for these vectors to become chiral: 
\bea
 -\alpha\frac{rs}{2} +\beta = 0 \,, \qquad
\gamma+2\delta-\alpha\frac{rs}{2} = 0 \,.
\eea
As before we analyse whether these conditions are satisfied for the curves 
${\cal C}^+$ or ${\cal C}^-$. One finds that both conditions hold for the
curve ${\cal C}^+$. Hence, $\kc_{r,s}^{(-1)Q}$ 
becomes chiral, i.e. of type $\kc_{r,s}^{(-1)G,Q}$ instead, on the curve 
${\cal C}^+$. However, for ${\cal C}^-$ these conditions
are generically not satisfied and therefore $\kc_{r,s}^{(-1)Q}$ does
not become chiral. 
The only exceptions to the generic case are again the discrete points 
$t_{s,n}=-\frac{s}{n}$ for $n=1,\ldots ,r$ on ${\cal C}^-$ where 
$\kc_{r,s}^{(-1)Q}$ also becomes chiral. Finally, 
in the limit $t \rightarrow 0$ along curve ${\cal C}^+$ the \svec s 
$\kc_{r,s}^{(-1)Q}$ are well defined, and chiral also consequently.

We can summarise these results as follows: for the values  
$\htop=\htop^{(+)}_{r,s}$, given by curve ${\cal C}^+$, the singular vectors 
$\kc_{r,s}^{(-1)Q}$ become chiral for all 
values of $t$ whilst $\kc_{r,s}^{(0)G}$ stay generically
\Gn-closed. Only for the values $t=-\frac{s}{n}$, $n=1,\ldots, r$, on
${\cal C}^+$ (corresponding to $\ctop = {3(n+s) \over n}$)
both types of \svec s become chiral. Similarly, 
for the values $\htop=\htop^{(-)}_{r,s}$, given by curve ${\cal C}^-$, 
the singular vectors $\kc_{r,s}^{(0)G}$ become chiral for all 
values of $t$ whilst $\kc_{r,s}^{(-1)Q}$ only become chiral  
for the values $t=-\frac{s}{n}$, $n=1,\ldots, r$.

As an example, the chiral \svec s $\kc_{r,s}^{(0)G,Q}$ and
$\kc_{r,s}^{(-1)G,Q}$ at level 1 are given by: 

\BE
\kc_{1,\ket{-1,-1}^G}^{(0)G,Q}=(-2\cL_{-1} + \cG_{-1}\cQ_0 )\ket{-1,-1}^G ,\EE
\BE
\kc_{1,\ket{-1,{6-\ctop\over3}}^G}^{(-1)G,Q}=(\cL_{-1}\cQ_0+\cH_{-1}\cQ_0 +
\cQ_{-1}) \ket{-1,{6-\ctop\over3}}^G , \EE

Observe that for $\ctop=9$ ($t=-2$) both chiral \svec s are together
in the same Verma module $V(\ket{-1,-1}^G)$, in agreement with our 
analysis.

\vskip .17in

{\bf No-label singular vectors}

\vskip .14in

Now we will investigate the appearance of no-label singular vectors at
level $l$ in generic Verma modules $V(\ket{-l,\htop}^G)$. Let us start 
with the uncharged no-label singular vectors 
$\kc_{l,\ket{-l,\htop}^G}^{(0)}\,$ which will be denoted simply as 
$\kc_{l}^{(0)}\,$. An uncharged no-label singular 
vector $\kc_{l}^{(0)}\,$ is necessarily accompanied at the same 
level $l$ in the same Verma module by three secondary \svec s which 
cannot `come back' to this one by acting with the algebra: one charged 
\Gn-closed singular vector $\kc_l^{(1)G}=\cG_0\kc_{l}^{(0)}\,$, one charged 
\Qn-closed singular vector $\kc_l^{(-1)Q}=\cQ_0\kc_{l}^{(0)}\,$, and
one uncharged chiral singular vector 
$\kc_l^{(0)G,Q}=\cG_0\cQ_0\kc_{l}^{(0)}=-\cQ_0\cG_0\kc_{l}^{(0)}\,$. 
Observe that the \svec\ $\kc_l^{(-1)Q}$ must be non-chiral necessarily
since otherwise $\cG_0\cQ_0\kc_{l}^{(0)}=0$ while 
$\cQ_0\cG_0\kc_{l}^{(0)} \neq 0$, as the \svec s $\kc_l^{(1)G}$ 
(built on $\ket{\D,\htop}^G$) never
become chiral. The fact that two charged singular vectors of types 
$\kc_{r,s}^{(-1)Q}$ and $\kc_l^{(1)G}$ exist at the same level in the same
generic Verma module $V(\ket{\D,\htop}^G)$ requires $f_{r,s}(\D, \htop, t)=
g_l^+ (\D, \htop, t) = 0$, given by eqns. \req{frs} and \req{gk},
where $l={\frac{rs}{2}}$. For the general case $\ctop \neq 3$ $\, (t \neq 0)$
the corresponding equation $\Delta_{r,s}=\Delta^+_{l}$ 
is satisfied by two curves of solutions for $\htop_{r,s}(t)$: 
\bea
{\cal C}^1: && \htop_{r,s}(t)=\frac{1}{2}(s-t(r+rs-1)) \,, \\
{\cal C}^2: && \htop_{r,s}(t)=-\frac{1}{2}(s-t(r-rs+1)) \,,
\eea
whereas for $\ctop = 3$ $\, (t = 0)$ one simply has
$\Delta^+_{\frac{rs}{2}}= \pm \frac{rs^2}{4}$.
The charged singular vector $\kc_l^{(1)G}$ is accompanied by the
uncharged singular vector $\kc_l^{(0)Q}=\cQ_0\kc_l^{(1)G}$ with the 
following leading coefficients for $\ctop \neq 3$ $\, (t \neq 0)$:
\bea
\kc_l^{(0)Q}=(2,-1, -2\frac{(l+1)(\htop+1)}{1-t},
\frac{(l+1)(\htop+1)}{1-t})^{0,Q} \,, \label{eq:psirs0Q}
\eea
where we choose the same notation as for $\kc_{r,s}^{(0)G}$, 
eqn. \req{X0rs}. (Observe that $\kc_l^{(0)Q}$ is \Qn-closed while the h.w. 
vector $\ket{-l,\htop}^G$ is \Gn-closed). The charged \svec\ 
$\kc_{r,s}^{(-1)Q}$, in turn, is accompanied by the uncharged \svec\ 
$\kc_{r,s}^{(0)G}=\cG_0\kc_{r,s}^{(-1)Q}$, the latter one
with coefficients given in eqn. (\ref{eq:psirs0G}). 

Thus, following curve ${\cal C}^1$ or curve ${\cal C}^2$ by 
varying the parameter $t$ we can assume that we have at level 
$\frac{rs}{2}$ the two uncharged singular vectors
$\kc_{\frac{rs}{2}}^{(0)Q}$ and $\kc_{r,s}^{(0)G}$, which
are generically different (i.e. not proportional) except in the
case when both of them become chiral:
$\kc_{\frac{rs}{2}}^{(0)G,Q}=\kc_{r,s}^{(0)G,Q}$. The reason is that two
chiral \svec s at the same level with the same charge never span
a two dimensional singular vector space, as we proved in ref. \cite{DB2}. 
When the \svec s have zero conformal weight, i.e. 
in the case $\Delta_{r,s}=\Delta^+_{\frac{rs}{2}}=-\frac{rs}{2}$ 
the \svec s $\kc_{\frac{rs}{2}}^{(0)Q}$ always become chiral,
as explained in section 3, whereas the \svec s $\kc_{r,s}^{(0)G}$ 
become chiral only under certain conditions that we have 
deduced a few paragraphs above. Observe that the
existence of no-label \svec s requires that these two uncharged \svec s 
become chiral, and therefore proportional, whereas the charged \svec\ of
type $\kc_{r,s}^{(-1)Q}$ should remain non-chiral. These are necessary, 
although not sufficient, conditions to guarantee the existence of the
no-label \svec s. 

Let us therefore analyse the different possibilities when
$\kc_{r,s}^{(0)G}$ becomes chiral and $\kc_{r,s}^{(-1)Q}$ stays
non-chiral. Let us start with the general case $\ctop \neq 3$ $\, (t \neq 0)$.
First of all, the condition $\Delta_{r,s}=-\frac{rs}{2}$, 
necessary for chiral \svec s $\kc_{r,s}^{(0)G,Q}$ to exist, leads
to the solutions given by curves ${\cal C}^+$ and ${\cal C}^-$, eqns. 
\req{curv1} and \req{curv2}. 
The solutions along curve ${\cal C}^+$, however,
involve chiral \svec s of type $\kc_{r,s}^{(-1)G,Q}$. As a consequence
we only need to investigate the solutions to the conditions 
$\Delta_{r,s}=\Delta^+_{\frac{rs}{2}}=-\frac{rs}{2}$ given by the 
intersections of curve ${\cal C}^-$ (for which $\kc_{r,s}^{(0)G}$
always becomes chiral) with curves ${\cal C}^1$ and ${\cal C}^2$. 
In the first case the only intersection points correspond to
$t=\frac{2}{r}$ whereas in the second case there are solutions 
for all $t$ provided $s=2$. A closer look at these solutions reveals
that no-label \svec s only exist for $t=\frac{2}{r}$, in agreement 
with the low level computations of these vectors given in refs. 
\cite{BJI6} and \cite{DB1}. The argument goes as follows. 

Along curve ${\cal C}^1$, varying $t$, there are two uncharged \svec s at 
level $l=\frac{rs}{2}$: $\kc_{\frac{rs}{2}}^{(0)Q}$ and $\kc_{r,s}^{(0)G}$. 
Since these correspond to vectors in the kernel of the inner product 
matrix for the Verma module, one finds that for the case of both
becoming chiral, and thus proportional, the rank of the inner product 
matrix would rise for these particular points on the curve
${\cal C}^1$. But the inner product 
matrix of the Verma module contains only rational 
functions of $t$, $\Delta$, and $\htop$ as entries and 
therefore its rank is upper semi-continuous. As a consequence the rank
cannot rise for particular values of $t$ and therefore for 
$t=\frac{2}{r}$ at least one additional uncharged null vector 
$\ket{\Upsilon}_{\frac{rs}{2}}^{(0)}$ needs to exist
at level\footnote{See ref. \cite{DB5} 
where this mechanism is explained in more detail.} $\frac{rs}{2}$. 
For the solution $s=2$ for all $t$, however, one finds that curves
${\cal C}^2$ and ${\cal C}^-$ are identical.  
Therefore, the vectors $\kc_{\frac{rs}{2}}^{(0)Q}$ and $\kc_{r,s}^{(0)G}$
are both chiral and proportional all along the curve ${\cal C}^2$. 
As a result the corresponding space of uncharged \svec s is just 
one-dimensional and consequently an additional null vector is not required.

The additional null vectors found for $t=\frac{2}{r}$ cannot be subsingular
vectors as there are no singular vectors at lower levels than 
$\ket{\Upsilon}_{\frac{rs}{2}}^{(0)}$ themselves. Neither
they can be \Qn-closed as this would result in two-dimensional
spaces of uncharged singular vectors annihilated by $\cQ_0$, which  
do not exist, as we proved in ref. \cite{DB2}. 
Finally, in ref. \cite{Doerr2} conditions were
given for Verma modules containing two-dimensional spaces of
uncharged \svec s annihilated by $\cG_0$.
Comparing these conditions with the solutions
$t=\frac{2}{r}$ on ${\cal C}^1$ shows that
$\ket{\Upsilon}_{\frac{rs}{2}}^{(0)}$ are neither \Gn-closed. The only 
possibility left is therefore that $\ket{\Upsilon}_{\frac{rs}{2}}^{(0)}$ 
are uncharged no-label singular vectors.

Finally, for the case $\ctop = 3$ $\, (t = 0)$ the two conditions 
$\Delta^+_{\frac{rs}{2}}=-\frac{rs}{2}$ and
$\Delta^+_{\frac{rs}{2}}= \pm \frac{rs^2}{4}$ lead to the unique solution
$s=2$, $\,\Delta^+_{\frac{rs}{2}}= -r$. For this solution, however,
the corresponding space of uncharged \svec s is just one-dimensional
so that no-label \svec s do not exist. (This solution can be viewed in 
fact as the case described above where ${\cal C}^-$ and ${\cal C}^2$
are identical, in the limit $t \to 0$).

These results can easily be transferred to the other types of
no-label singular vectors using the mappings analysed in ref. \cite{BJI6}.
Since these mappings do not modify the value of $t$, one deduces
that no-label \svec s only exist for generic Verma modules with 
$t=\frac{2}{r}$. In particular, no-label \svec s of types $\kc_l^{(0)}$ and
$\kc_l^{(-1)}$, built on \Gn-closed h.w. vectors $\ket{\D_{r,s},\htop}^G$, 
appear at level $l = \frac{rs}{2}$ for $t=\frac{2}{r}\,$
(corresponding to $\ctop = {3r - 6 \over r}$), 
$\Delta_{r,s}=-\frac{rs}{2}$, and $\htop=-1-\frac{s}{2}+\frac{1}{r}$ and 
$\htop= 1+\frac{s}{2}+\frac{1}{r}$, respectively.

As an example, let us write the no-label \svec\ $\kc_l^{(0)}$
built on a \Gn-closed h.w. vector $\ket{\D_{r,s},\htop}^G$,
at level 1, together with the three secondary \svec s that it generates
at level 1 by the action of \Gn\ and \Qn :
\BE \kc_{1,\ket{-1,-1, \, t=2}^G}^{(0)}\, = 
(\cL_{-1} - \cH_{-1})\ket{-1,-1, \, t=2}^G ,\EE
\BE \kc_{1,\ket{-1,-1, \, t=2}^G}^{(1)G}=
\cG_0\kc_{1,\ket{-1,-1, \, t=2}^G}^{(0)}\, =
 2 \cG_{-1} \ket{-1,-1, \, t=2}^G ,\EE
\BE \kc_{1,\ket{-1,-1, \, t=2}^G}^{(-1)Q}=
\cQ_0\kc_{1,\ket{-1,-1, \, t=2}^G}^{(0)}\,
 = (\cL_{-1}\cQ_0 - \cH_{-1}\cQ_0 - \cQ_{-1})\ket{-1,-1, \, t=2}^G ,\EE 
\BE  
\kc_{1,\ket{-1,-1, \, t=2}^G}^{(0)G,Q} = 
 \cG_0\cQ_0\kc_{1,\ket{-1,-1, \, t=2}^G}^{(0)} = 
 2 (-2\cL_{-1} + \cG_{-1}\cQ_0 )\ket{-1,-1, \, t=2}^G .\EE
\noi
The no-label \svec\ only
exists for $t=2$ ($\ctop=-3$) whereas the three secondary \svec s are just
the particular cases, for $t=2$, of the one-parameter families of \svec s
of the corresponding types, which exist for all values of $t$. (The  
\svec s of the Topological N=2 algebra at level 1 were given in
ref. \cite{BJI6}). Moreover,
the \svec\ $\kc_{1,\ket{-1,-1}^G}^{(1)G}$ is always primitive for any value 
of $t$, except for $t=2$, the \svec\ $\kc_{1,\ket{-1,-1}^G}^{(-1)Q}$ is also
primitive, except for $t=2$, even when it becomes chiral (for $t=-2$), 
and the \svec\ $\kc_{1,\ket{-1,-1}^G}^{(0)G,Q}$ is always secondary,
except for the value $t=-2$ ($\ctop=9$), for which it becomes primitive
together with the other chiral \svec\ $\kc_{1,\ket{-1,-1}^G}^{(-1)G,Q}$.

\vskip .17in
\setcounter{equation}{0}
\def\theequation{B.\arabic{equation}}

\subsection*{Appendix B}\lvm

In what follows we present the complete set of \svec s at level 1 in Verma 
modules of the Topological N=2 algebra with zero conformal weight $\D=0$. 
That is, the \svec s at level 1 in: generic Verma modules 
$V(\ket{0,\htop}^G)$ and $V(\ket{0,\htop}^Q)$, chiral Verma modules 
$V(\ket{0,\htop}^{G,Q})$, and no-label Verma modules $V(\ket{0,\htop})$. 
Several of these \svec s were written down in ref. \cite{BJI6} 
(for any value of $\D$ in the case of generic Verma modules).

\vskip .17in
\noi
{\bf Generic Verma modules}

\vskip .15in

In generic Verma modules $V(\ket{0,\htop}^G)$ and $V(\ket{0,\htop-1}^Q)$
one finds \svec s at level 1 for the following values of $\htop$, in
agreement with eqns. \req{h0rs}, \req{hhrs} and \req{hpml}: 
$\htop^{(0)}_{1,2}=t-1=-{\ctop\over 3}$, $\hat\htop_{1,2}=1$,
$\htop^+_1=0$ and $\htop^-_1=t={3-\ctop\over3}$. The corresponding
\svec s in $V(\ket{0,\htop}^G)$ are

\BE
\kc_{1,\ket{0,-{\ctop\over3}}^G}^{(0)G}=({\ctop+3\over3}\cL_{-1}+
{\ctop+3\over3}\cH_{-1}-\cG_{-1}\cQ_0)\ket{0,-{\ctop\over3}}^G , \EE
\BE
\kc_{1,\ket{0,-{\ctop\over3}}^G}^{(-1)Q}=((\ctop-3)\cL_{-1}\cQ_0+
(\ctop+3)\cH_{-1}\cQ_0+(\ctop+3)\cQ_{-1})\ket{0,-{\ctop\over3}}^G ,\EE
\BE
\kc_{1,\ket{0,1}^G}^{(0)G}=\cG_{-1}\cQ_0\,\ket{0,1}^G , \qquad
\kc_{1,\ket{0,1}^G}^{(-1)Q}=\cL_{-1}\cQ_0\,\ket{0,1}^G , \EE
\BE
\kc_{1,\ket{0,0}^G}^{(1)G}=\cG_{-1}\,\ket{0,0}^G ,\qquad 
\kc_{1,\ket{0,0}^G}^{(0)Q}= \Qz \cG_{-1}\,\ket{0,0}^G , \EE
\BE
\kc_{1,\ket{0,{3-\ctop\over3}}^G}^{(-1)G}=(\cL_{-1}\cQ_0+\cH_{-1}\cQ_0)
\ket{0,{3-\ctop\over3}}^G , \qquad
\kc_{1,\ket{0,{3-\ctop\over3}}^G}^{(-2)Q}=\cQ_{-1}\cQ_0\,
\ket{0,{3-\ctop\over3}}^G . \EE

\noi
The \svec s in $V(\ket{0,\htop-1}^Q)$ are

\BE
\kc_{1,\ket{0,-{\ctop+3\over3}}^Q}^{(0)Q}=
\cQ_{-1}\cG_0\,\ket{0,-{\ctop+3\over3}}^Q , \qquad 
\kc_{1,\ket{0,-{\ctop+3\over3}}^Q}^{(1)G}=(\cL_{-1}\cG_0+\cH_{-1}\cG_0)
\ket{0,-{\ctop+3\over3}}^Q , \EE 
\BE
\kc_{1,\ket{0,0}^Q}^{(0)Q}=
({\ctop+3\over3}\cL_{-1}-\cQ_{-1}\cG_0)\ket{0,0}^Q , \EE
\BE
\kc_{1,\ket{0,0}^Q}^{(1)G}=({\ctop-3\over6}\cL_{-1}\cG_0-\cH_{-1}\cG_0+
{\ctop+3\over6}\cG_{-1})\ket{0,0}^Q , \EE
\BE
\kc_{1,\ket{0,-1}^Q}^{(1)Q}=\cL_{-1}\cG_0\,\ket{0,-1}^Q ,\qquad
\kc_{1,\ket{0,-1}^Q}^{(2)G}=\cG_{-1}\cG_0\,\ket{0,-1}^Q , \EE
\BE
\kc_{1,\ket{0,-{\ctop\over3}}^Q}^{(-1)Q}=\cQ_{-1}\,\ket{0,-{\ctop\over3}}^Q ,
\qquad \kc_{1,\ket{0,-{\ctop\over3}}^Q}^{(0)G}=
\Gz\cQ_{-1}\,\ket{0,-{\ctop\over3}}^Q . \EE
\vskip .2in 

\noi
All these \svec s also apply for $\ctop=3$ ($t=0$) and no additional 
\svec s appear for this value.
Chiral singular vectors and no-label \svec s at level 1 require $\D=-1$,
therefore they are absent for $\D=0$.

\vskip .17in
\noi
{\bf Chiral Verma modules}

\vskip .15in
In chiral Verma modules $V(\ket{0,\htop}^{G,Q})$ 
one finds \svec s at level 1 for the following values of $\htop$, in
agreement with eqns. \req{hrs0} and \req{h1rs}:  
$\htop^{(0)}_{1,2}=t-1=-{\ctop\over 3}$ and $\htop^{(1)}_{1,2}=0$.
The corresponding \svec s, for all values of $\ctop$, are

\BE        
 \ket\chi_1^{(0)G} = (\cL_{-1}+\cH_{-1})\ket{0,-{\ctop\over3}}^{G,Q},
  \qquad \ket\chi_1^{(-1)Q} = \cQ_{-1} \ket{0,-{\ctop\over3}}^{G,Q}, \EE  
\BE   
\ket\chi_1^{(1)G} = \cG_{-1}\ket{0,0}^{G,Q} , \qquad 
  \ket\chi_1^{(0)Q} = \cL_{-1}\ket{0,0}^{G,Q}.  \EE

\vskip .17in
\noi
{\bf No-label Verma modules}

\vskip .15in

In no-label Verma modules $V(\ket{0,\htop})$ 
one finds \svec s at level 1 for the following values of $\htop$, in
agreement with our analysis in subsection 3.3: 
$\htop^{(0)}_{1,2}=\htop^-_1-1=t-1=-{\ctop\over 3}$, 
$\hat\htop_{1,2}=1$, $\htop^{(0)}_{1,2}-1=t-2=-{\ctop+3 \over 3}$, 
$\htop^{(1)}_{1,2}=\htop^+_1=0$, $\htop^+_1-1=-1$ and 
$\htop^-_1=t={3-\ctop\over3}$. For $\ctop=9$ ($t=-2$) there are even 
two-dimensional singular spaces. The corresponding \svec s, for  
all values of $\ctop$, are

\BE
\kc_{1,\ket{0,-{\ctop\over3}}}^{(0)G}=(\cL_{-1}\cG_0\cQ_0+\cH_{-1}\cG_0\cQ_0)
\ket{0,-{\ctop\over3}} , \qquad
\kc_{1,\ket{0,-{\ctop\over3}}}^{(-1)Q}=
\cQ_{-1}\cQ_0\cG_0\,\ket{0,-{\ctop\over3}} , \label{ssvnl1} \EE
\BE \kc_{1,\ket{0,1}}^{(0)G}=({3-\ctop\over6}\cL_{-1}\cG_0\cQ_0+
\cH_{-1}\cG_0\cQ_0-{\ctop+3\over6}\cG_{-1}\cQ_0)\ket{0,1} , \EE
\BE \kc_{1,\ket{0,1}}^{(-1)Q}=({\ctop+3\over3}\cL_{-1}\cQ_0+
\cQ_{-1}\cQ_0\cG_0\,\ket{0,1} , \EE
\BE
\kc_{1,\ket{0,-{\ctop+3\over3}}}^{(0)Q}=((\ctop-3)\cL_{-1}\cQ_0\cG_0+
(\ctop+3)\cH_{-1}\cQ_0\cG_0+(\ctop+3)\cQ_{-1}\cG_0)
\ket{0,-{\ctop+3\over3}} , \EE
\BE
\kc_{1,\ket{0,-{\ctop+3\over3}}}^{(1)G}=({\ctop+3\over3}\cL_{-1}\cG_0+
{\ctop+3\over3}\cH_{-1}\cG_0+\cG_{-1}\cG_0\cQ_0)\ket{0,-{\ctop+3\over3}} ,\EE
\BE
\kc_{1,\ket{0,0}}^{(1)G}=\cG_{-1}\cG_0\cQ_0\,\ket{0,0} ,\qquad
\kc_{1,\ket{0,0}}^{(0)Q}=\cL_{-1}\cG_0\cQ_0\,\ket{0,0} ,\label{ssvnl2} \EE
\BE
\kc_{1,\ket{0,-1}}^{(1)Q}=\Qz\cG_{-1}\cG_0\,\ket{0,-1} ,\qquad 
\kc_{1,\ket{0,-1}}^{(2)G}=\cG_{-1}\cG_0\,\ket{0,-1} , \EE
\BE
\kc_{1,\ket{0,{3-\ctop\over3}}}^{(-1)G}=
\Gz\cQ_{-1}\cQ_0\,\ket{0,{3-\ctop\over3}} , \qquad
\kc_{1,\ket{0,{3-\ctop\over3}}}^{(-2)Q}=
\cQ_{-1}\cQ_0\,\ket{0,{3-\ctop\over3}} . \EE

\vskip .2in
\noi
In addition, for $\ctop=9$ ($t=-2$) one finds the two-dimensional 
singular spaces generated by the \svec s
\BE
\kc_{1,\ket{0,-3}}^{(0)G}=(\cL_{-1}+\cH_{-1})\cG_0\cQ_0\,\ket{0,-3} , \qquad  
\hat\kc_{1,\ket{0,-3}}^{(0)G}=(\cG_{-1}\Qz-2\Gz\cQ_{-1}-\cH_{-1}\Gz\Qz)
\ket{0,-3} , \EE

\BE 
\kc_{1,\ket{0,-3}}^{(-1)Q}= \cQ_{-1}\cQ_0\cG_0\,\ket{0,-3} , \qquad
\hat\kc_{1,\ket{0,-3}}^{(-1)Q}=(\cL_{-1}\Qz+2\cH_{-1}\Qz+2\cQ_{-1})
\ket{0,-3} , \EE

\BE
\kc_{1,\ket{0,0}}^{(1)G}=\cG_{-1}\cG_0\cQ_0\,\ket{0,0} ,\qquad
\hat\kc_{1,\ket{0,0}}^{(1)G}=(\cL_{-1}\Gz-\cH_{-1}\Gz+2\cG_{-1}) \ket{0,0} ,
\EE

\BE
\kc_{1,\ket{0,0}}^{(0)Q}=\cL_{-1}\cG_0\cQ_0\,\ket{0,0} ,\qquad
\hat\kc_{1,\ket{0,0}}^{(0)Q}= (\cQ_{-1}\Gz-2\Qz\cG_{-1}+\cH_{-1}\Qz\Gz)
\ket{0,0} .
\EE
For each of these pairs the \svec\ on the left is the particular case,
for $\ctop=9$, of the one-parameter family of \svec s that exists for all 
values of $\ctop$, given in eqns. \req{ssvnl1} or \req{ssvnl2}. 
The \svec s on the right, however, 
are the particular cases of a one-parameter family of 
subsingular vectors that turn out to be singular just for $\ctop=9$.

\end{document}